\documentclass[apj]{emulateapj}
\usepackage{natbib}
\usepackage{times}
\usepackage{color}
\usepackage{epsf}
\usepackage{graphicx}
\usepackage{ulem}

\def\be{\begin{equation}}
\def\ee{\end{equation}}
\def\bea{\begin{eqnarray}}
\def\eea{\end{eqnarray}}

\newcommand{\simgt}{\lower.5ex\hbox{$\; \buildrel > \over \sim \;$}}
\newcommand{\simlt}{\lower.5ex\hbox{$\; \buildrel < \over \sim \;$}}
\newcommand{\oii}{{[{\sc O\,ii}]}}
\newcommand{\lya}{Ly \ensuremath{\alpha}}
\newcommand{\msun}{\ensuremath{M_{\odot}}}
\newcommand{\halpha}{H\ensuremath{\alpha}}

\newcommand{\spitzer}{{\it Spitzer}}
\def\gax{{$\mathrel{\hbox{\rlap{\hbox{\lower4pt\hbox{$\sim$}}}\hbox{$>$}}}$}}
\def\kms{\ensuremath{{\rm km s}^{-1}}}

\begin{document}

\title[Subaru PFS]{
Extragalactic Science,
Cosmology 
and Galactic Archaeology
with the Subaru Prime Focus Spectrograph (PFS)}
\author{Masahiro Takada$^1$, 
Richard Ellis$^2$, 
Masashi Chiba$^3$, 
Jenny E. Greene$^4$, 
Hiroaki Aihara$^{1,5}$,
Nobuo Arimoto$^6$,
Kevin Bundy$^1$,
Judith Cohen$^2$, 
Olivier Dor\'e$^{2,7}$,
Genevieve Graves$^{4}$,
James E. Gunn$^4$,
Timothy Heckman$^8$,
Chris Hirata$^2$, 
Paul Ho$^9$,
Jean-Paul Kneib$^{10}$, 
Olivier Le F\`evre$^{10}$,
Lihwai Lin$^9$, 
Surhud More$^1$,
Hitoshi Murayama$^{1,11}$,
Tohru Nagao$^{12}$, 
Masami Ouchi$^{13}$,  
Michael Seiffert$^{2,7}$,
John Silverman$^1$, 
Laerte Sodr\'e Jr$^{14}$, 
David N. Spergel$^{1,4}$,
Michael A. Strauss$^4$, 
Hajime Sugai$^1$,
Yasushi Suto$^{5}$,
Hideki Takami$^{6}$, 
Rosemary Wyse$^8$}
\altaffiltext{1}{
Kavli Institute for the Physics and Mathematics of the Universe (Kavli
IPMU, WPI), The University of Tokyo, Chiba 277-8583}
\altaffiltext{2}{California Institute of Technology, Pasadena, CA 91125,
U.S.A}
\altaffiltext{3}{Astronomical Institute, Tohoku University, Sendai,
980-8578}
\altaffiltext{4}{Department of Astrophysical Sciences,
Princeton University, Princeton, NJ 08544, U.S.A}
\altaffiltext{5}{Department of Physics, 
The University of Tokyo, Tokyo, 113-0033 }
\altaffiltext{6}{National Astronomical Observatory of Japan, 
Tokyo, 181-8588}
\altaffiltext{7}{Jet Propulsion Laboratory, California Institute of Technology, Pasadena, CA, U.S.A.}
\altaffiltext{8}{Department of Physics \& Astronomy, Johns Hopkins University, Baltimore, MD 21218, U.S.A.}
\altaffiltext{9}{Institute of Astronomy and Astrophysics, Academia
Sinica, Taipei 10617, Taiwan}
\altaffiltext{10}{Laboratoire d'Astrophysique de Marseille, 
F-13388 Marseille Cedex 13, France}
\altaffiltext{11}{
Physics Department, University of California, Berkeley and Lawrence Berkeley National Laboratory, Berkeley, California 94720, U.S.A
}
\altaffiltext{12}{
The Hakubi Center for Advanced Research, Kyoto University, Kyoto 606-8302}
\altaffiltext{13}{Institute for Cosmic Ray Research, The University of
 Tokyo, Chiba 
277-8582}
\altaffiltext{14}{
Instituto de Astronomia, 
Geof\'isica e Ci\^{e}ncias Atmosf\'ericas 
S\~ao Paulo, 05508-090, Brazil}

\begin{abstract}
The Subaru Prime Focus Spectrograph (PFS) is a massively-multiplexed
fiber-fed optical and near-infrared 3-arm 
spectrograph ($N_{\rm fiber}$=2400,
 380$\le \lambda\le 1260$nm, 1.3 degree diameter hexagonal field), 
offering
unique opportunities in survey astronomy. Following a successful
external design review the instrument is now under construction with
first light anticipated in late 2017. Here we summarize the science case
for this unique instrument in terms of provisional plans for a Subaru
Strategic Program of $\simeq$300 nights. We describe plans to constrain
the nature of dark energy via a survey of emission line galaxies
spanning a comoving volume of 9.3$h^{-3}$Gpc$^3$ in the redshift range
$0.8<z<2.4$. In each of 6 independent redshift bins, 
the cosmological
distances will be measured
to 3\% precision
via the baryonic acoustic oscillation scale,
and redshift-space distortion measures will be used  to constrain 
structure growth to 6\%
precision.
In the near-field cosmology program,
radial velocities and chemical abundances
of stars in the Milky Way and M31 will be used to infer the past
assembly histories of spiral galaxies and the structure of
their dark matter halos.
Data will be secured for $10^6$ stars in the Galactic thick-disk,
halo and tidal streams as faint as $V\sim 22$, including stars with $V < 20$
to complement the goals of the Gaia mission.
A medium-resolution mode with $R = 5,000$ to be implemented
in the red arm will allow the measurement of multiple $\alpha$-element
abundances and more precise velocities for Galactic stars, elucidating
the detailed chemo-dynamical structure and evolution of each of the
main stellar components of the Milky Way Galaxy and of its dwarf
spheroidal galaxies.
The M31 campaign will target red giant branch stars with 21.5$<V<$22.5, 
obtaining radial velocities and metallicities over an
unprecedented area of 65 deg$^2$.
For the extragalactic program, 
our simulations suggest the wide
wavelength range of PFS will be particularly powerful in probing the
galaxy population and its clustering over a wide redshift range. We
propose to conduct a color-selected survey of $1<z<2$ galaxies and AGN
over 16 deg$^2$ to $J\simeq$23.4, yielding a fair sample of galaxies
 with stellar masses above $\sim 10^{10}M_\odot$ at $z\simeq 2$.
A two-tiered survey of higher redshift
Lyman break galaxies and Lyman alpha emitters will quantify the
properties of early systems close to the reionization epoch. PFS will
also provide unique spectroscopic opportunities beyond these
currently-envisaged surveys, particularly in the era of Euclid, LSST and TMT.
\end{abstract}

%\vspace*{\stretch{3}}
%\begin{center}
%{\bf\Huge  Scientific Program and Requirements}\\
%%\vspace*{\stretch{2}}
%\vspace{2em}
%{\bf \large Richard Ellis \& Masahiro Takada: Co-Chairs of PFS Survey
% Committee}
%\vspace*{\stretch{2}}
%\vspace*{\stretch{2}}
%\vspace*{\stretch{2.5}}
%\end{center}

%\thispagestyle{empty}
%\include{member}
%\thispagestyle{empty}
%\tableofcontents
%\listoffigures
%\listoftables
%\newpage

\section{Introduction}

There is currently a major expansion in survey imaging capability via
the use of CCD and near-infrared detector mosaics on a wide range
of ground-based telescopes. Such imaging surveys provide accurate
photometric and other data to enable the study of gravitational lensing
signals which trace the distribution of dark matter and to conduct
census studies of Galactic structures and distant star-forming
galaxies. For over a decade it has been recognized that a similar
revolution would be provided by a massively-multiplexed
spectrograph on a large aperture telescope. Spectra provide precise radial velocities, metallicities
and emission line properties for faint and distant sources and enable
additional probes of cosmology. The main challenge in realizing this
second revolution has been access to a wide field telescope, essential
for efficient multi-object spectroscopy of panoramic fields, and the
cost of implementing the appropriate instrumentation.

A proposal to construct a Subaru Prime Focus Spectrograph (PFS) emerged
following the cancellation in May 2009 of the Gemini-sponsored
Wide-Field Multi-Object Spectrometer (WFMOS). WFMOS was envisaged as a
facility instrument on the Subaru telescope sharing the optics designed
for the new prime focus camera, Hyper Suprime-Cam (HSC). Two teams received
Gemini funding for a conceptual design study of WFMOS and, prior to
cancellation, a team led by Caltech and the Jet Propulsion Laboratory
(JPL) secured preliminary approval. Soon after, however, the
Gemini Board indicated they did not have sufficient funding to proceed
and the WFMOS project was terminated.

The
Kavli Institute for the Physics and Mathematics of the Universe (Kavli IPMU) at
the University of Tokyo submitted a proposal for stimulus funding to the
Japanese government in September 2009 using design concepts pioneered in
the WFMOS study led by Caltech and JPL. The successful outcome of this
proposal in early 2010 initiated the present PFS partnership which now
includes Caltech/JPL, Princeton and Johns Hopkins Universities, the
Laboratoire d'Astrophysique de Marseille,
Academia Sinica Institute of
Astronomy \& Astrophysics (ASIAA) Taiwan, 
the University of S\~ao Paulo and the Laboratorio Nacional de Astrof\'isica
in Brazil.

In addition to the leadership funding provided by Kavli IPMU, four 
important milestones 
have enabled progress and led to the decision to commence 
construction. Firstly, in January 2011 the Subaru Users Meeting endorsed the PFS project 
as a next-generation instrument for the Subaru Prime Focus, recognizing
the international PFS team and its responsibilities. This decision led to
the establishment of a PFS project office at Kavli IPMU in early 2011 and 
an allocation of funds and manpower by the National Astronomical Observatory of Japan (NAOJ) 
towards integration, commissioning and survey operations. 
A second milestone followed the MOU in December 2011 between the Director-General of NAOJ and the Director of Kavli IPMU, that anticipates a Subaru Strategic Program for PFS providing up to 300 nights of observing time for the PFS team in collaboration with the Japanese astronomical community.
These developments provided the essential impetus for the science plans and
technical requirements defined in this article. The third milestone was a
successful Conceptual Design Review (CoDR) held in March 2012 which triggered
the decision to commence construction with first light anticipated in
2017. The fourth milestone was a successful Preliminary Design Review (PDR)
held in Feb 2013. 
The CoDR and PDR documentations 
included a detailed science case for PFS and a list 
of technical requirements. This article reproduces these for general
interest.

\begin{figure*}[t]
\begin{center}
\includegraphics[width=15cm, angle=0]{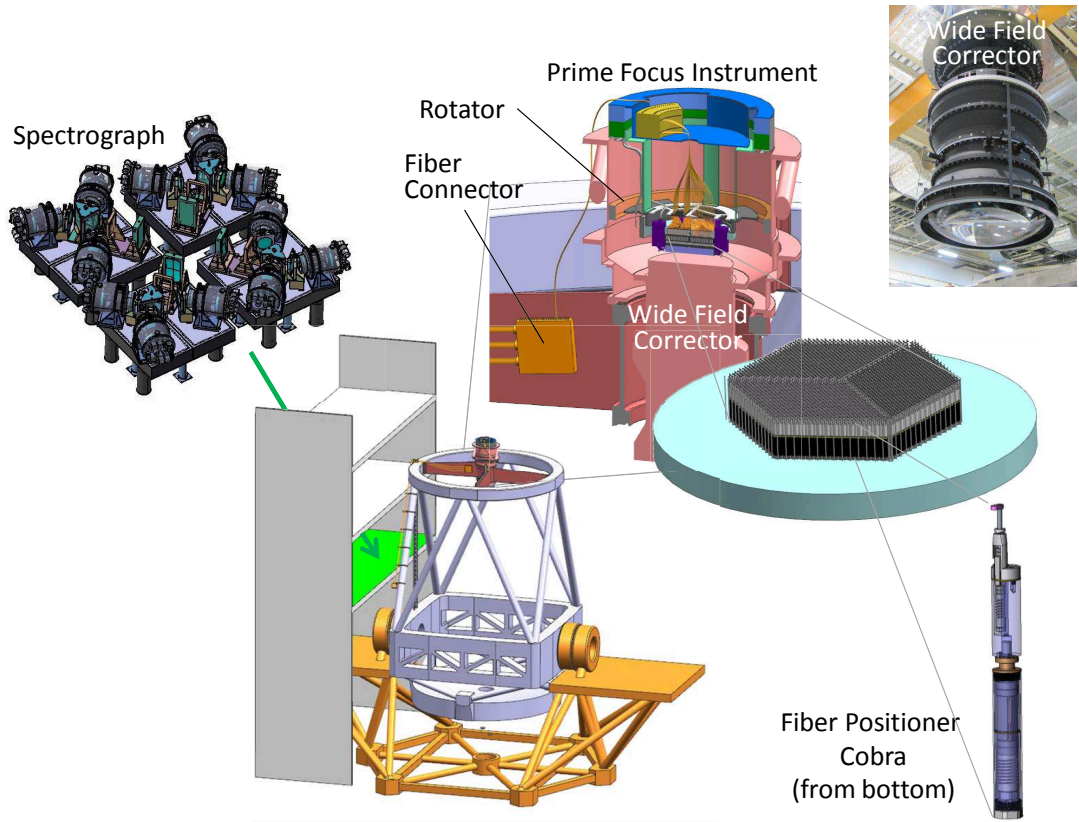}
\caption{A brief overview of the baseline design of 
PFS instruments, which consist of
 components 
{\bf Wide
 Field Corrector}, {\bf Field Rotator}, {\bf Prime Focus Unit}, and {\bf
 Fiber Positioner}. A {\bf Fiber Connector} relays light to four identical
fixed-format 3-arm twin-dichroic all-Schmidt {\bf Spectrographs} providing continuous wavelength 
coverage from 380nm to 1.26$\mu$m. 
} 
  \end{center}
\label{fig:design}
\end{figure*}
 
This article is not intended to provide a technical description of PFS but
a brief overview is helpful (see Fig.~\ref{fig:design}). 
Further technical details
of the instrument and its current design can be found at 
{\sf http://sumire.ipmu.jp/en/2652} \citep{Sugaietal:12}.
%{\it Further technical details
%of the instrument and its current design may be found at...??}.
PFS is designed to allow simultaneous low and intermediate-resolution spectroscopy of 2400 astronomical
targets over a 1.3 degree diameter hexagonal field. It shares the {\it Wide Field Corrector} and associated 
Field {\it Rotator} and Hexapod already constructed for the HSC. An array of 2400 optical fibers 
is in the {\it Prime Focus Instrument} and each fiber tip position is controlled in-plane by a two-stage
piezo-electric {\it Fiber Positioner} Cobra system. Each fiber can be positioned within a particular patrol region such that 
these patrol regions fully sample the 1.3 degree field.  A {\it Fiber Connector} relays light to four identical
fixed-format 3-arm twin-dichroic all-Schmidt {\it Spectrographs} providing continuous wavelength 
coverage from 380nm to 1.26$\mu$m. The blue and red channels will use two Hamamatsu 2K$\times$4K 
edge-buttable fully-depleted CCDs (as in HSC). The near-infrared channel will use a
new Teledyne 4RG 4K$\times$4K HgCdTe 1.7$\mu$m cut-off array.
We also plan to have a medium resolution mode with $R\simeq 5000$ for
the red channel, which is feasible by using a simple
grating/grism exchange mechanism at the red-channel spectrograph.

The present article describes the detailed scientific case for PFS in the context of a Subaru
Strategic Program (SSP) of $\simeq$300 nights of observing time. Since such a program
would not be implemented until 2017 at the earliest, the main motivation in formulating
the team's plans at this stage is in providing a list of key science requirements for the
technical design. In Sections \ref{chap:cosmology} -- \ref{chap:galaxy} we describe 3 key
components of the science case for PFS that will likely form the basis of the Subaru Strategic 
Program. For each of these 3 cases we provide a summary, a detailed science justification
and survey strategy as well as the flow-down from these to the technical requirements for the 
instrument. We summarize these science requirements for PFS in more detail in Section \ref{chap:sr} and discuss 
some outstanding issues in Section \ref{chap:issues}.

\section{Cosmology}
\label{chap:cosmology}

{\it Summary:  PFS will be remarkably powerful in spectroscopic surveys
of faint galaxies because of its large multiplex gain and the 8.2 meter 
aperture of the Subaru telescope. The extended wavelength coverage 
provided by the red and near-infrared spectrograph arms (650 -- 1260~nm) 
will permit a unique survey of \oii\ emission-line galaxies extending over 
the redshift range $0.8\le z\le 2.4$. As large-scale structures remain 
in the linear regime at high redshift, such a survey will give detailed new 
information on the cosmological parameters as well as the growth rate 
of structure formation. This combination will provide a valuable test
of alternative models of gravity on large scale which may provide a possible 
explanation for dark energy. Multi-color data planned to arrive from the 
HSC imager will be used to select target galaxies for 
spectroscopy and the expected high throughput should yield a $\simeq$75\% 
success rate of detecting \oii\  emission at $S/N>8.5$. Herein, we propose 
to conduct a 100~ night cosmological survey over 1400 deg$^2$, sampling 
galaxies within a comoving volume of 9 $({\rm Gpc}/h)^3$ over $0.8\le z\le 2.4$. This
will complement the 
lower redshift survey being 
undertaken by the SDSS BOSS collaboration.

The primary goals of the PFS cosmology survey are to: (1) measure the
Hubble expansion rate and the angular diameter distance to 3\%
fractional accuracies in each of 6 redshift bins over $0.8<z<2.4$ via
the baryonic acoustic oscillation (BAO) method, (2) use the distance
measurements for determining the dark energy density parameter
$\Omega_{\rm de}(z)$ to about 7\% accuracy in each redshift bin, when
combined with lower redshift BAO measurements, (3) use the geometrical 
constraints to determine the curvature parameter $\Omega_{K}$ to $0.3\%$ accuracy, 
and (4) measure the redshift-space distortion (RSD) in order to reconstruct the growth
rate of large-scale structure to 6\% accuracy since a redshift $z=2.4$.
These PFS measurements of the large scale galaxy distribution can be combined
with complementary weak lensing information from the HSC survey in order
to significantly improve the cosmological and structure growth constraints 
and reduce uncertainties arising from galaxy bias and nonlinear effects
that are otherwise major sources of systematic error in spectroscopic
surveys.}

\subsection{Cosmology Objectives}

The accelerated expansion of the Universe is the most intriguing problem in
cosmology. It either requires the introduction of a mysterious form of energy,
``dark energy'', or it could signal a breakdown of Einstein theory of General Relativity 
on cosmological scales. To distinguish between these and other possibilities
requires precise observational constraints on both the expansion history of 
the universe and the growth rate of large-scale structure. 

Measurements of galaxy clustering statistics are
 one of the most powerful means of
addressing the nature of dark energy. The tight coupling between
baryons and photons prior to the decoupling epoch of $z\simeq 1100$ leaves a
characteristic imprint on the pattern of galaxy clustering on large scales -- the 
so-called baryonic acoustic oscillation (BAO) scale. As the BAO 
length scale is precisely constrained to be $
147\pm 0.6~$Mpc from cosmic microwave background (CMB) experiments
\citep[][]{WMAP7,PlanckCosmo:13}, it offers a standard ruler by which we can infer the angular 
diameter distance and the Hubble expansion rate from the observed 
correlation function of the galaxy distribution.
The BAO scale is in the linear or weakly-nonlinear density regime and 
thus provides a robust geometrical test. Furthermore, if uncertainties
arising from galaxy bias can be removed or accurately modeled, we can use
the amplitude and shape information of the galaxy correlation function in
order to constrain cosmological parameters as well as the growth rate of
structure formation.

Recognizing this, the main scientific questions we seek to address with 
the PFS cosmology survey are:
\begin{enumerate}
\item {\it Is the cosmic acceleration caused by dark energy or does
       it represent a failure of Einstein's theory of gravity on
      cosmological length scales?}
\item {\it What is the physics of the early universe that generates the
      primordial fluctuations as the seed of large-scale structures?}
%\item {\bf What is the absolute mass scale of neutrinos?}
\end{enumerate}
To address these fundamental questions, the main goals for the
PFS cosmology survey are to:
\begin{itemize}
\item Constrain the angular diameter distance and the Hubble expansion
      rate via the BAO experiment to a precision comparable with, or
      better than, existing, ongoing or planned BAO surveys.
\item Derive the BAO constraints in a redshift range that is
      complementary to those probed by the existing or planned BAO
      surveys on the time scale of the PFS survey. 
\item Utilize the unique capabilities of the 8.2m Subaru Telescope and
      the PFS spectrograph for maximizing cosmological science.
\item Use the shape and amplitude of galaxy correlation
      function in order to constrain cosmological parameters as well as
      the growth rate of structure formation.
\item Combine the weak lensing information, delivered from the HSC
      survey, with the PFS cosmology survey in order to improve the
      cosmological constraints by calibrating systematic uncertainties
      that cannot be resolved by either of the PFS and HSC surveys alone.
\end{itemize}

\subsection{PFS Cosmology Survey}
\label{cosmology:sec:survey}

\begin{deluxetable*}{l|l|l|l}
\tablewidth{0pt}
\tabletypesize{\footnotesize}
\tablecaption{Instrumentation parameters}
\startdata \hline\hline 
Number of fibers & \multicolumn{3}{l}{2400 (600 for each spectrograph)} \\
Field of view & \multicolumn{3}{l}{1.3 deg (hexagonal -- diameter of circumscribed circle)} \\ 
Field of view area & \multicolumn3l{1.098 deg$^2$} \\
Fiber diameter & \multicolumn{3}{l}{1.13$''$ diameter at the field center; 1.03$''$
     at the edge}\\ \hline
&Blue arm & Red arm & IR arm\\ 
\cline{1-4}
Wavelength coverage [nm] & 380--670 & 650--1000  & 970--1260\\
Spectral resolution $\lambda/\Delta \lambda$ & 1900 & 2400 & 3500 \\
Pixel scale [\AA/pix]& 0.71 & 0.85 & 0.81\\
Read-out noise [e$^{-}$ rms/pix]& 3 & 3 & 4$^{\rm a}$ \\
Detector type/read-out mode & CCD & CCD & HgCdTe/SUTR \\
Thermal background [e$^{-}$/pix/sec] & None & None & 0.013
%(0.01 in beam plus 0.003 from dewar)
\\
Dark current [e$^{-}$/pix/sec]& $3.89\times 10^{-4}$ & $3.89\times 10^{-4}$ & 0.01\\
Spectrograph image quality [$\mu$m rms/axis] & 14 & 14$^{\rm b}$ & 14 \\
\hline
Sky continuum & \multicolumn3l{21.56 mag AB/arcsec$^2$ @ 1 $\mu$m at zenith$^{\rm c}$} \\
OH line brightness & \multicolumn3l{16.6 mag AB/arcsec$^2$ @ $J$ band at zenith$^{\rm c}$} \\
Moonlight & \multicolumn3l{None (dark time)} \\
Atmospheric extinction & \multicolumn3l{Variable; continuum is 0.05 mag/airmass @ $\lambda>0.9 \mu$m} \\
Instrument throughput & \multicolumn{3}{l}{Based on the Preliminary
 Design Review studies}\\
Grating wings & \multicolumn{3}{l}{Lorentzian with $\frac13$ of the true number of lines}\\
Diffuse stray light & \multicolumn{3}{l}{2\% of the total light reaching the detector} \\
Sky subtraction residuals & \multicolumn{3}{l}{2\% per pixel$^{\rm c}$}
\enddata
\tablenotetext{a}{
Per sub-exposure; if the NIR channel is not reset between exposures, the
 actual assumption corresponds to $4\sqrt2=5.6$ $e^-$ rms.}
\tablenotetext{b}{
Defocus within the thick CCD is considered separately.}
\tablenotetext{c}{
Equivalent to 1\% per 4-pixel resolution
 element.} 
\tablecomments{
PFS instrumentation parameters and associated assumptions used for estimating an expected
 signal-to-noise ratio for an observation of emission-line galaxies. 
\label{tab:pfs_spec}
}
\end{deluxetable*}

Here we describe the parameters of the PFS cosmology survey that
are required to meet the above scientific goals.

Firstly, we will consider which type of galaxies to target with PFS. 
Given the optical and near infrared wavelength coverage of PFS, 
\oii\ emission-line galaxies (ELG; \oii\ =3727\AA) are particularly 
useful tracers allowing an efficient survey out to high redshift beyond
$z=1$, a redshift range that is difficult to probe with 4m-class telescopes. Luminous red
galaxies (LRGs) are a further potentially-useful tracer of large-scale structure as
studied by the SDSS survey, but at $z\simgt 1.4$ they reveal weaker spectral 
features and are less abundant per unit volume. In the following we focus on
ELGs to explore an optimal survey design. We may retain LRGs in future 
considerations of our survey plans, but their study is not required to meet 
the PFS cosmology objectives defined above.

\subsubsection{Sensitivity of the PFS spectrograph}
\label{cosmology:sec:sensitivity}

\begin{figure*}
\begin{center}
\includegraphics[width=9cm, angle=-90]{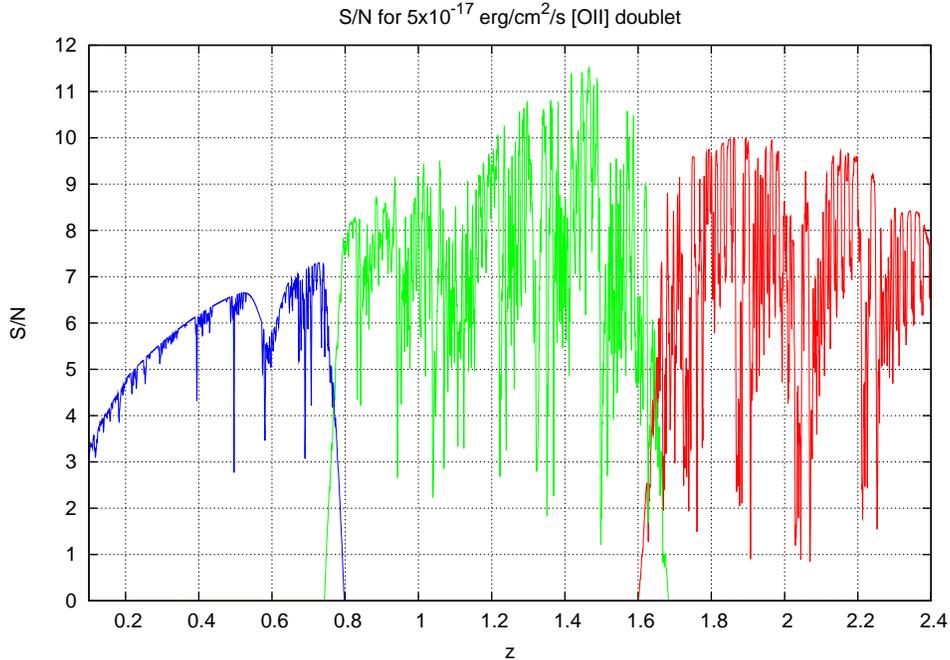}
  \end{center}
\caption{ Expected signal-to-noise ($S/N$)
ratio for measuring the \oii\ emission line as a function of 
redshift; the blue, green and red curves show the results for the PFS
blue, red and IR arms in Table~\ref{tab:pfs_spec}, respectively, for an
total emission line flux of $5\times 10^{-17}$ erg/cm$^2$/s. To
properly account for the uncertainties, we assumed the
instrumentation parameters of the current baseline design listed in
Table~\ref{tab:pfs_spec}, an observation at the edge of the focal plane, and included
the sky emission/absorption and the Galactic dust extinction of
$E(B-V)=0.05$ and 26 degrees for the zenith angle of the telescope.
This computation assumes 15 min total exposure (split into two
exposures; $450{\rm sec}\times 2$), $\sigma_v=70$ km/s for the velocity
dispersion (the intrinsic line width), and 0.8$''$ for the seeing FWHM. We
also accounted for the finite galaxy size relative to the seeing
profile and the fiber size, assuming an exponential profile with scale
radius 0.3$''$ for the emission-line region
(about 3.5~kpc$/h$ for a
 galaxy at $z=1$). 
 Note that $S/N$ is estimated
by the root-sum-square of the spectral pixels (i.e. it is a matched filter combining 
both doublet members). The current design allows a significant detection
of  \oii\  emission line over a wide range of redshift, up to $z\simeq 2.4$
with near-equal sensitivities of the red and NIR arms.
\label{fig:sn}} 
\end{figure*}

To estimate the feasibility of PFS for a wide-field survey of
ELGs, we have studied the expected performance of
measuring a \oii\ line of a galaxy in our targeted redshift range for a
representative exposure time during the dark lunar phase. In doing so, we properly account 
for the sky emission (continuum plus OH emission lines) and absorption 
as well as the instrumentation parameters for the current baseline 
design (as listed in Table~\ref{tab:pfs_spec}). 

Fig.~\ref{fig:sn} shows the expected signal-to-noise ratio of 
the \oii\ line as a function of 
redshift, measured with each of the blue, red and near-infrared (NIR) arms of PFS.
As a working example, here we assume $f_{\rm [OII]}=5\times 10^{-17}~{\rm erg/s/cm}^2$
for the total flux of the 
\oii\ doublet, 15min for the exposure time, 0.8$''$ seeing size and 
$E(B-V)=0.05$ for the Galactic dust extinction, respectively. The galaxy radial profile 
is assumed to be an exponential disk with a half-light radius of 0.3$''$
(about 3.5~kpc$/h$ for a
 galaxy at $z=1$). 
Note that for the galaxy
yield forecasts, we use half-light radii from the COSMOS Mock Catalog
\citep[][]{Jouveletal:09}, and re-compute the fiber aperture correction for each
galaxy.\footnote{This assumes that the \oii\ emission traces the $i$-band
continuum in which the galaxy sizes were measured.}
We have assumed the 15 minute integration is split into 2 sub-exposures for 
cosmic ray (CR) detection in the CCD channel.
The NIR channel will 
perform CR rejection by processing of the frames acquired during sample-up-the-ramp 
(SUTR) mode. The cosmology ETC assumes 4 $e^-$ read noise per sub-exposure 
(appropriate for $\sim 90$ samples along a 450sec ramp).
We will probably {\it not} reset the NIR channel in between
sub-exposures, so 
we assume an 
overall read noise of $4\sqrt 2 = 5.6$ $e^-$ per pixel for the following study.

In addition to throughput and sky brightness considerations, we have considered
several other potential limitations. Their amplitude is difficult to estimate, but they have been
important for previous spectrographs and so we make an explicit allowance
for them so as to adopt a conservative approach. The systematic 
sky subtraction residuals and small-angle stray light are very important factors 
in the study of galaxy spectra where \oii\ is partially blended with a sky line.  
Diffuse stray light is a concern when \oii\ lies in a cleaner part of the NIR spectrum.

\begin{itemize}
\item {\it Systematic sky subtraction residuals} -- These are modeled by adding a ``noise'' term 
corresponding to some percentage of the sky counts in each spectral pixel. We currently set this to 2\% 
of the brightest of the pixel and its neighbor on either side (equivalent to 1\% sky subtraction accuracy 
on a 4-pixel resolution element).
\item {\it Small-angle stray light} -- We assign to the grating an
 effective number of lines that is 
%$\frac13$ 
$1/3$ of the actual number.
\item {\it Diffuse stray light} -- We take 2\% of the OH line flux incident on the detector and uniformly spread 
it over all pixels. (This may be appropriate for a detector that reflects 10\% of the incident radiation, and then 
there are many surfaces that could potentially reflect this radiation back. Refining this parameter will be a 
priority since the $S/N$ forecasts degrade rapidly if it gets worse.)
\end{itemize}

Continuing our conservative approach, we assumed the instrumentation
throughput at the edge of the focal plane and 26 degrees for the zenith
angle. (The latter corresponds to observations at declination 5$^\circ$S,
the southern boundary of the HSC survey region, and $\pm0.5$ hours away
from transit.) 

Fig.~\ref{fig:sn}
shows that the current design of PFS allows a
significant detection of the \oii\ line over a wide range of redshift,
up to $z\simeq 2.4$. Most importantly, the baseline design provides
near-equal sensitivity of the red and NIR arms for measuring the
\oii\ line for the same exposure time.  Hence PFS can execute a 
cosmological survey very efficiently over a wide range of redshift, 
provided sufficiently bright ELGs are available for study (see below). 
  
In the following analysis, we set a threshold of $S/N=8.5$ (matched
filter) for detection of an ELG. In principle, it may be possible to
accept less significant detections. However given the uncertainties 
in the airglow and the early stage of the instrument design, we consider
it prudent to leave some margin in $S/N$.

\subsubsection{Target selection of emission-line galaxies}
\label{sec:target}

\begin{figure*}
\begin{center}
\includegraphics[angle=-90,width=16cm]{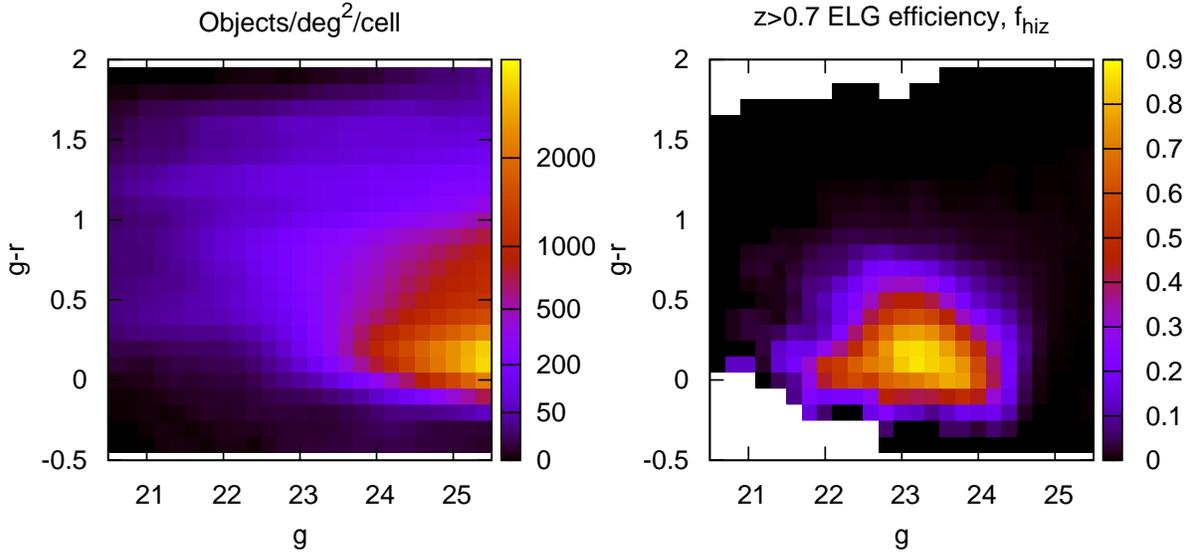} 
\end{center}
\caption{\label{fig:grplot}
 {\it Left panel}: The distribution of objects in the COSMOS
Mock Catalog in the color-magnitude diagram. {\it Right}: The fraction
of objects in each cell that are $z>0.7$ ELGs with \oii\ doublets
detectable at $\ge 8.5\sigma$ in PFS in $2\times 7.5$ min exposures. 
The third dimension ($r-i$) 
%cannot be displayed 
is not shown 
on this 2D plot, but
allows us to select lower or higher redshift galaxies within the PFS survey range.}
\end{figure*}
\begin{figure}
\begin{center}
\includegraphics[angle=-90,width=0.48\textwidth]{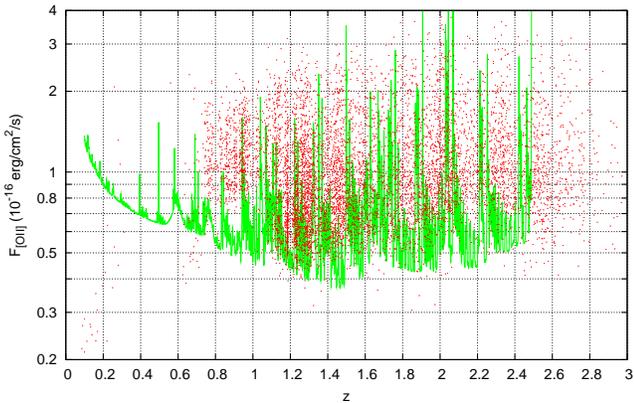} 
\end{center}
\caption{\label{fig:snplot2}
The sensitivity of PFS (green curve; 8.5$\sigma$, $2\times 450$s exposures, dark time) to the \oii\ 
doublet at $r_{\rm eff}=0.3''$ and 1:1 line ratio, versus the selected
 targets (red points). 
Note that the redshift $z=2.4$ corresponds to
 the long wavelength end of the NIR arm. 
Most targets will yield successful redshifts,
but some are lost within the atmospheric emission or absorption lines, a few are at $z\ge2.5$,
and there is a small number of faint blue nearby objects (lower-left
 corner) for which we cannot detect \oii.
Note that the line ratio ($F_{3726}:F_{3729}$) and effective radius are re-computed for each galaxy 
in the COSMOS Mock Catalog, and hence the sensitivity curve drawn does not correspond to an exact boundary
between detections and non-detections.}
\end{figure}

We now address how to optimally select
ELGs as suitable targets in the proposed redshift range. We will assume 
that we can use the
multi-color imaging data of the planned HSC survey
which will be executed ahead of the PFS survey. The currently-planned 
HSC survey will reach $i\simeq 26$
($5\sigma$ for a point source and 2'' aperture), in the 5 passbands
$grizy$ over $\sim1500$ square degrees.

As seen in Fig.~\ref{fig:sn}, if we target ELGs over the wide redshift range $0.8\le z\le 2.4$,
the wide wavelength coverage of red and
NIR arms allows a very efficient selection of [OII] emission-line
galaxies. A $g-r$ color cut is ideal for selecting galaxies in this redshift range:
if an object is blue ($g-r<0.3$), then it likely has no spectral breaks in the
$g$ and $r$ bands -- this means the redshift
is high enough for the Balmer/4000\AA\ break to have redshifted beyond the $r$
band, but the Lyman break has not yet entered the $g$ band. Furthermore, $g-r<0.3$
implies a blue rest-frame UV slope, which has a strong correlation with
the star-formation activity that produces \oii\ emission.

To estimate the efficiency of various target selection algorithms, 
we used the COSMOS
Mock Catalog \citep[][]{Jouveletal:09}, where fluxes of various
emission lines of each galaxy are estimated based on physical parameters
(SFR, stellar mass and metallicity) using the COSMOS 30 passband
photometric data and zCOSMOS spectroscopic data. We have chosen the
preliminary target selection cuts:
\begin{eqnarray}
&&22.8\le g \le 24.2\mbox{ AND } -0.1<g-r<0.3 \nonumber\\
&&\mbox{ AND NOT } 
(g>23.6 \mbox{ AND } r-i>0.3). 
\label{eq:color-cut}
\end{eqnarray}
The HSC depth ($g,r,i\simeq 26$ mag AB at 5$\sigma$) is sufficient
to find the target galaxies {\it and} to provide accurate $g-r$ and $r-i$
colors. The ELGs in the redshift range $0.8\simlt z\simlt 2.4$ are primarily selected from the
color cut $-0.1<g-r<0.3$, and the $g$-magnitude cut gives preference to
bright objects while reducing low-redshift contamination, as can be
seen in Fig.~\ref{fig:grplot}. The condition on $r-i$ for fainter magnitudes is designed to tilt the
redshift distribution in favor of more objects at $z>1$.
We can further divide the targets into a ``bright subsample'' ($g\le 23.9$) and
a ``faint subsample'' ($g>23.9$), with the brighter targets prioritized
when 
we wish to increase the success rate.

In the COSMOS Mock Catalog there are $7847$ target galaxies available per PFS
field-of-view (1.098 deg$^2$ for the 1.3$^\circ$ FoV diameter).  Hence
there are a sufficient number of target galaxies compared to the number
of fibers ($N$=2400) for the baseline design.  The green-solid curve 
in Fig.~\ref{fig:snplot2}
shows the redshift dependence of \oii\ flux with $S/N\ge 8.5$ for 
a 15 minutes exposure. To estimate the expected $S/N$ for each galaxy, we 
employed the same method used in Fig.~\ref{fig:sn} and also used the galaxy 
size information and \oii\ doublet ratio available from the COSMOS mock catalog. 

\begin{figure}[t]
\begin{center}
\includegraphics[width=0.36\textwidth, angle=-90]{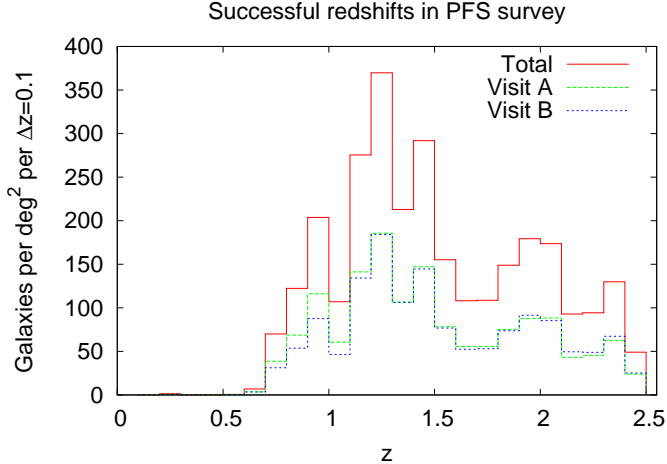} 
 \caption{The distribution of successful redshifts (\oii\
detected at $S/N>8.5$) for the proposed PFS cosmology survey, including
breakdown into the two visits.  The jagged features in the curves
reflect the effect of sampling variance of large-scale structures in the COSMOS
field due to the finite survey area (the mock is based on the data of
1.24 square degrees). We refer to the two visits as ``Visit A'' and ``Visit
B'' (see text for details), respectively, where we preferentially select
brighter targets with $g<23.9$ in Visit A in order to have some flexibility 
between dark/grey nights.\label{fig:dNdz}}
  \end{center}
\end{figure}

Assuming 2400 fibers in the focal plane as in Table~\ref{tab:pfs_spec}
and using the results in  Fig.~\ref{fig:grplot}, we can estimate a success
rate of finding \oii\ emission-line galaxies among the target galaxies.
We show the redshift and \oii\ flux distributions of the targets in Fig.~\ref{fig:snplot2},
and the redshift histogram of the successful \oii\ detections in Fig.~\ref{fig:dNdz}.
Of the targets, 74\%\ have successful \oii\ detections, 6\%\ are faint blue local ($z\le 0.3$) objects,
6\%\ fail due to an \oii\ feature that falls off the red end of the spectrograph, and
the remaining 14\%\ fail due to some combination of too faint \oii\ feature or
overlap with  an atmospheric emission or absorption complex.
To have a sufficiently dense sampling of galaxies to trace large-scale
structures in each redshift slice, we will need multiple visits of each field;
our BAO forecasts found that 2 visits gave the best constraints.

To obtain a reliable estimate of the number of observable targets, we took into account the fiber
allocation efficiency assuming a Poisson distribution of target
galaxies on the sky, which should be a good approximation for a given
wide redshift coverage. We conservatively assume non-overlapping patrol
zones 
between the different fibers.\footnote{That is, we allow a galaxy only to be observed
by the nearest fiber, even though 21\%\ of galaxies are in overlaps and could
potentially be observed by either of two fibers.} The fiber assignment algorithm is
designed to put the easier, i.e. brighter,
targets in one of the visits (``Visit A'') and then the harder
targets that require better conditions in another (``Visit B''). We
divide our targets into two tiers -- the bright ($g\le 23.9$) and faint ($g>23.9$) subsamples.
The fiber assignment logic within each patrol zone is then:
\begin{itemize}
\item If at least two bright targets are available, one is assigned to Visit A and another to Visit B.
\item If one bright target and at least one faint target are available, then the bright target is observed in Visit A and one of the faint targets is observed in Visit B.
\item If one bright target and no faint targets are available, then the bright target is observed in Visit A and the fiber becomes a sky fiber in Visit B.
\item If more than two bright targets, although unlikely, are available,
      then the brightest target is observed in Visit A and the next
      brightest target is observed in Visit B.
\item If no bright targets are available and there are at least two faint targets, then a faint target is observed in both Visits A and B.
\item If no bright targets are available and there is only one faint target, then the fiber becomes a sky fiber in Visit A and the faint target is observed in Visit B.
\item If no targets are available, then the fiber is a sky fiber in both Visits A and B.
\end{itemize}
This algorithm produces a roughly balanced fraction of sky fibers in the two visits.
The predicted allocations of the fibers are:
\begin{itemize}
\item Visit A: 85\% bright targets, 6\% faint targets, 9\% sky.
\item Visit B: 56\% bright targets, 33\% faint targets, 11\% sky.
\end{itemize}
Thus this leaves about 240 sky fibers in each visit for calibrating
 the sky spectrum. 

Note that the two visits could be scheduled in either order. 
Since Visit A has the brighter targets, it can achieve a high success fraction under
worse conditions than Visit B. We have therefore assumed that Visit B
takes place during dark time, whereas Visit A is scheduled
on a night of 7 days from the New Moon (but at least $45^\circ$ away  
from the moon).
The exposure times in both cases are kept at $2\times 450$sec. The predicted redshift success rate for a 
$S/N>8.5$ threshold is 75\% (Visit A)\footnote{This would rise to 79\% if Visit A were during dark time.} or 73\% (Visit B).

An alternative to the baseline (using gray time for Visit A) would be to use dark time only for the cosmology survey, 
and shorten the exposure time for Visit A, thereby reducing the total number of nights required but using time 
that may be in high demand by other programs. This trade will be made when we design an integrated observing 
schedule for PFS.
% (see ).
%Section 6).

\begin{deluxetable*}{l|cccccc}
\tablewidth{0pt}
\tabletypesize{\footnotesize}
\tablecaption{\label{tab:survey}PFS Cosmology Survey Parameters}
\startdata \hline\hline 
%\begin{table*}
%\begin{center}
%{PFS Cosmology Survey Parameters}\\
%\begin{tabular}{l|cccccc}\hline\hline
 & $V_{\rm survey}$ & $N_g$ & $\bar{n}_g$ & bias & $\bar{n}_gP_g(k)$ & $\bar{n}_gP_g(k)$ \\
redshift & $[{\rm Gpc}/h]^3$ & per field
& $[10^{-4}(h/{\rm Mpc})^3]$ & $b_g$ & $k=0.1h/$Mpc 
& $k=0.2h/$Mpc 
\\ \hline
$0.6<z<0.8$ & 0.59 & 85 & 1.9  &1.18 & 0.74 & 0.25\\
$0.8<z<1.0$ & 0.79 & 358 & 6.0 &1.26 & 2.23 & 0.74\\
$1.0<z<1.2$ & 0.96 & 420 & 5.8 &1.34 & 2.10 & 0.68\\
$1.2<z<1.4$ & 1.09 & 640 & 7.8 &1.42 & 2.64 & 0.87\\
$1.4<z<1.6$ & 1.19 & 491 & 5.5 &1.50 & 1.78 & 0.59\\
$1.6<z<2.0$ & 2.58 & 598 & 3.1 &1.62 & 0.95 & 0.31\\
$2.0<z<2.4$ & 2.71 & 539 & 2.7 &1.78 & 0.76 & 0.25
\enddata
\\ \hline\hline
%\end{tabular}
\tablecomments{ 
The leftmost column shows the redshift range of each
slice, and the other columns show the comoving volume ($V$), the number
of [OII] galaxies per field ($N_g$), the mean comoving number density
($\bar{n}_g$), the linear bias parameter ($b_g$) and the values of
$\bar{n}_gP_g(k)$ at $k=0.1$ and $0.2h/{\rm Mpc}$ for each slice,
respectively.  The survey volume is for a survey area of $1464$ square
degrees, which is estimated assuming 15 min of open-shutter time per
visit, 2 visits per field, 3 min overhead per visit, and 100 clear
nights. For comparison, the BOSS BAO survey has the survey parameters:
$10000$ sq. degrees area coverage over $0.4<z<0.7$, $V_{\rm survey}=4.4 ~({\rm
Gpc}/h)^3$, $\bar{n}_g=3\times 10^{-4}(h/{\rm Mpc})^{3}$, $b_g=2.3$ and
$\bar{n}_gP_{g}(k=0.1~h/{\rm Mpc})\simeq 5$.
% \label{tab:survey}
}
%}}
\end{deluxetable*}
%\end{center}
%\end{table*}

\subsubsection{Survey Strategy}

Using the results of target selection in Fig.~\ref{fig:dNdz}, we 
have adopted parameters for the PFS cosmology survey summarized in
Table~\ref{tab:survey}.  Since our primary observable is the galaxy
two-point correlation function or the galaxy power spectrum, the key
factors that govern the results are the geometrical volume surveyed 
and the ratio of clustering power to shot noise, $\bar n_gP_g$.  To have a galaxy power
spectrum measurement that reaches the sampling variance limit for our
volume and is not degraded by shot noise, the number density of galaxies 
must satisfy $\bar{n}_gP_g\simgt 1$ at BAO scales.
As given by the
columns of Table~\ref{tab:survey}, the PFS survey we are proposing has
$\bar{n}_g P_g\simgt \mbox{ a few}$ and slightly less than 1 at $k=0.1$ and
$0.2~h/{\rm Mpc}$, respectively, over the entire target redshift range. 
With only one visit per field, these numbers are about a factor 2
smaller than in Table~\ref{tab:survey}. On the other hand, if we have
more than two visits, the survey area we can cover for a given number of
nights becomes smaller. Incorporating multiple visits ensures more flexibility 
in optimizing the survey, for example in including a mixture of targets in 
different magnitude ranges.  

We have assumed that the bias factor for the ELGs is given by $b_g(z)=0.9+0.4z$. 
This specific function was a fit to semi-analytic models \citep[][]{Orsietal:10}, but 
compares very well to real data: e.g. the DEEP2 ``main blue'' sample has a 
measured bias of $b_g = 1.28\pm 0.04$
at $z=0.9$ \citep[][]{Coiletal:08}.\footnote{This was normalized to a
$\sigma_8=0.9$ model; with the lower $\sigma_8$ now favored, the bias
would be higher, i.e. ``better'' for the BAO analysis.}  We have much
less information about clustering of ELGs at redshifts beyond the DEEP2
survey, but H$\alpha$ emitters at $z=2.23$ have a correlation length of
$r_0 = 3.1\pm 0.7$ Mpc$/h$ \citep[][]{Sobraletal:10}, implying a bias of $b_g
= 1.87^{+0.24}_{-0.26}$.

The power spectrum measurement accuracy depends also on the area
coverage. In order for the PFS survey to have a constraining power on
cosmological parameters comparable with the existing or planned BAO
surveys, we need a sufficiently large area coverage. We have found that,
if about 100 clear nights are allocated to the PFS cosmology survey, it
can meet our scientific goals. Hence we assume 100 clear nights for the
following analysis, and the total area covered is estimated as
\begin{eqnarray}
&&\hspace{-2em}\frac{
100~[{\rm nights}]\times 8~[{\rm hours}]\times 60~[{\rm min}]
}{2~[{\rm visits}]\times (15~[\rm min]+3~[\rm min])}
\nonumber\\
&&\hspace{2em}
\times 1.098
~[\mbox{sq. degrees per FoV}] = 1464~\mbox{sq. degrees}.
\end{eqnarray}

Here we conservatively assumed 3min overhead for each new pointing, which
covers readout, slewing and the time for accurate fiber positioning, 
and assumed that 8 hours per night are available for observation on
source. The comoving volume in each redshift slice is given in
Table~\ref{tab:survey}. 
The total volume 
is about $9.3~({\rm Gpc}/h)^3$, a factor 2 larger volume than the SDSS
BOSS galaxy survey, which is about $4.4~{({\rm Gpc}/h)^3}$.  A notable strength
of the PFS survey is that it probes large-scale structure in higher
redshifts, where the fluctuations are largely in the linear regime and
therefore allow a cleaner estimation of cosmological parameters. In fact
the genuine cosmological power comes from the volume in Fourier
space; the effective volume at each wavenumber is given as $V_{\rm
eff}(k)=\left[\bar{n}_gP_g(k)/(1+\bar{n}_gP_g(k))\right]^2V_{\rm
survey}$, where
$V_{\rm survey}$ is the comoving volume.  The total number of the
Fourier modes usable for constraining cosmology is estimated by
integrating the effective volume in Fourier space up to the maximum
wavenumber $k_{\rm max}$ which is determined such that the theoretical
model to be compared with the measurement is reliable up to $k_{\rm
max}$. The larger the redshift, the higher the wavenumber we can
use, because the nonlinear scale becomes smaller (higher $k$).  
Hence, the proposed PFS survey offers much more than a factor 2
improvement compared to the BOSS constraint (e.g., Anderson et al. 2012).
Although Table~\ref{tab:survey} also gives estimates for the lower redshift slice 
$0.6<z<0.8$, which partially overlaps the ongoing BOSS and WiggleZ surveys, 
the cosmological constraining power of this slice is not as great due to the
smaller areal coverage and reduced galaxy number density. However,
we consider it important to retain this redshift slice as it can give a useful 
benchmark in comparison with other surveys such as the BOSS and WiggleZ 
surveys, particularly for calibrating systematic issues. 

\subsubsection{Expected cosmological constraints}
We now can estimate the power of the PFS cosmology survey in
Table~\ref{tab:survey} for constraining cosmological parameters. To
ensure a fair comparison with other surveys, we primarily assess the
power of PFS survey in terms of its BAO geometrical constraints.

\bigskip
\noindent{\it Geometrical constraints:}\\
The galaxy two-point correlation function is measured as a function of the
separation between paired galaxies. The position of each galaxy
needs to be inferred from the measured redshift and angular position.
Then the separation lengths perpendicular and parallel to the line-of-sight
direction from the measured quantities are given as $r_\perp\propto
\Delta \theta$ and $r_\parallel\propto \Delta z$, where $\Delta \theta $
and $ \Delta z$ are the differences between the angular positions and
the redshifts of the paired galaxies. For this conversion, we need to
assume a reference cosmological model to relate the observables ($\Delta
\theta$, $\Delta z$) to the quantities $(r_\perp,r_\parallel)$. Thus,
the wavenumbers are given as
\begin{equation}
k_{\perp,{\rm ref}}= \frac{D_{A}(z)}{D_{A, {\rm
 ref}}(z)}
k_{\perp},
\hspace{1em}
k_{\parallel, {\rm ref}}=\frac{H_{\rm ref}(z)}{H(z)}k_\parallel.
\end{equation}
The quantities with subscript ``ref'' are the quantities estimated from
the observables assuming a ``reference'' cosmological model, and the
quantities without the subscript are the underlying true values. Since
the reference cosmological model assumed generally differs from the
underlying true cosmology, it causes an apparent distortion in the
two-dimensional pattern of galaxy clustering. In principle, the
distortion  could be
measured using only the anisotropy of clustering statistics \citep[][]{AP:79},
but a more robust measurement can be obtained using features in the power
spectrum, particularly if they are at a known scale so that we can measure both
$D_A(z)$ and $H(z)$. In particular, the CMB-inferred BAO scale of 150 Mpc
gives a powerful standard ruler for this geometrical test
\citep[][]{Eisensteinetal:05, Percivaletal:07, Blakeetal:11,
HuHaiman:03, SeoEisenstein:03}.

The use of this scale without edge effects in a survey field requires
the survey to be contiguous on a scale large compared to the BAO length;
at our minimum redshift ($z=0.8$), 2.5 BAO lengths corresponds to 7.5
degrees on the sky, so we set this as our minimum width. This
requirement will be refined further by simulations.

In more detail, the galaxy power spectrum in redshift space
 is given in the linear regime as
\begin{eqnarray}
&&P_{g,s}(k_{\perp,{\rm ref}},k_{\parallel,{\rm ref}}; z)=\frac{D_{A,{\rm
 ref}}(z)^2H(z)}{H_{\rm ref}(z)D_A(z)^2}
\left[1+\beta(z)\frac{k_{\parallel}^2}{k^2}\right]^2
\nonumber\\
&&\hspace{10em}\times 
b_g^2P^L_m(k;z)+P_{\rm
sn},
\label{eq:Pg}
\end{eqnarray}
where $b_g$ is the linear bias parameter, $\beta$ is the linear
redshift-space distortion (RSD) parameter, defined as $\beta\equiv
(1/b_g)\left.d\ln D/d\ln a\right|_z$ \citep[][]{Kaiser:87}, $D$ is the linear
growth rate, $P^L_m$ is the linear mass power spectrum, and $P_{\rm sn}$
is a parameter to model the residual shot noise.  We can use the BAO
features in the linear power spectrum $P^{L}_m$ as a standard ruler in
order to constrain $D_A(z)$ and $H(z)$. The BAO constraints are
relatively robust against the galaxy bias uncertainty and the other
nonlinearity effects, because none of the systematic effects introduces
any particular length scale comparable with the BAO scale.  Further, if
we can use the shape and amplitude information in the galaxy power
spectrum, we can constrain the growth rate as well as other cosmological
parameters such as the neutrino mass and the primordial power spectrum
parameters \citep[][]{Takadaetal:06}, as we will discuss below.

To make the parameter forecast, we have used the method developed in Seo
\& Eisenstein (2007). In this method, we include
the smearing effect of the BAO features due to the bulk flow of galaxies
in large-scale structure
\citep[][]{Matsubara:08,Taruyaetal:09,NishimichiTaruya:11}. For the BAO
survey of multiple redshift bins, the Fisher information matrix of model
parameters can be computed as
\begin{eqnarray}
F_{\alpha\beta}&=&\sum_{z_i}\int_{-1}^1\!d\mu\int^{k_{\rm max}}_{k_{\rm
 min}}\!\frac{2\pi k^2dk}{2(2\pi)^3}
\frac{\partial \ln P_{g, s}(k,\mu; z_i)}{\partial
 p_\alpha}
\nonumber\\
&&\hspace{4em}\times 
\frac{\partial \ln P_{g ,s}(k,\mu; z_i)}{\partial
 p_\beta}
V_{\rm eff}(k; z_i)\nonumber\\
%\frac{\partial \ln P_{g, s}(k,\mu)}{\partial p_\beta}
%\left[\frac{\bar{n}_gP_g}{1+\bar{n}_gP_g}\right]
&&\hspace{4em}
\times\exp\left[-k^2\Sigma_\perp^2 -k^2\mu^2(\Sigma_\parallel^2-\Sigma_\perp^2)\right],
\label{eq:fisher}
\end{eqnarray}
where $\mu$ is the cosine between the wavevector and the line-of-sight
direction, $\mu\equiv k_\parallel/k$; $\sum_{z_i}$ is the sum over
different redshift bins; $\partial P_{g, s}/\partial p_\alpha$ is the
partial derivative of the galaxy power spectrum (Eq.~\ref{eq:Pg}) with
respect to the $\alpha$-th parameter around the fiducial cosmological
model; the effective survey volume $V_{\rm eff}$ and the Lagrangian
displacement fields $\Sigma_\parallel$ and $\Sigma$ to model the
smearing effect are given
as
\begin{eqnarray}
V_{\rm eff}(k,\mu;z_i)&\equiv&
 \left[\frac{\bar{n}_g(z_i)P_{g, s}(k,\mu;z_i)}
{\bar{n}_g(z_i)P_{g, s}(k,\mu; z_i)+1}\right]^2V_{\rm
 survey}(z_i)\\
\Sigma_{\perp}(z)&\equiv & c_{\rm rec}D(z)\Sigma_0, \\
\Sigma_{\parallel}(z)&\equiv & c_{\rm rec}D(z)(1+f_g)\Sigma_0.
\label{eq:sigma}
\end{eqnarray}
Here $V_{\rm survey}(z_i)$ is the comoving volume of the redshift slice
centered at $z_i$; the present-day Lagrangian displacement field is
$\Sigma_0=11h^{-1}{\rm Mpc}$ for $\sigma_8=0.8$ \citep[][]{Eisensteinetal:07}; 
$D(z)$ is the growth
rate normalized as $D(z=0)=1$; $f_g=d\ln D/d\ln a $. The parameter
$c_{\rm rec}$ is a parameter to model the reconstruction method of the
BAO peaks (see below). 
In
Eq.~(\ref{eq:fisher}), we take the exponential factor of the smearing
effect outside of the derivatives of $P_{\rm g,s}$. This is equivalent
to marginalizing over uncertainties in $\Sigma_\parallel$ and
$\Sigma_\perp$.
The growth rate in $\Sigma_{\parallel}$ or $\Sigma_\perp$
takes into account the smaller smearing effect at higher redshift due to
the less evolved large-scale structure. 
For the parameters, we included the cosmological
parameters, the distances in each redshift slice, and the nuisance parameters:
\begin{eqnarray}
p_{\alpha}&=&\{\Omega_{\rm m0}, A_s, n_s, \alpha_s, \Omega_{\rm m0}h^2,
 \Omega_{\rm b0}h^2, D_A(z_i), H(z_i), 
\nonumber\\
&&\hspace{2em}
b_g(z_i), \beta(z_i), P_{\rm
 sn}(z_i)  \}, 
\label{eq:parameters}
\end{eqnarray}
where $A_s$, $n_s$ and $\alpha_s$ are parameters of the primordial power
spectrum; $A_s$ is the amplitude of the primordial curvature
perturbation, and $n_s$ and $\alpha_s$ are the spectral tilt and the
running spectral index.  The set of cosmological parameters determines
the shape of the linear power spectrum. By using the method above, we
can estimate the cosmological distance information solely from the BAO
peaks, including marginalization over modeling uncertainties in 
%not from 
the broad-band shape of the power spectrum. For the
$k$-integration, we set $k_{\rm min}=10^{-4}h/{\rm Mpc}$ and $k_{\rm
max}=0.5~h/{\rm Mpc}$ for all the redshift slices, but the exponential factor in
Eq.~(\ref{eq:fisher}) suppresses the information from the nonlinear
scales. The Fisher parameter forecasts depend on the fiducial
cosmological model, for which we assumed the model consistent with the
WMAP 7-year data \citep[][]{WMAP7}.

Further, we assume that we can implement the promising method of
Eisenstein et al. (2007) for improving the BAO measurements. Since the
peculiar velocity field of galaxies in large-scale structure can be
inferred from the measured galaxy distribution, the inferred velocity
field allows us to pull back each galaxy to its position at an earlier
epoch and then reconstructing the galaxy distribution more in the linear
regime. 
As a result, one can correct to some extent the smearing effect
in Eq.~(\ref{eq:fisher}) and sharpen the BAO peaks in the galaxy power
spectrum. Recently, Padmanabhan et al. (2012) implemented this method
with  real data from the SDSS DR7 LRG catalog, and showed that the reconstruction
method can improve the distance error in the BAO scale
by  a factor of 2. The
improvement was equivalent to reducing the nonlinear smoothing scale from
$8.1$ to $\Sigma_{\rm nl}=4.4~h^{-1}{\rm Mpc}$, about a factor 2 reduction
in the displacement field. To implement this reconstruction method
requires a sufficiently high number density of the sampled galaxies in
order to reliably infer the peculiar velocity field from the measured
galaxy distribution.  Each redshift slice of the PFS survey (see
Table~\ref{tab:survey}) satisfies the requirement; the number density of
galaxies in each redshift slice is higher than that of both the SDSS DR7 LRGs
($\bar{n}\simeq 10^{-4}(h/{\rm Mpc})^3$)  and
the BOSS LRGs ($\bar{n}\simeq 3\times 10^{-4}(h/{\rm
Mpc})^3$). Hence we can safely assume that the reconstruction method can
be applied to the PFS BAO experiment.  In the Fisher matrix
calculation, we used $c_{\rm rec}=0.5$ for an implementation of the
reconstruction method\footnote{With these assumptions, 
we found we can roughly reproduce
the distance measurement accuracy for the SDSS LRGs as found in
Padmanabhan et al. (2012).}.

Finally, we have used the CMB information expected from the Planck
satellite experiment \citep{PlanckCosmo:13}, which gives a precise constraint on the sound
horizon scale in order for us to use the BAO scale as a standard
ruler. In addition to the cosmological parameters ($\Omega_{\rm m0},
A_s, n_s, \alpha_s, \Omega_{\rm m0}h^2,\Omega_{\rm b0}h^2$), we
included, in the CMB Fisher matrix, $\tau$ (the optical depth to the
last scattering surface) and $D_A(z_{\rm CMB})$ (the angular diameter
distance to the last scattering surface). Then we can compute the Fisher
matrix for the BAO experiment by adding the galaxy and CMB Fisher
matrices; $\bm{F}_{\alpha'\beta'}= \bm{F}^{\rm
galaxy}_{\alpha'\beta'}+\bm{F}^{\rm Planck}_{\alpha'\beta'}$. The
dimension of the Fisher matrix for the PFS survey in combination with
the Planck information is $38\times 38$ (see
Eq.~\ref{eq:parameters}). When further combined with the SDSS and BOSS
BAO information \citep{Eisensteinetal:05,Andersonetal:12}, the dimension of the Fisher matrix increases
accordingly.

\begin{figure*}[t]
\begin{center}
\includegraphics[width=15cm, angle=0]{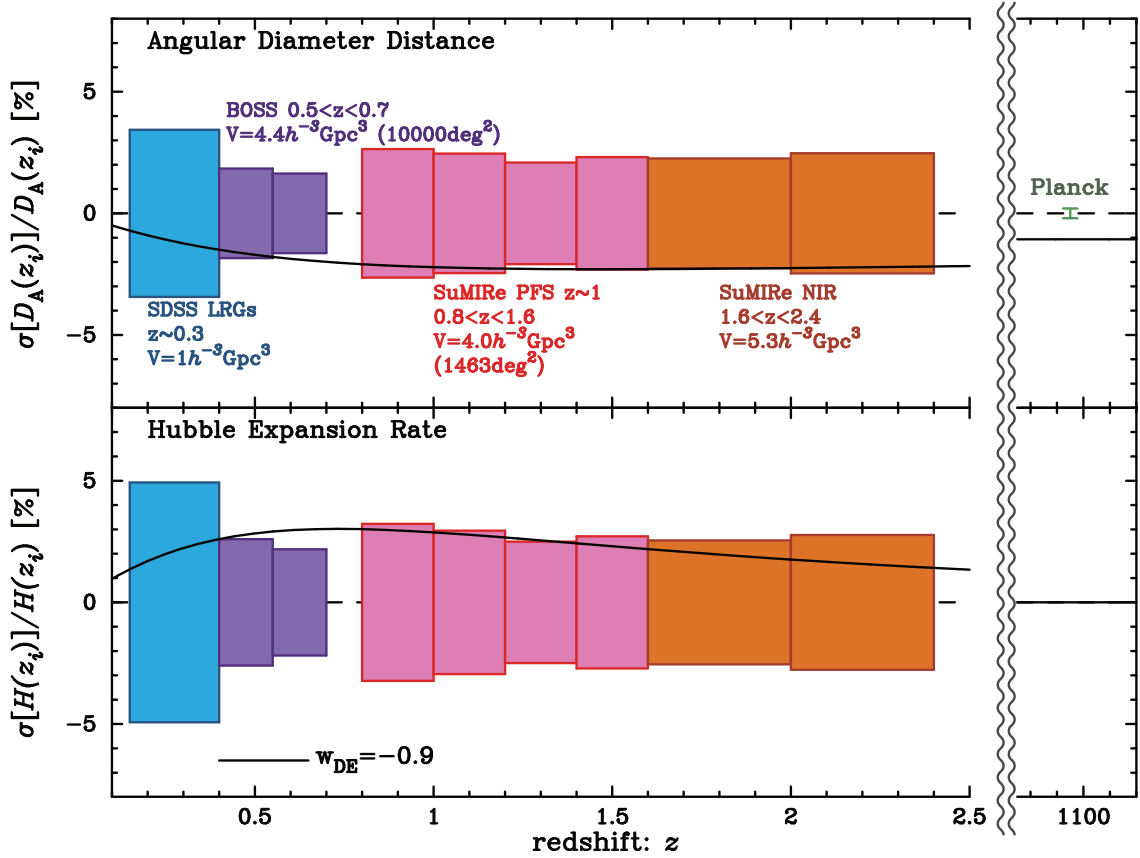}
\caption{ {Fractional errors in the angular diameter
distance and the Hubble expansion rate via the PFS BAO experiment  (see
Table~\ref{tab:survey}) including marginalization over uncertainties of
other parameters. The expected accuracies are compared to the existing
and ongoing SDSS and BOSS surveys. The PFS survey will provide
geometrical constraints to higher redshift than the SDSS and BOSS
 surveys, but with comparable precision. The 
 solid curve in each panel shows the fractional difference when changing
 the dark energy equation of state parameter from the fiducial model 
$w=-1$ to $w=-0.9$. 
} } \label{fig:da-h}
  \end{center}
\end{figure*}

Fig.~\ref{fig:da-h} shows the expected accuracies of determining the
angular diameter distance and the Hubble expansion rate in each redshift
slice with the PFS cosmology survey (Table~\ref{tab:survey}).  The
errors include marginalization over uncertainties of other
parameters. 
The PFS forecasts can be compared with the accuracies of the existing
and ongoing SDSS/BOSS surveys. As can be clearly
seen, the PFS cosmology survey can constrain $D_A(z)$ and $H(z)$ over a
wider range of redshift, yet with similar precision to, the SDSS
and BOSS surveys. Even though the PFS area coverage is smaller than that of
SDSS or BOSS surveys by a factor of 7 (1400 vs. 10000 sq. degrees), the
PFS survey covers a factor 10 and 2 times 
larger comoving volume than the SDSS and 
BOSS surveys, respectively. 

\medskip
\noindent{\it Cosmological implications:}\\
It is worth noting that the BOSS survey will measure 
the clustering statistics of
the Lyman-$\alpha$ forest over $2.1\simlt z\simlt 3.4$ \citep{Slosaretal:13}.  PFS thus serves to
fill a natural `gap' in-between the galaxy and Lyman-$\alpha$ BAO
experiments, allowing us to probe the expansion history over the entire
range of redshifts, $0\simlt z\simlt 3$, i.e. through the period
where
it is believed that the cosmic expansion went from
decelerated to accelerated phases.
If we model the expansion history as parametrized by the dark
energy model ($w_0, w_a$) \citep[][]{Linder:03} 
and the curvature parameter ($K$),
\begin{eqnarray}
H^2(z)&=&H_0^2\left[\Omega_{\rm m0}(1+z)^3-\frac{K}{H_0^2}(1+z)^2
\right.
\nonumber\\
&& \hspace{2em}
\left.
+
\Omega_{\rm de, 0} a^{-3(1+w_0+w_a)}e^{
3w_a(a-1)}\right],
\end{eqnarray}
we can propagate the distance measurement errors into the accuracies of
estimating the parameters.
To be more explicit, we can do
this, based on the Fisher information matrix formalism, by projecting
the BAO Fisher matrix onto different parameter space:
\begin{equation}
F_{a'b'}\equiv \frac{\partial \tilde{p}_{a}}{\partial p_{a'}}
F^{\rm sub}_{ab}\frac{\partial \tilde{p}_{b}}{\partial p_{b'}}.
\label{eq:project}
\end{equation}
Here the new set of parameters $\tilde{p}_{a'}$ is given as
$\tilde{p}_{a'}=(\Omega_{\rm de0},\Omega_{\rm m0}h^2, \Omega_K, w_0,
w_a)$, which specifies the cosmic expansion history given by the above
equation, and $F^{\rm sub}_{ab}$ is the sub-matrix computed by inverting
the sub-matrix of the inverse of the full BAO matrix,
$[\bm{F}^{-1}]_{\alpha\beta}$, containing only the parts of the geometrical
parameters, $p_{a}=\{\Omega_{\rm m0},\Omega_{\rm m0}h^2, D_A(z_i),
H(z_i)\}$. Hence the derived constraints on $\tilde{p}_{a'}$ include
marginalization over other parameters such as the galaxy bias and the
$\beta$ parameters. Table~\ref{tab:paras} shows the expected accuracies
of the dark energy parameters and the curvature parameter for the PFS
survey. Here $w_{\rm pivot}$ is the dark energy equation state at the
``pivot'' redshift, at which the dark energy equation of state is best
constrained for a given survey. The quantity ${\rm FoM}_{\rm de}$ is the
dark energy figure-of-merit defined in the Dark Energy Task Force Report
\citep[][]{DETF}, which quantifies the ability of a given survey to
constrain both $w_0$ and $w_a$; ${\rm FoM}_{\rm de}\equiv
1/[\sigma(w_{\rm pivot})\sigma(w_a)]$, which is proportional to the area
of the marginalized constraint ellipse in a sub-space of
$(w_0,w_a)$. Table~\ref{tab:paras} clearly shows that the PFS BAO can
significantly tighten the parameter constraints over the SDSS and BOSS
surveys. Most interestingly, the PFS has the potential to constrain the
curvature parameter to a precision of 0.3\%. If we can detect a non-zero
curvature, this would represent a {\it fundamental discovery} giving
 critical
constraint into the physics of the early universe, for example insight into different
inflation scenarios \citep[][]{Efstathiou:03,Contaldietal:03,inflation:06,
KlebanSchillo:12,GuthNomura:12}.
%\mt{(some
%papers should be cited)}.

\begin{figure*}[t]
\begin{center}
\includegraphics[width=0.5\textwidth, angle=-90]{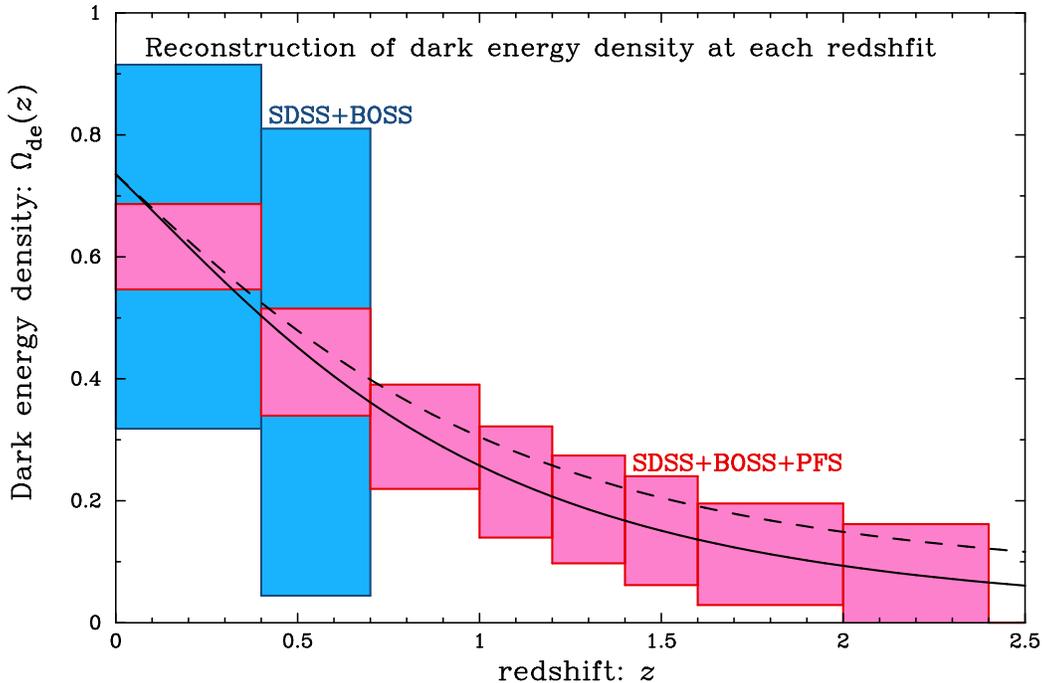}
\caption{ {Expected accuracy of reconstructing the dark energy
 density parameter at each redshift, $\Omega_{\rm de}(z)\equiv \rho_{\rm
 de}(z)/[3H^2(z)/8\pi G]$ from the BAO-measured $D_A(z)$ and $H(z)$ in
 Fig.~\ref{fig:da-h}. Here we considered the cosmological constant
 ($\rho_{\rm de}(z)=\rho_{\rm de 0}=$constant) and the flat universe
 ($\Omega_K=0$) as the fiducial
 model. Adding the PFS BAO constraints to the SDSS and BOSS constraints
 enables reconstruction of the dark energy density to $z\simeq 2$, and also
 significantly improves the precision at low redshifts, as the comoving distance 
at the high redshift  arises from an integration of $H(z)$. 
The solid curve shows the energy density parameter for the
 fiducial 
 $\Lambda$CDM model, while the dashed curve shows the redshift evolution
for an early dark energy model in Droan \& Robbers (2006), where we
 employed $w_0=-1$ and $\Omega_d^e=0.05$ for the model parameters (see
 text for details).
} } \label{fig:rhode}
  \end{center}
\end{figure*}

\begin{deluxetable*}{l|lllll|ll}
\tablewidth{0pt}
\tabletypesize{\small}
\tablecaption{
Forecasted accuracies of cosmological parameters}
\startdata  \hline \hline
%\begin{table*}
%\begin{center}
%Forecasted accuracies of cosmological parameters\\
%\begin{tabular}{l|lllll|ll}\hline\hline
Survey & $\Omega_{\rm de0}$ & $w_{\rm pivot}$ & $w_a$ & ${\rm FoM}_{\rm de}$
\hspace*{1em} & $\Omega_K$ 
%& $\gamma_g$
& $m_{\nu,{\rm tot}}$~[eV] & $f_{\rm NL}$ \\ \hline 
SDSS+BOSS & 0.0061 & 0.076 & 1.2 & 11 & 0.0071 & 0.188 & 16\\
SDSS+BOSS+PFS & 0.0051 & 0.059 & 0.36& 47 & 0.0030 & 0.133 & 11
\enddata
\\ \hline\hline
%\end{tabular}
\tablecomments{
%\caption{
The constraints on $\Omega_{\rm de0}, w_{\rm pivot}, w_a$ and
 $\Omega_K(\equiv -K/H_0^2)$ are from the BAO distance measurements in
 Fig.~\ref{fig:da-h}, i.e. not including the information on the
 broad-band shape of the galaxy power spectrum. Note that $w_{\rm
 pivot}$ is the dark energy equation state at the ``pivot'' redshift, at
 which the dark energy equation state parameter is best constrained for
 the given PFS BAO measurements. 
The constraints on 
%the growth rate parameter $\gamma_g$, 
the neutrino mass $m_{\nu,{\rm tot}}$
 and $f_{\rm NL}$ are derived by including the broad-band shape
 information. See the text for details. 
\label{tab:paras}}
\end{deluxetable*}
%\end{center}
%\end{table*}

\medskip
\noindent{\it Nature of dark energy:}\\
The parametrization $(w_0,w_a)$ adopted for the dark energy equation of
state samples only a narrow range of dark energy models.  Given that there
is no well-accepted model for dark energy we seek to interpret our PFS
data in a more model-independent way. The wide redshift coverage of the PFS 
survey, in combination with the SDSS and BOSS surveys, allows us to do this
by directly reconstructing
 the dark energy density as a function of redshift solely 
based on the geometrical BAO constraints. To illustrate this, we use the Hubble expansion history
parametrized in terms of dark energy density parameters in each redshift
bins:
\begin{equation}
H^2(z)=H_0^2\left[\Omega_{\rm m0}(1+z)^3-\frac{K}{H_0^2}(1+z)^2+\frac{\rho_{{\rm de},
	     z_i}(z\in z_i)}{\rho_{\rm cr0}}\right],
\end{equation}
where $\rho_{{\rm de}, z_i}$ is the dark energy parameter in the
redshift bin centered at $z_i$. For the combined BAO survey of SDSS,
BOSS and PFS, we include the 9 dark energy densities, $\rho_{\rm
de}(z_i)$, given in 9 redshift bins ($8$ redshift bins of the galaxy
surveys plus the redshift bin from $2.4$ to $z_{\rm CMB}$). Then, similarly
to the method described around Eq.~(\ref{eq:project}), we can propagate
the BAO-measured distance errors into the accuracies of reconstructing the
dark energy densities in each redshift bin. Fig.~\ref{fig:rhode} shows
the result, where we assumed the cosmological constant, $\rho_{\rm
de}={\rm constant}$, as the fiducial model. The figure clearly shows that
the PFS BAO survey is capable of reconstructing dark energy densities up to high
redshift, $z\simeq 2$ for a model in which $\rho_{\rm de}(z)\simeq {\rm
constant}$, thereby testing various types of early dark energy models.
 For comparison, the dashed curve shows 
the dark energy density parameter for an early dark energy
(EDE) model proposed in Doran \& Robbers (2006), which gives a more
significant contribution to the cosmic expansion at higher redshift than
in the cosmological constant: 
\begin{equation}
 \Omega_{\rm de}^{\rm EDE}(a)=\frac{\Omega_{\rm
  de0}-\Omega_{\rm e}(1-a^{-3w_0})}
{\Omega_{\rm de0}+\Omega_{\rm m0}a^{3w_0}}+\Omega_{\rm e}(1-a^{-3w_0}),
\end{equation}
where we fixed the parameters to $\Omega_{\rm e}=0.05$ and $w_0=-1$ as
the fiducial model. Note that we set the present-day energy density of
EDE to be the same as that of the cosmological constant model. 
The figure clearly shows that PFS can put a more
stringent constraint on such an early-dark energy model thanks to its
wide redshift coverage, e.g. compared to the BOSS alone.  To be more
precise, the combined BOSS and PFS geometrical constraints 
can achieve a precision of $\sigma(\Omega_{\rm
e})=0.048$ even under the conservative setup of our forecasts, while the
BOSS alone  cannot constrain the parameter well
($\sigma(\Omega_{\rm e})=1.2$).

Further, we should emphasize that adding the PFS BAO measurements to the
SDSS and BOSS information can significantly improve the reconstruction
of dark energy density at low redshift, partly because the SDSS+BOSS BAO
alone cannot break degeneracies between parameters ($\Omega_{\rm m0},
\Omega_{\rm m0}, K, \rho_{\rm de}(z_i)$) given the distance measurements
to the three redshifts ($z_{\rm SDSS}, z_{\rm BOSS}, z_{\rm CMB}$), and
also because the angular diameter distances for the PFS redshifts are all
sensitive to dark energy densities at low redshift via the integration
relation between the angular diameter distance and the Hubble expansion
rate. However, we should note that the constraint on the curvature
parameter is significantly degraded in this case to
$\sigma(\Omega_K)=0.057$ from the result in Table~\ref{tab:paras}
because of the larger number of free parameters. Thus the accuracy of the
curvature parameter is sensitive to which dark energy model we use for
marginalization.

\begin{figure}[t]
\begin{center}
\includegraphics[width=0.5\textwidth, angle=0]{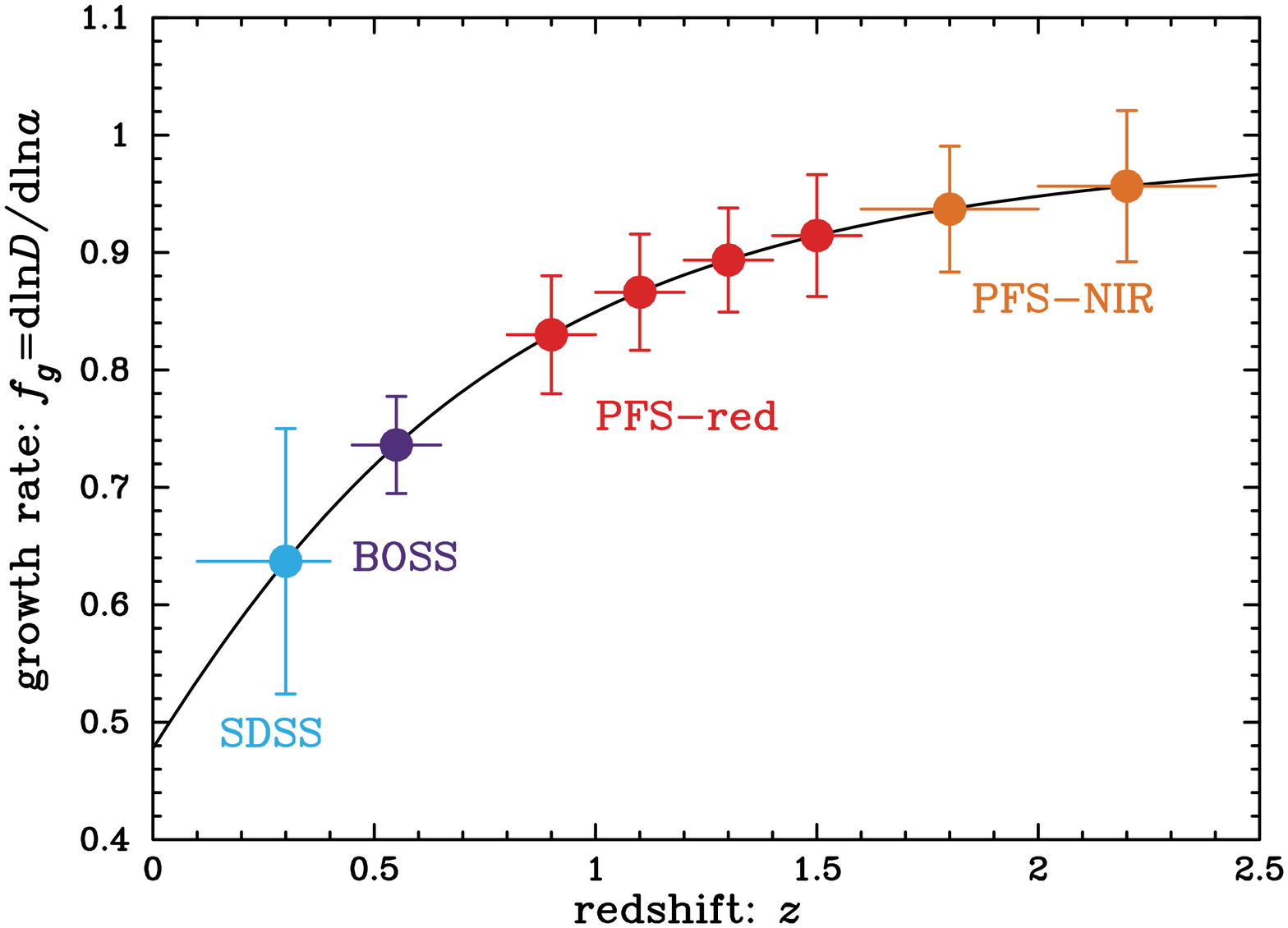}
\caption{ {Marginalized errors in reconstructing 
the growth rate, $f_g\equiv
 d\ln D/\ln a$, in each redshift slice. 
} } \label{fig:fg}
  \end{center}
\end{figure}

\medskip
\noindent{\it Testing the growth of structure:}\\
We now turn to the utility of measuring the {\it redshift-space distortion (RSD) effect} 
and the broad-band shape of the galaxy power spectrum. If we can reliably
model the RSD and the shape of the power spectrum in the weakly nonlinear
regime including a possible scale-dependent bias, we can use this
information not only to improve the cosmological constraints
\citep[][]{Takadaetal:06}, but also to constrain the growth rate which is
sensitive to the theory of gravity on cosmological scales. Encouraging progress
is being made via many efforts to develop a more
accurate model of the redshift-space power spectrum in the weakly
nonlinear regime as we discuss below.
\citep[][]{Matsubara:08,Taruyaetal:09,NishimichiTaruya:11,Tangetal:11,Hikageetal:12a,Hikageetal:12b}.

To estimate the power of the PFS survey, we use the linear theory
prediction for the amplitude of the RSD effect, $\beta(z)=f_g(z)/b_g(z)$, in
Eq.~(\ref{eq:Pg}), where $f_g$ is defined by the growth rate as
$f_{g}\equiv d\ln D/\ln a$.  Then we can include the RSD effect in the
Fisher matrix formalism by using $f_g$ in each redshift slice instead of
treating $\beta$ as parameters (see Eqs.~\ref{eq:Pg} --
\ref{eq:parameters}). With this implementation, we can break
degeneracies between the RSD effect $f_g/b_g$ and the galaxy bias
uncertainty $b_g$ from the measured anisotropic modulations in the
redshift-space galaxy power spectrum. Then we can in turn use the
amplitude and shape information of the underlying linear power spectrum.

Fig.~\ref{fig:fg} shows the expected accuracies of constraining 
the growth rate, $f_g(=d\ln D/d\ln a)$, 
in each redshift slice via the RSD measurements.
The PFS survey can constrain the growth rate in each redshift to a 6\%
accuracy. 
In particular, PFS will provide accurate constraints on the growth rate at redshifts
beyond $z=1$, when the cosmic expansion is in its decelerated phase. Such
constraint are very important for testing whether dark energy is an illusion
caused by an incomplete understanding of General Relativity.

\medskip
\noindent{\it Other constraints:}\\
With the growth rate constraints and the information on the shape of the
galaxy power spectrum, we can also constrain other interesting parameters,
such as the sum of neutrino masses ($m_{\nu,{\rm tot}}$) and the degree of
primordial non-Gaussianity ($f_{\rm NL}$). Primordial
non-Gaussianity induces a characteristic scale-dependent biasing effect
on the galaxy distribution at very large scales \citep[][]{Dalaletal:08} that
are well in the linear regime and cannot be explained by other
nonlinearity effects. Hence we can use the largest-scale signal of
galaxy clustering to explore the signature of the primordial
non-Gaussianity. Table~\ref{tab:paras} shows the expected accuracy of
constraining $f_{\rm NL}$ to an accuracy of $\sigma(f_{\rm NL})\simeq
11$ if systematic errors are under control 
(see below for discussion on possible systematic errors). The Planck 
experiment showed a more stringent upper limit on $f_{\rm NL}$ such as 
$f_{\rm NL}\simlt 5$ (68\% C.L.) \citep{PlanckNG:13}; PFS is limited by
its relatively small area coverage to access the largest-length scales.

On the other hand, the massive neutrinos, as found by
terrestrial experiments, suppress the galaxy clustering
power on scales smaller than the neutrino free-streaming scale, which
imprints a characteristic scale-dependent effect on the galaxy power
spectrum \citep[][]{Takadaetal:06}. The amount of the suppression 
scales with the sum of neutrino mass as $\Delta P_g/P_g \simeq -8
\Omega_{\nu0}/\Omega_{\rm m0} \simeq -8 m_{\nu,{\rm tot}}/(94.1~{\rm
eV}\Omega_{\rm m0}h^2)$ at the scales smaller than the neutrino free-streaming
scale. Thus
the neutrinos of $m_{\nu,{\rm tot}}=0.1~$eV,
close to the lower bound of the inverted neutrino mass hierarchy, lead
to about 6\% suppression in the galaxy power spectrum compared to the
case without the massive neutrinos. Hence, we can use the measured
clustering amplitude to constrain the neutrino mass. However, the
achievable precision of neutrino mass depends on the level of our
understanding of the nonlinear power spectrum including the galaxy bias
uncertainty \citep[][]{Saitoetal:08,Saitoetal:09}. Here, by assuming that an
accurate model of the galaxy power spectrum is available, we estimate
the power of PFS to constrain the neutrino mass. To be more precise,
we assumed that the following set of parameters, instead of
Eq.~(\ref{eq:parameters}), can model 
the redshift-space galaxy power spectrum based on the extended
perturbation theory based method in combination with numerical simulations: 
\begin{eqnarray}
p_{\alpha}&=&\{\Omega_{\rm m0}, A_s, n_s, \alpha_s, \Omega_{\rm m0}h^2,
 \Omega_{\rm b0}h^2, \Omega_{K}, w_0, w_a,
\nonumber\\
&&\hspace{2em}
m_{\nu, {\rm tot}}, 
b_g(z_i), P_{\rm
 sn}(z_i)\}. 
\label{eq:parameters_nu}
\end{eqnarray}
In this parameter estimation we did not use the reconstruction method
(i.e., we  set $c_{\rm rec}=1$ for the reconstruction parameter in
Eq.~\ref{eq:sigma}), because the reconstruction method of BAO peaks
alters the shape and amplitude of the power spectrum.  With this
implementation, we can include the shape and amplitude information of
the power spectrum for constraining the cosmological parameters,
marginalized over uncertainties of the nuisance parameters.  Also note
that, for the parameter estimation, we included a broader range of
cosmological parameters such as the curvature $\Omega_K$ and the dark
energy parameters $(w_0,w_a)$, which also cause a suppression in the
growth rate of structure formation as do the massive neutrinos.
However, we assumed linear bias parameters for each redshift slice, but
instead included parameters to model the residual shot noise
contamination, which mimic a scale-dependent bias.  As can be found from
Table~\ref{tab:paras}, the PFS survey can achieve a precision of
$\sigma(m_{\nu, {\rm tot}})\simeq$0.13~eV. However, we again note that,
even though we included marginalization over other parameters, the
neutrino mass constraint is sensitive to the level of our understanding
of nonlinearity effects such as nonlinear bias or nonlinear evolution of
matter clustering as we will discuss below in \S~\ref{sec:sys_cosmo}. 
If we restrict the
parameter forecast to the power spectrum information up to $k_{\rm
max}=0.1~h/{\rm Mpc}$, where the linear theory assumption is considered
to be valid, the expected constraint of neutrino masses is degraded to
$\sigma(m_{\nu, {\rm tot}})\simeq 0.55~{\rm eV}$. Thus an understanding
of the nonlinearity effects is important to obtain a higher-precision,
robust constraint on the neutrino mass.

\subsubsection{Discussion of  other systematic errors}
\label{sec:sys_cosmo}

\begin{figure}[t]
\begin{center}
\includegraphics[width=0.5\textwidth, angle=0]{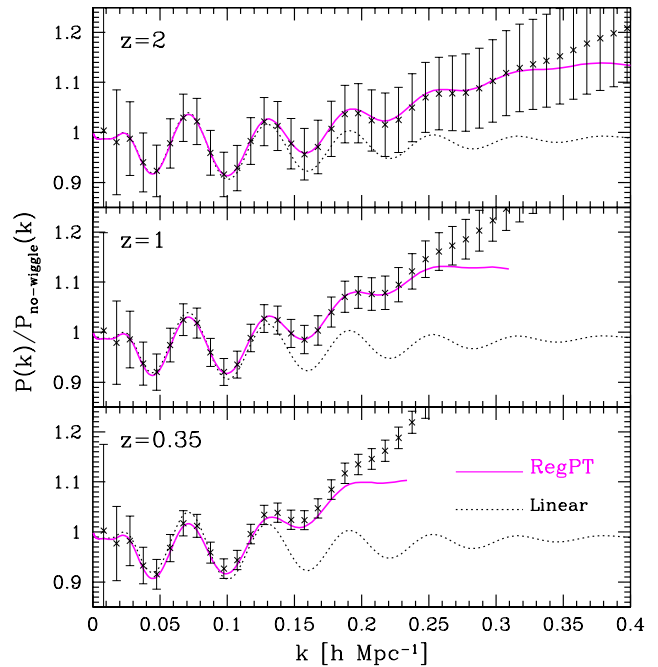}
\caption{
Nonlinear matter power spectra at different redshifts,
 $z=0.35$, 1 and $2$ in the bottom, middle and upper panels,
 respectively. For illustrative purposes, the power spectrum is divided
 by the linear power spectrum with BAO features not included. 
The data points with error bars in each panel are the
 power spectrum estimated from $N$-body simulations, where the error bars
 are the expected $1\sigma$ statistical errors of the power spectrum
 measurement at each $k$ bin, assuming the survey
volume of $3~({\rm Gpc}/h)^3$ and the comoving number density of
 $3\times 10^{-4}~(h/{\rm Mpc})^3$. 
The survey parameters 
roughly correspond to those for
the $z=1$ and 2 slices
of the PFS cosmology survey (see Table~\ref{tab:survey}).
The solid curve 
shows the analytical prediction based on  the refined
 perturbation theory \citep{Taruyaetal:12}, which shows a 
satisfactory
 agreement with the simulation result, up to a higher wavenumber for
 higher redshift slices. 
 \label{fig:pk_diffzs}}
\end{center}
\end{figure}

\begin{figure}[t]
\begin{center}
\includegraphics[width=0.33\textwidth, angle=-90]{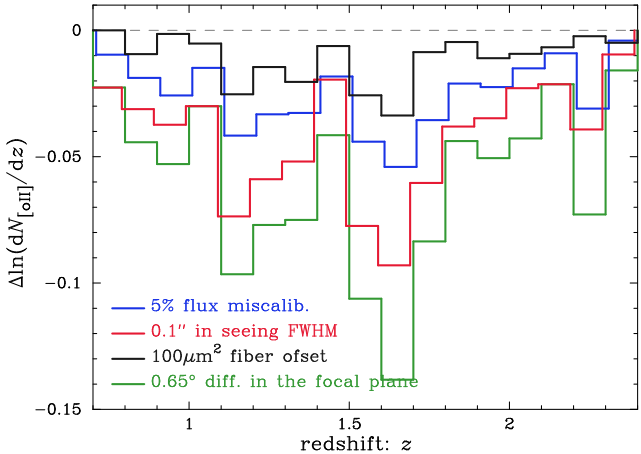}
\end{center}
\caption{How observational systematic errors affect the
selection function of the detected \oii\ emission-line galaxies as a
function of redshift, computed by using our exposure time calculator in
combination with the COSMOS mock catalog (see Section~\ref{sec:target}).
Here we assumed the fixed threshold $S/N=8.5$ for the \oii\ line
detection. As for working examples, we consider the fiber positioning
offset by 10~$\mu{\rm m}$ 
radius from the true centroid), 
misestimation of PSF FWHM by 0.1$^{\prime\prime}$, and $5\%$
error in flux calibration. For comparison, we also show the difference
between the numbers of detected \oii\ emitters at the center and edge
(0.65 degrees in radius) of the focal plane. These observational effects
 change the number of detected \oii\ emitters. 
 \label{fig:dndz_sys} }
\end{figure}

There remain theoretical and observational systematic errors that may
affect our PFS cosmology program. 
To realize the full potential of the PFS cosmology survey, we must
carefully consider and account for these errors.

\medskip

\noindent{\it Uncertainty in the modeling of galaxy power spectrum}:\\
The linear theory of structure formation 
is not sufficiently accurate to be compared
with the measured power spectrum of galaxies even at BAO scales, given
the statistical precision of the PFS cosmology survey. In
other words, interpreting the the galaxy power spectrum is complicated
by non-linear effects such as 
those relating to large scale structure and
the connection between the galaxy and dark matter distributions.
An advantage of the PFS cosmology survey
is its focus on higher redshift; 
the rate of evolution in large scale structure, and hence the effect
of non-linearities, is reduced at higher redshift.
There are promising developments
towards an accurate modeling of the galaxy power spectrum based on a
suite of high-resolution simulations as well as refined perturbation
theory, at BAO scales that are still in the mildly nonlinear
regime \citep{Matsubara:08,
Taruyaetal:09,Matsubara:11,Taruyaetal:12,SugiyamaFutamase:12}. Fig.~\ref{fig:pk_diffzs}
compares the nonlinear matter power spectra computed from the refined
perturbation theory and from $N$-body simulations for different redshift
slices, $z=0.35, 1$ and $2$ \citep{Taruyaetal:12}. The figure clearly
shows that non-linearities have a reduced
effect at higher redshift and also that the refined perturbation
theory provides a better match to the simulation results.
A gain in the maximum
wavenumber ($k_{\rm max}$) up to which to include the power spectrum
information for the cosmological analysis is equivalent to having a larger
survey volume; a factor 2 gain in $k_{\rm max}$ is equivalent to a
factor 8 larger survey volume in the sampling variance limited
regime\footnote{The cumulative statistical precision, ($S/N)^2$, for the power
spectrum measurement scales with the Fourier-space volume as $\propto
k_{\rm max}^3$.}.  
Thus the PFS survey, in combination with the refined theoretical
model, provides a more robust and powerful cosmological
result than one based on a survey at lower redshift.

We also stress that there will be many synergistic opportunities 
enabled by the fact that the PFS survey will be undertaken in the same
area of sky as the HSC imaging survey. Weak lensing lensing information 
from HSC will be very effective in correcting and calibrating systematic 
effects inherent in the galaxy clustering analysis,  nonlinear 
redshift-space distortion and the galaxy bias uncertainty, up to 
the $z\sim 2$ slice \citep{Hikageetal:12a,Hikageetal:12b,Nishizawaetal:12}. The spectroscopic 
data from the PFS survey can likewise be used to calibrate
the photo-$z$ errors and the redshift distribution of HSC imaging
galaxies, which is one of the major uncertainties in the HSC
cosmology analysis. Thus, by combining the HSC imaging and PFS spectroscopic
surveys, we can significantly improve the cosmological constraints, making 
the joint HSC and PFS experiments comparable to a Stage-IV Dark Energy
experiment in the parlance of recent US studies.

\medskip

\noindent{\it Observational systematic errors}: \\ The PFS cosmology
survey relies on the use of \oii\ emission-line galaxies, detected with $S/N$
greater than a given threshold ($S/N=8.5$ assumed throughout this
document). 
A variety of observational factors will affect the detection
efficiency of this line, causing apparent density fluctuations 
in the observed galaxy distribution. These include
\begin{itemize}
 \item Offset of the fibers from its expected position, e.g.
due to a systematic error in the astrometric solution
and/or imperfect positioning.
\item Variation in the throughput over the field angle (e.g. due to the
      vignetting).
\item A misestimation in the seeing FWHM. The PSF misestimation
      causes a biased estimate of the intrinsic \oii\ flux.
\item A flux miscalibration such as an error in the magnitude zero
       point.
\end{itemize}
In Fig.~\ref{fig:dndz_sys}, we use our exposure time calculator to
estimate how the systematic errors mentioned above change the number of
detected \oii\ emitters in each redshift bin, where we employed the same
threshold $S/N=8.5$. Here we consider some typical values for each of
the systematic errors as indicated by the legend. 
Thus, the systematic
errors 
might alter
the number of detected galaxies in each pointing, and the effects
generally vary with different pointings, causing apparent density
fluctuations of the detected galaxies on the sky. 

Thus the systematic errors need to be well corrected in order for us not
to obtain a biased constraint on  the growth rate or 
cosmological parameters from the
measured power spectrum amplitudes. 
A calibration of the systematic errors requires
an adequate strategy of the PFS cosmology survey; e.g., frequent observations
of standard stars, and a large-angle dithering (tiling) offset between
different pointings.
In particular, we are planning to use, for the calibration, 
the sample of \oii\ emission-line
galaxies taken from the PFS galaxy survey (see Section~\ref{chap:galaxy}
for details).
 With its much deeper depth
and much higher completeness in the PFS galaxy survey, the sample allows
us to understand intrinsic properties of \oii\ emission-line galaxies
such as the luminosity function, the line width and the line ratio of
\oii\ doublet. We will in turn use the calibration sample of \oii\
emitters in the PFS galaxy survey to estimate the selection function of
\oii\ emitters in
the PFS cosmology survey as a function of observational conditions and
redshift.

\subsection{Scientific Requirements for PFS Cosmology Survey}

As discussed above, PFS has the unique capability to execute
a very powerful cosmology survey across a wide range of redshifts, 
considerably extending
current and planned BAO surveys on $\le 4$ m-class
telescopes. 
Here Tables~\ref{cosmology:tab:re1} and \ref{cosmology:tab:re2}
summarize requirements on 
the survey parameters and the instrument parameters
for the PFS cosmological
applications. 
\begin{deluxetable*}{l|l}
\tablewidth{0pt}
\tabletypesize{\footnotesize}
\tablecaption{Cosmology Survey Requirements
\label{cosmology:tab:re1}}
%\begin{table*}
%{\begin{center}
%\caption{Requirements for PFS Cosmology Survey
%\label{cosmology:tab:re1}
%}
%\begin{tabular}{l|l} \hline\hline
\startdata \hline\hline
\multicolumn2c{Science yield requirements} \\
\hline
Distance measurements & $\simlt 3\%$ measurement of $D_A(z)$ and $H(z)$ 
in each of 6 redshift bins \\
 & \hspace{1em}via BAO (0.8--1.0, 1.0--1.2, 1.2--1.4, 1.4--1.6, 1.6--2.0, and 2.0--2.4) \\
Dark energy reconstruction 
& $\simlt 7\%$ measurement of $\Omega_{\rm de}(z)$ 
in each of 6 redshift bins via BAO\\
Curvature & Measure $\Omega_K$ to $\simlt 0.3\%$ via BAO\\
%\\
Growth of structure & $\simlt 6\%$ measurement of the growth rate of structure in each of 6 bins via RSD \\
\hline
\multicolumn2c{Galaxy catalog requirements} \\
\hline
Redshift range & $0.8\le z\le 2.4$ ($0.8\le z\le 1.6$ minimum)\\
Number density of galaxies & $\ge 2900$ deg$^{-2}$ \\
$dN/dz$ of ELGs &  $\bar{n}_gP_g>1$ ($0.8<z<1.6$) or $\bar{n}_gP_g>0.5$ ($1.6<z<2.4$) at $k= 0.1h$/Mpc \\
Total survey area & $\ge$1400 deg$^2$ \\
Incorrect redshift fraction & $<1\%$\\
Redshift precision, accuracy& $\Delta z/(1+z)<0.0007$, $1\sigma$ ($\sigma_v < 200~{\rm km/s}$)  \\
Survey geometry & Width $>7.5$ degrees; $\le$ 4 contiguously-connected survey regions\\
\hline
\multicolumn2c{Survey implementation requirements} \\
\hline
Total nights & $\simeq 100$ clear nights \\
Lunar phase & Dark (1 of 2 visits) or age $<7$ days (other visit) \\
Imaging survey & HSC $gri$ data to $\approx 26$th magnitude AB $(5\sigma)$ 
%%\hline\hline
\enddata
\end{deluxetable*}
%\begin{tabular}{l|l} \hline\hline
\begin{deluxetable*}{l|l}
\tablewidth{0pt}
\tabletypesize{\footnotesize}
\tablecaption{PFS Instrument Requirements for the PFS Cosmology Survey
\label{cosmology:tab:re2}}
%}\label{tab:re2}
%\begin{tabular}{l|l} \hline\hline
\startdata \hline\hline
Wavelength coverage  & $ 650\le\lambda\le 1260$nm ($650\le \lambda\le 1000$nm minimum)\\
Number of fibers & $2400$ \\
%Total galaxies & $\simeq 4$M (to cover 1400 sq. degrees in the total     survey area)\\
Overhead & $<0.2\times$ the open-shutter time \\
Throughput & Average $>22\%$ (red) or $>24\%$ (NIR) \\
& Worst part of band, $>20\%$ (red) or $>18\%$ (NIR) \\
& (Excludes atmosphere, central obscuration + WFC vignetting, and fiber aperture \\ & effect.) \\
Fiber aperture factor & Encircled energy in fiber is $\ge59\%$ (point source) or $\ge45\%$ (galaxy, $r_{\rm eff}=0.3''$) \\
 & (Equivalent to 0.8$''$ FWHM seeing + 11 $\mu$m rms/axis additional aberration
 \\ & + 0.12$''$ fiber offset.) \\
Spectrograph image quality & $\le 14$ $\mu$m rms per axis (excluding fiber geometric size, pixel tophat, and \\
 & internal defocus due to thickness of red CCD) \\
Spectral resolution & Red: $R\sim 3000$; NIR: $R\sim 4000$ (to resolve out OH lines but limit read noise) \\
Stray light & Near OH lines: Lorentzian wings at $\le 3\times$ amplitude of perfect grating \\
 & Diffuse: equivalent to $\le 2$\% of total sky brightness spread over detector \\
Read noise & $\le 3$ (red) or $\le 4$ (NIR) $e^-$ rms per pixel \\
 & (If the NIR channel is not reset between exposures, $\le4\sqrt2$ $e^-$ rms is acceptable.) \\
Sky subtraction accuracy & $<1$\% of sky background per 4-pixel
 resolution element 
\enddata
\end{deluxetable*}

\section{PFS Galactic Archaeology}
\label{chp:GA}

{\it Summary: 
The Subaru PFS offers an unprecedented
opportunity to investigate the assembly histories of both baryonic and
dark matter components of two typical, large galaxies, namely the
Milky Way and M31. This will be achieved through the measurements of
radial velocities and elemental abundances for a large sample ($\sim$ a
million) of their member stars. We here describe our Galactic
Archaeology (GA) science case for a PFS survey designed to incorporate the envisaged
medium-resolution (MR) mode, adopting $R = 5,000$ over the spectral
range of $7,100$--$8,850$~\AA\ in the red arm, combined with the baseline
low-resolution (LR) mode of $R = 2,000$--$3,000$ in the blue
arm. The MR mode will allow the measurement of  multiple $\alpha$-element
abundances for stars in the Milky Way, elucidating  the detailed chemical evolution
history for each of the Galactic components. This mode will further enable us
to study the global chemo-dynamical evolution and dark-matter distribution of Galactic
dwarf spheroidal galaxies, which are the possible remaining building blocks of
the Milky Way.  This fundamental science is possible only with the $\alpha$-element
abundance measurements and more precise radial velocities offered by the
planned medium-resolution mode of PFS. 
}

\subsection{Galactic Archaeology Objectives}

Our understanding of how galaxies like the Milky Way and M31 formed
and evolved in the expanding Universe remains limited. The current
paradigm of cosmic structure formation based on 
the $\Lambda$CDM
model proposes that galaxy formation is driven by hierarchical
assembly of dark matter halos, starting from sub-galactic scales. The
repeated merging and overall clustering of many small dark halos over
cosmic time forms a larger, galaxy-sized halo. The associated cooling
and collapse of baryonic matter confined in the dark halos at each
stage of the hierarchy forms the visible galaxies, with the morphology
of the final galaxy set by the physics of star formation and of the
merging process.  Indeed, mergers are predicted to play a key role in
the creation of stellar halos, bulges, thick disks and even some parts
of thin disks.

\begin{figure}
\begin{centering}
\leavevmode\epsfxsize=7.0cm \epsfbox{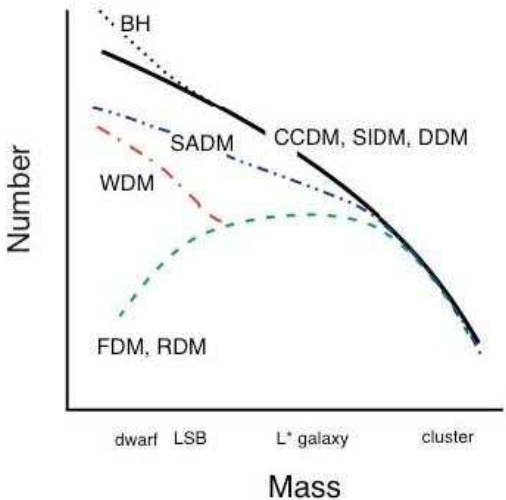}
\caption{
An illustration of how the number of objects of a given type
depends on their present day mass for different
dark matter models \citep[][]{OstrikerSteinherdt:03} including
%(Ostriker \& Steinhardt 2003).
Collisionless Cold dark matter (CCDM),
Strongly Self-Interacting dark matter (SIDM),
Warm dark matter (WDM), Repulsive dark matter (RDM),
Fuzzy dark matter (FDM), Self-Annihilating dark matter (SADM),
Decaying dark matter (DDM) and Massive Black Holes (BH).
Observations on small scales are key to distinguishing these.
}
\label{fig:CDM}
\end{centering}
\end{figure}

However, CDM models have encountered several fundamental difficulties
in explaining observations on galactic scales; the model predicts an
excess of small scale clustering power than observed. 
One of the most serious issues is the prediction of many more
subhalos in a Milky Way-sized system than the modest number
of visible satellites. 
This discrepancy can only be reconciled if the vast majority of small halos are dark.
Our understanding of the star formation
process on these small scales, which correspond to dwarf satellite
galaxies, remains limited, so there may well be a large population of
dark satellites.  
They would then only reveal their presence through 
their dynamical effects on visible systems.
Alternatively, the assumption that CDM is collisionless
may not be correct \citep[][see  Fig.~\ref{fig:CDM} for the predictions of various dark
matter models]{OstrikerSteinherdt:03}.
The merging history of the `building
blocks' of  galaxies  is determined by  the power spectrum, i.e. by the 
nature of dark matter, and this we  plan to investigate through
the study of the stellar populations in both the  Milky Way and M31. 

\begin{figure}
\begin{centering}
\leavevmode\epsfxsize=5.5cm \epsfbox{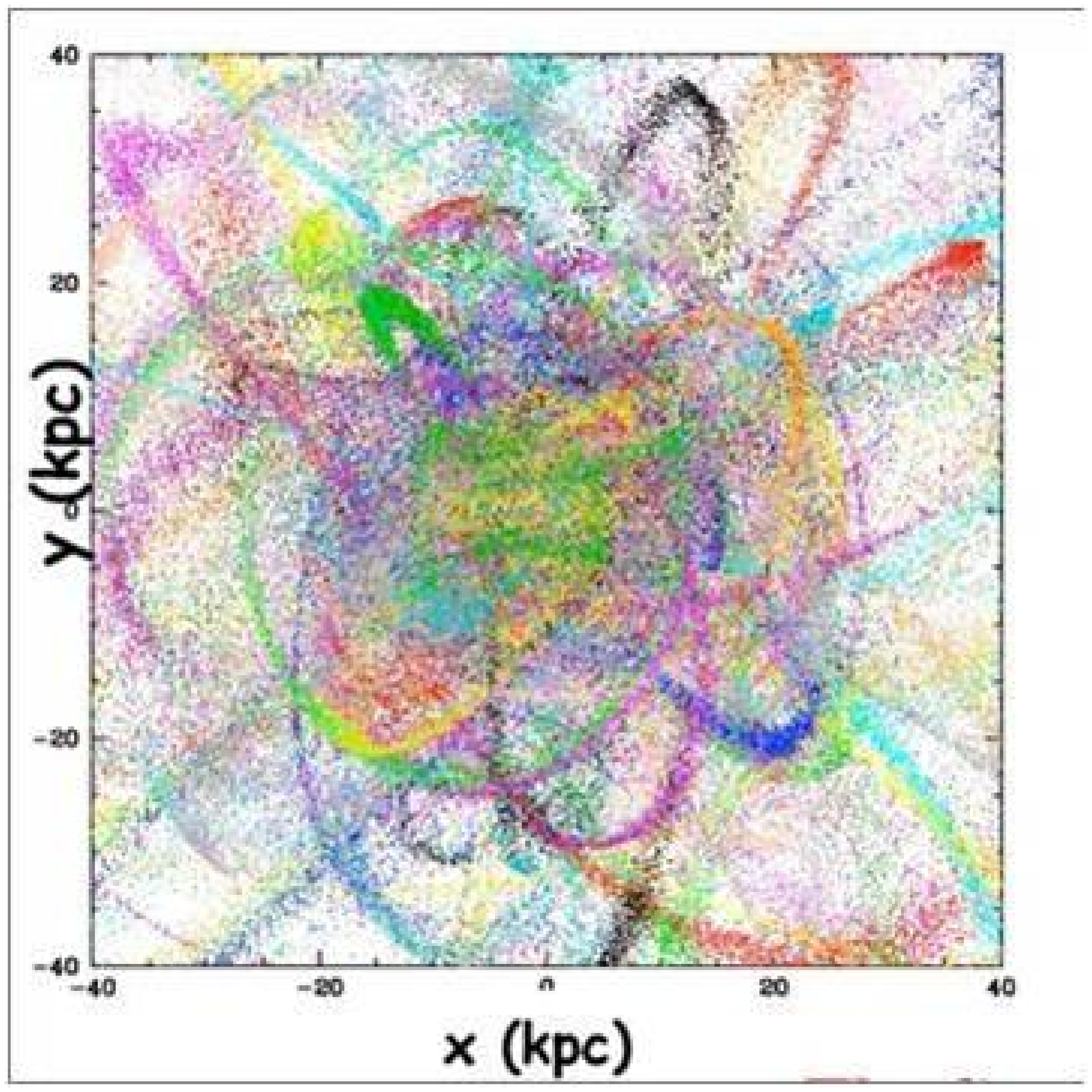}
\leavevmode\epsfxsize=9.cm \epsfbox{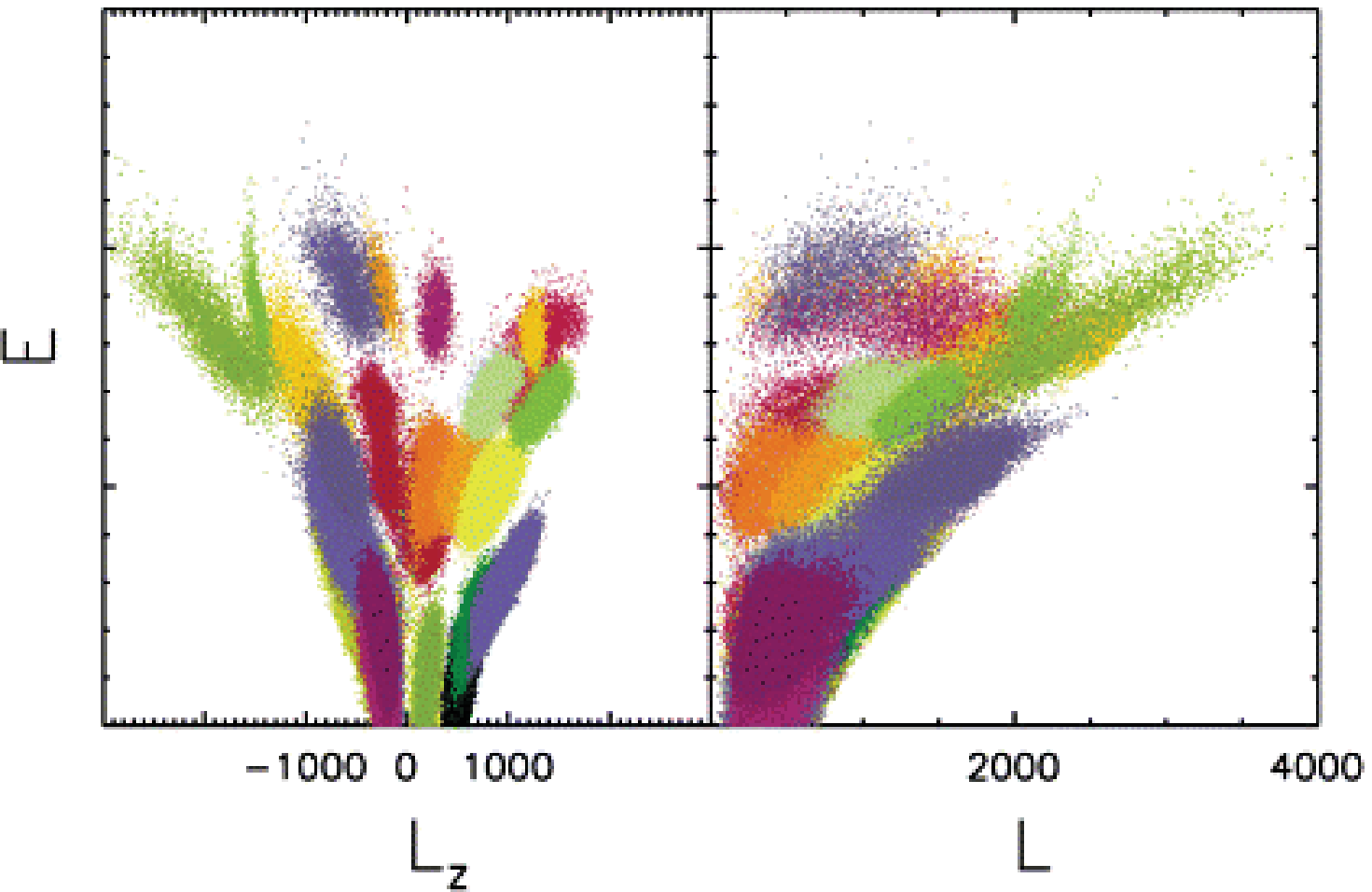}
\caption{
Top: Model distribution of tidal streams in 
a Milky Way-like galaxy
in spatial coordinates where the different
colors represent different satellites \citep[from][]{FreemanBlandHawthorn:02}.
These stream-like features disappear after several dynamical times.
Bottom: Model distribution of nearby stars in the integrals of motion space, i.e., $E$ vs. $L_z$
and $E$ vs. $L$, based on numerical simulations of satellites
falling into the Milky Way
\citep[from][]{HelmiZeeuw:00}.
The different colors represent different satellites. Shown is the final 
distribution of stars after 12 Gyr within about 6 kpc from the Sun,
after convolution with the errors expected for Gaia. It is clear that 
each of the progenitor galaxies can be traced via the current phase space distribution.
}
\label{fig:phase}
\end{centering}
\end{figure}

Our primary science goal with Subaru/PFS is thus to constrain the assembly
process of dark matter halos on Galactic scales, and in particular test
the predictions of CDM, by dedicated observations of
old stars nearby, formed at high look-back times. 
Resolved ancient stars in the
Milky Way, M31 and other Local Group galaxies are ideal targets because they 
offer us our most detailed views of galactic structure and evolution 
through their kinematics and chemical
abundances \citep[][]{FreemanBlandHawthorn:02}.
 Kinematics of stars reflect 
past galaxy collapse and/or merging events. Their distribution in phase space as 
defined by integrals of motion such as angular momentum remains basically unchanged
\cite[e.g., see Fig.~\ref{fig:phase} and][]{HelmiZeeuw:00}. 
The chemical abundances of stars reflect their
past star-formation history and chemical evolution, 
possibly in association with the dynamical state of proto-galactic clouds such as 
their collision and merging \citep[e.g.,][]{DeSilvaetal:07,NissenSchuster:10}.
All of these processes are predicted to be controlled by hierarchical 
clustering of CDM via self-gravity on galactic and sub-galactic scales
\citep[e.g.,][]{BullockJohnston:05,Cooperetal:10,Fontetal:11}.
Thus, to assess 
what CDM models predict, it is essential to derive the spatial distribution of dark matter 
subhalos in the Galaxy and M31
through their dynamical effects on visible stellar 
systems. Stars are indeed ideal tracers of a background gravitational field
dominated by dark matter.

To make progress  we propose to obtain spectra for about a million 
stars in the Milky Way and 
M31,
as one component within the framework of the proposed PFS Subaru
Strategic Program. These spectroscopic studies will be in strong synergy with the upcoming astrometric satellite, 
Gaia, that will provide very precise measurements of distances and proper motions for Galactic 
stars, and with the HSC survey which will provide the essential large imaging data set. 
Therefore, PFS, in combination with Gaia and HSC, will provide us with unique data in the area of  
near-field cosmology. With these data we hope to gain an ultimate understanding of the nature 
of dark matter and the associated galaxy formation processes.

The main questions we seek to address in our dedicated Galactic Archaeology (GA) survey
are summarized as follows.\\
\begin{enumerate}
\item{\it  What is the merging history of the Milky Way?
-- addressing the role and nature of dark matter in galaxy formation.} 
\item{\it  How did the old Galactic components (thick disk and stellar halo) form?
-- addressing baryonic physics at early epochs.} 
\item{\it How does M31 differ from the Milky Way?
-- contrast merging and baryonic processes on small scales in two
     systems.}
\end{enumerate}               

% 3.2
\subsection{Science goals with the planned medium-resolution mode}
\label{sec:GA_action}

The implementation of the planned MR mode of $R=5,000$
in the red arm will allow us to investigate further important aspects in
the GA science case. We describe here the two main science goals in GA with
the MR mode: $\alpha$-element to iron abundance ratios and precision radial velocities.

% 3.2.1
\subsubsection{Alpha elements}

\begin{figure*}[htpd]
\begin{centering}
%\leavevmode\epsfxsize=11.cm \epsfbox{alpha.eps}
\centerline{\includegraphics[width=11cm]{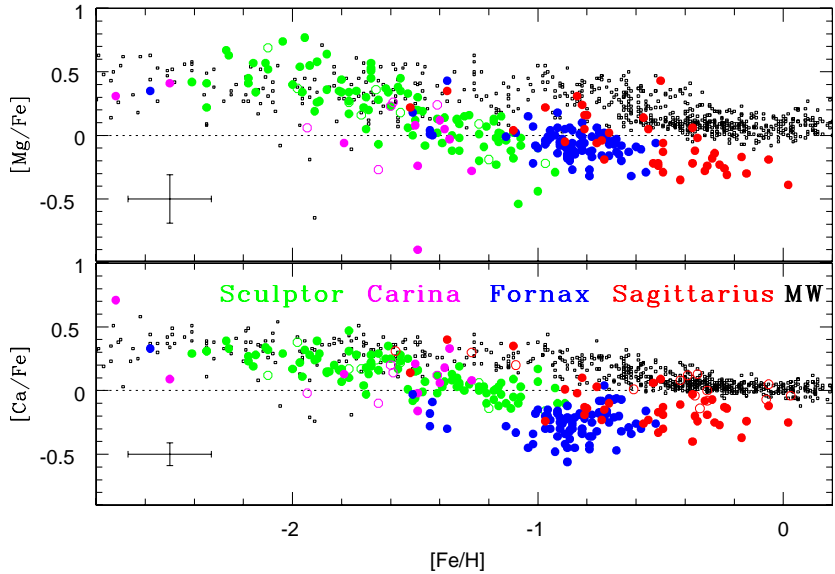}}
\caption{
Alpha-elements (Mg and Ca) in four nearby dwarf spheroidal galaxies:
Sgr, Fnx, Scl and Carina. The small black symbols denote the abundances of local MW disk
and halo stars 
\citep[from][]{Tolstoyetal:09}.
%(Tolstoy et al. 2009).
PFS will measure the $\alpha$-element abundances for a much large number of stars in the 
Galactic halo, thick disk and their substructures, in addition to  member stars across the radial extent of dSphs, identified by their kinematical and spatial distributions.
}
\label{fig:alpha}
\end{centering}
\end{figure*}

\begin{figure}[htpd]
\begin{centering}
%\leavevmode\epsfxsize=8.cm \epsfbox{elements.eps}
\centerline{\includegraphics[width=8.cm]{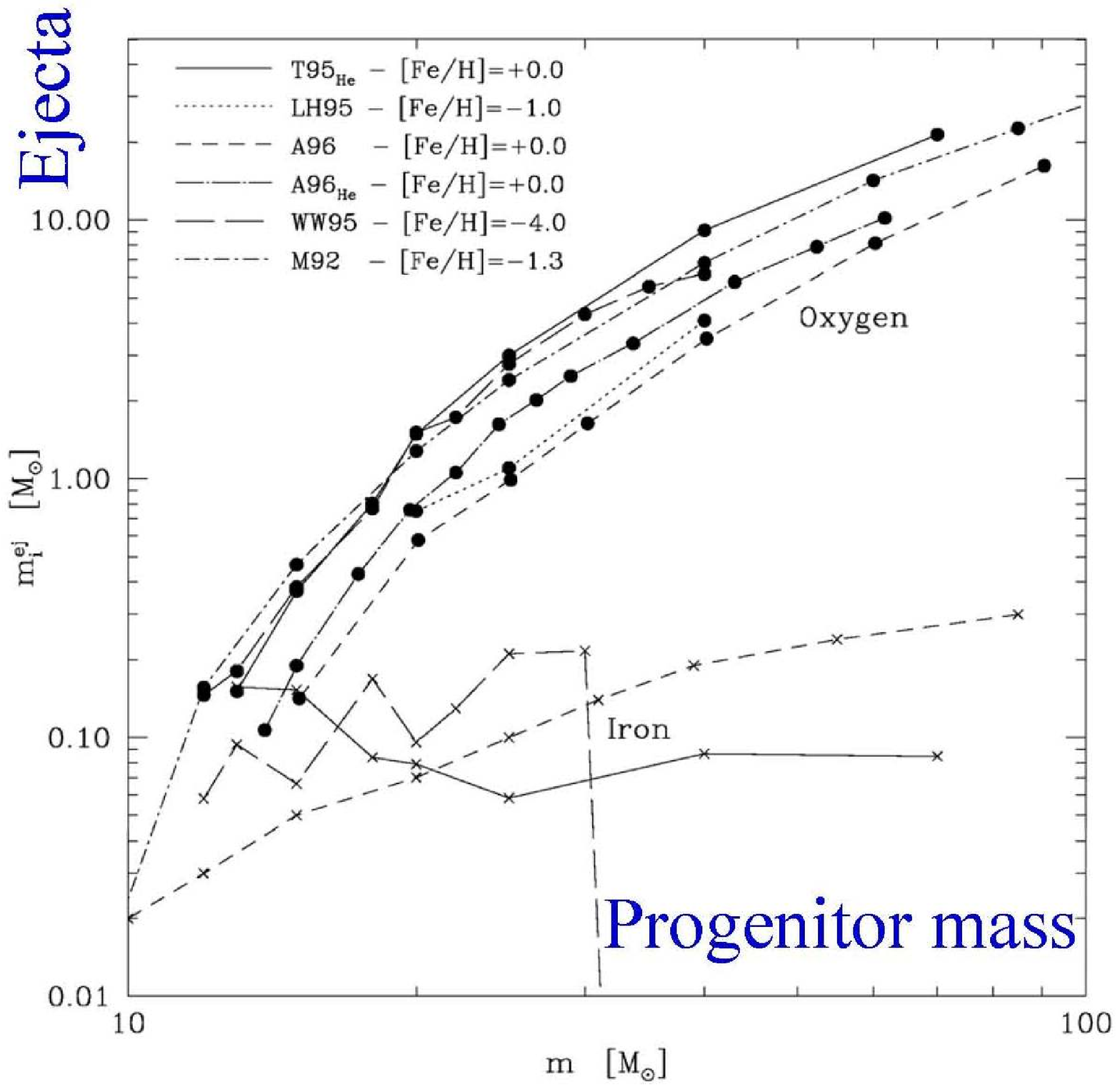}}
\caption{
Mass of ejecta from Type II SN in the form of oxygen (an $\alpha$-element; 
upper four curves) and iron (lower three curves) yields as a function of
progenitor mass 
\citep[from][]{Gibson98}.
%(Gibson 1998). 
Note that a higher oxygen yield is ejected 
from a more massive progenitor, while the iron yield is approximately constant.
}
\label{fig:oxygen}
\end{centering}
\end{figure}

The planned MR mode will enable us to unravel the temporal chemical
evolution of building blocks of the Galactic halo and thick disk, in
addition to surviving Galactic satellites, through their elemental
abundances. In addition to iron, the $\alpha$-elements observable in
the wavelength range $\lambda\simeq [700,900] $ (nm)
provide the ratios [X$/$Fe] with X$=${Mg, Si,
Ca, Ti} (Fig.~\ref{fig:alpha}).  These abundance ratios depend
sensitively on the relative contributions of core collapse (Type~II)
supernovae, from  short-lived massive stars (masses
greater than $\sim 8M_\odot$) and Type Ia SNe, which explode on longer
timescales.  Stars formed from gas that was enriched predominantly by
Type II SNe show a high [$\alpha/$Fe] ratio, reflecting
the yields of core-collapse SNe; such `high-alpha' stars will be expected
to form early in the enrichment
process, on timescales of 10$^7$ years after the onset of star formation,
a typical life time of
the SN progenitors.  Further, more massive progenitors produce higher
$\alpha$-element abundances (see Fig.~\ref{fig:oxygen} for the case
of oxygen, one of the $\alpha$-elements).  On the other hand, Type Ia
SNe produce very significant iron on longer timescales, typically
$\sim 10^{9}$ years after the birth of the progenitor stars (with $< 8
M_{\odot}$); stars formed out of gas that has been enriched by Type Ia
SNe show a low [$\alpha/$Fe] ratio. Thus, the plot of [$\alpha/$Fe]
vs. [Fe/H] shows a high [$\alpha/$Fe] plateau at low [Fe/H] where
short-lived Type II SNe dominate then a turndown at higher [Fe/H] --
but fixed delay time ($\sim $Gyrs) 
-- when Type Ia SNe start to explode and
contribute iron. For example, a system with lower star formation rate
(SFR) shows the turndown in [$\alpha/$Fe] vs. [Fe/H] at lower
[Fe/H]. Also, if the initial mass function (IMF) of core-collapse
progenitors were biased to more massive stars, the [$\alpha/$Fe]
`plateau' would have a higher value; a change of IMF slope of $\sim 1$
over the progenitor mass range of 10 to 100~$M_{\odot}$ predicts a
difference in [$\alpha/$Fe] of at least 0.3, which is thus detectable
if this ratio is measured to better than 0.2 dex 
\citep[][]{WyseGilmore:92}.

Therefore, in addition to the overall metallicity (provided by our
low-resolution (LR)
survey) which reflects the integrated star formation and chemical
evolution up to that star's birth, the measurements of
[$\alpha/$Fe] ratios provided by the MR mode will enable us to
identify substructures through chemical `tagging' with
elemental abundance patterns, establish timescales of enrichment and star formation and constrain the
IMF, for each of the major stellar components of the Milky Way and surviving  satellite galaxies.  Indeed, this is what Galactic
Archaeology should ultimately pursue as a goal, based on coordinated
observation of higher-resolution stellar spectra.

It is also worth emphasizing that Subaru/PFS will offer us a very large FoV that
is ideal for exploring Milky Way dwarf spheroidal (dSph) galaxies, such as Fornax, Ursa Minor
and Sextans, out to their tidal radii. In order to measure the enrichment 
histories of these systems, it is essential to measure their [$\alpha$/Fe]
distributions across large spatial areas, whereas present samples with elemental abundance
measurements are limited to the stars in the inner
regions \citep[e.g.,][]{Kirbyetal:09,Kirbyetal:10}. 
PFS's spectral coverage, resolution, and
field of view will make it the most powerful instrument to study
the putative building blocks of the Milky Way halo and to study
star formation at the smallest scales.

% 3.2.2
\subsubsection{Precision radial velocities}

The red-arm MR mode provides us with more precise radial velocities of stars than
the original spectrograph design can achieve. The MR mode will provide velocities with errors
$\sim 3$ km~s$^{-1}$ rather than
5--10 km~s$^{-1}$ with the LR mode. This improvement for radial velocity
measurements is essential for isolating cold kinematic substructures in 
the halo and thick disk. 
These substructures possibly originate
from building blocks of the Milky Way such as dSphs, with adiabatic evolution in phase space 'cooling' their internal kinematics. 
Arguably even more importantly, the MR mode will enable us to separate
and measure the internal kinematics of Galactic dSphs (having velocity
dispersion typically 
$<10$ km~s$^{-1}$) over much larger spatial areas---out to their tidal radii---than
previous spectroscopic studies have explored with comprehensive sampling.

\begin{figure}[htpd]
\begin{centering}
%\leavevmode\epsfxsize=5.cm \epsfbox{velsig1.eps}
%\leavevmode\epsfxsize=5.cm \epsfbox{velsig2.eps}
%\leavevmode\epsfxsize=5.cm \epsfbox{velsig3.eps}
\includegraphics[width=0.3\textwidth]{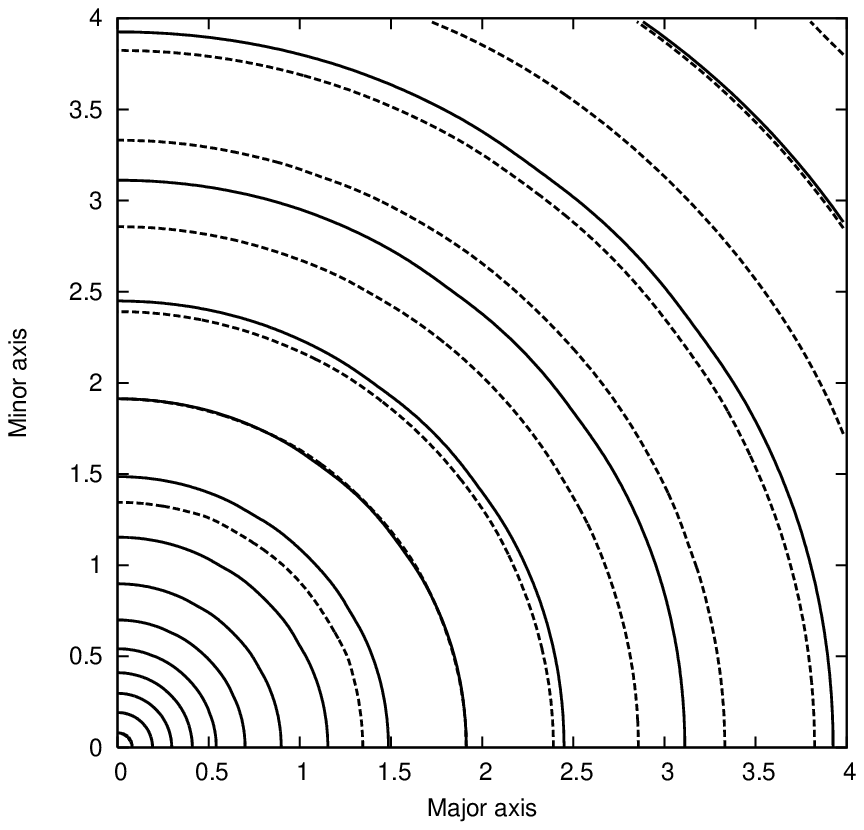}
\includegraphics[width=0.3\textwidth]{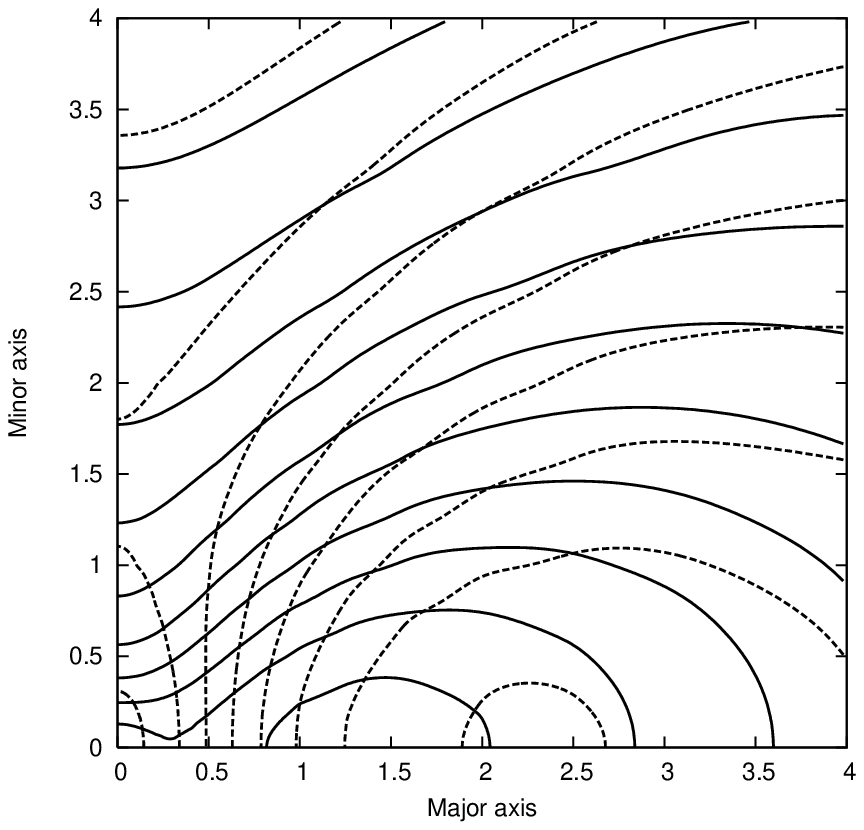}
\includegraphics[width=0.3\textwidth]{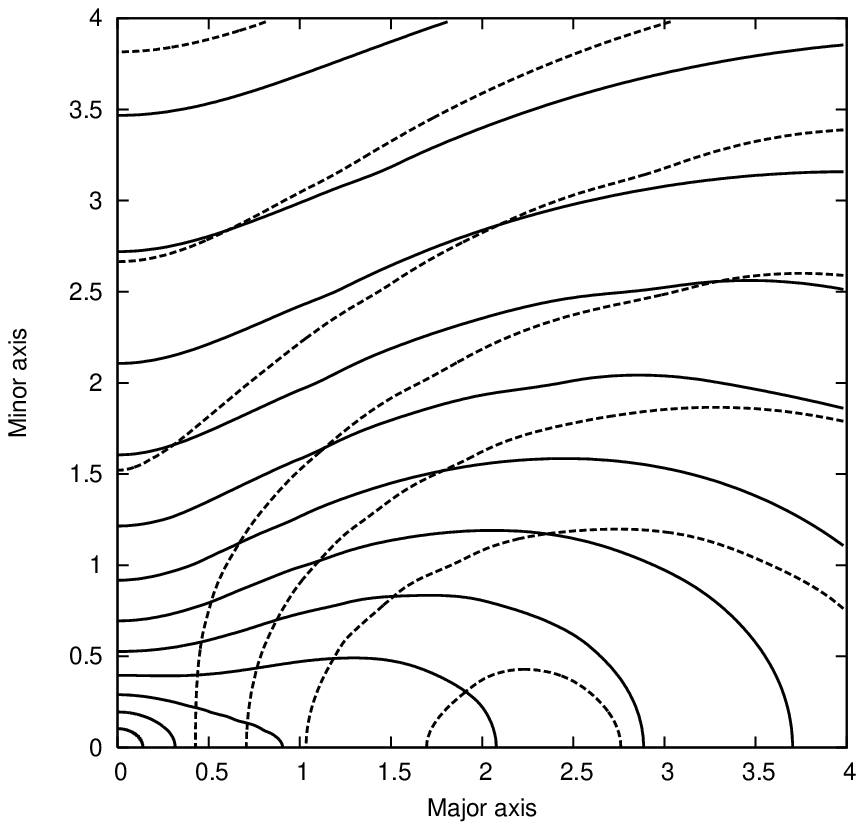}
\caption{
Contours of line of sight velocity dispersion of stars in the meridional plane
for models of dSphs,
where major and minor axes are normalized by a stellar scale length.
Solid and dashed lines show NFW and cored dark-halo models, respectively
\citep[from][]{HayashiChiba:12}.
Top: spherically symmetric model for both dark halo and luminous stellar components,
with their axial ratios denoted as $Q=1$ and $q=1$, respectively. Middle:
$(Q, q) = (1, 0.8)$. Bottom: $(Q, q) = (0.8, 0.8)$.
Note that the 2D distribution of velocity dispersion depends sensitively on
shapes and profiles of both dark halo and luminous stellar components.
PFS, with its large FoV, will play an essential role in these velocity measurements.
}
\label{fig:dSph}
\end{centering}
\end{figure}

Our present investigations of the dark-matter distribution in dSphs
are mostly limited to the inner regions, where abundant member stars
are available from spectroscopic observations even with a small FoV
\citep[e.g.,][]{Walker:09}.  Recently, \cite{WalkerPenarrubia:11}
develop a method for measuring the slopes of mass profiles in the
inner regions of dSphs and conclude that the Fornax and Sculptor dSphs
have `cores' of constant dark-matter density. This technique
relies upon the existence and identification of chemodynamically distinct sub-populations of member stars,
each defined by spatial distribution, kinematics and metallicity. 
This motivates the PFS
survey of these and other dSphs, to obtain the necessary velocity and metallicity data to  apply their method.  Further,
most existing mass models of dSphs are based on the simple assumption
of spherical symmetry, and in order to derive more detailed, realistic
spatial distributions of their dark halos, such as their global shape
and mass, we require line-of-sight velocity data for stars over much
larger areas, and full two-dimensional coverage (not just major/minor
axes), as demonstrated by recent mass models \cite[e.g., see
Fig.~\ref{fig:dSph} and][]{HayashiChiba:12}.
It is also of great importance
to have velocity data out to the tidal radii of dSphs, thereby
enabling us to estimate the true total mass of a dark halo (through
the abrupt change of the velocity dispersion profile in the edge) and
detecting any effects of Galactic tides in their outskirts.

\begin{figure}[htpd]
\begin{centering}
%\leavevmode\epsfxsize=9.cm \epsfbox{dSph_obs.eps}
\includegraphics[width=9.cm]{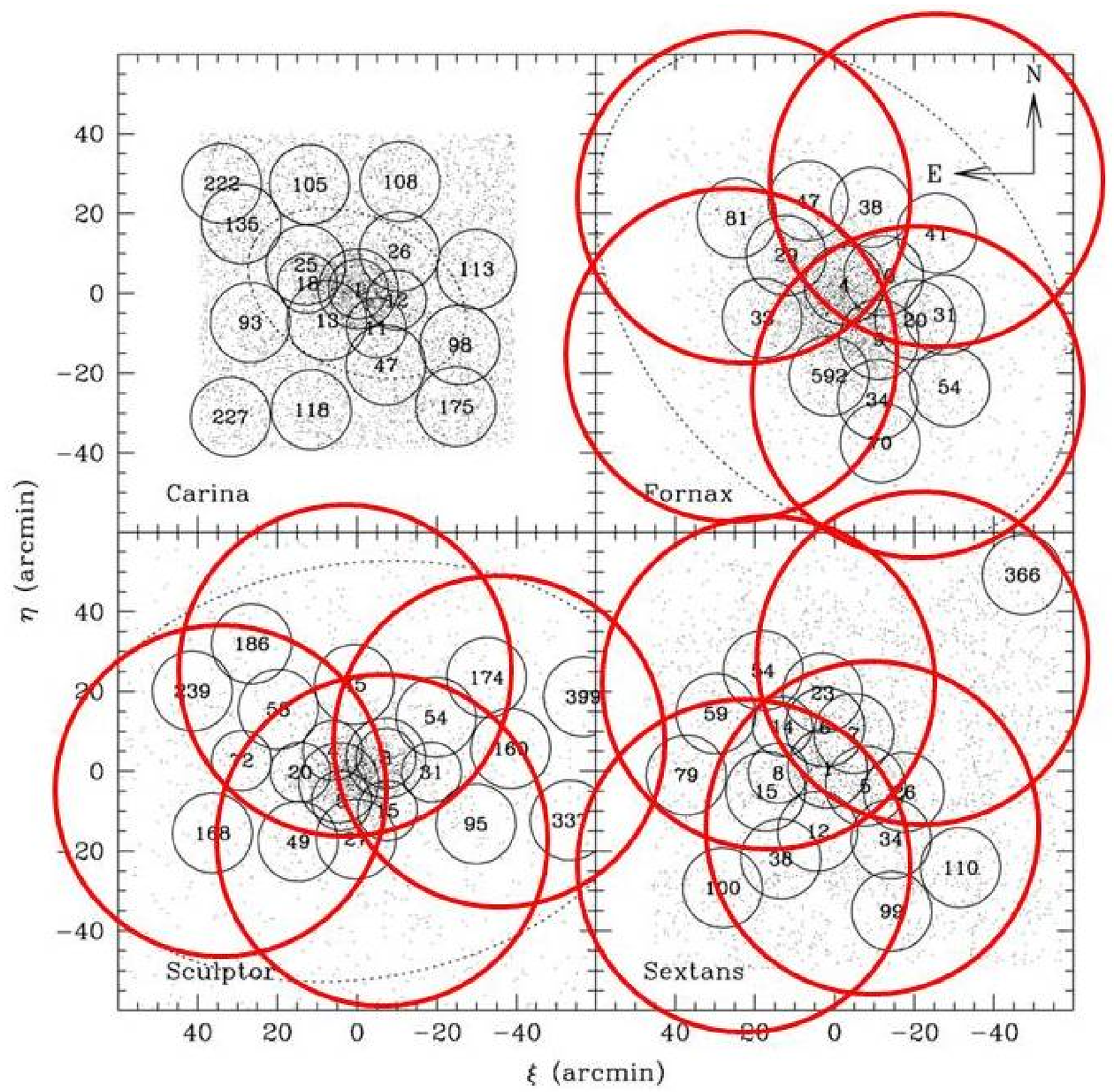}
\caption{
Proposed Subaru/PFS pointings (red circles) for the wide-field velocity measurements
of some Galactic dSphs (Fornax, Sculptor and Sextans) reaching their tidal radii
(dotted lines). Note that previous velocity measurements were limited only
to the central parts (black circles) 
\citep[adapted from][]{Walker:09}.
%(adapted from Walker et al. 2009).
}
\label{fig:dSph_obs}
\end{centering}
\end{figure}

The advent of the large FoV and fiber multiplexing of  Subaru/PFS is tremendously
advantageous in selecting member stars of dSphs from background/foreground
field stars out to their tidal radii 
(Fig.~\ref{fig:dSph_obs}). It is worth noting here that since 
the probability of
finding true member stars in the outer parts of dSphs is generally small,
a pre-imaging survey of target dSphs using HSC with a newly developed narrow-band
filter
(NB515 filter with CW = 5145 \AA \ and BW = 80 \AA, which is similar to the DDO51
filter\footnote{The DDO51 
filter is an intermediate filter in the Washington photometric
system, where the strong surface gravity sensitivity of the Mg~I triplet
and MgH band features near 5150 \AA \ allows discrimination between
dwarfs and giants of F-K spectral type.})
will be very useful for efficiently distinguishing red-giant branch (RGB)
stars in dSphs from foreground dwarf stars. Our experiments using Suprime Cam
with a proto-type NB515 filter suggest that the number of RGB candidates within
a fiducial RGB locus of a color-magnitude diagram for observed stars,
after NB515-imaging selection, is narrowed down to about 50~\% of that
without using this NB filter (Tanaka et al. 2013 in preparation).
Thus, the combination of HSC and PFS will allow us to explore
this important dark-matter science; dSphs are the ideal sites
for studying the nature of dark matter because these galaxies are largely
dark-matter dominated and their past star formation is not likely to have modified the dark-matter distribution. Furthermore, this selection of member stars in dSphs
is essential for the measurement of their $\alpha$-element abundances
with the same MR mode out to large radii, thereby enabling us
to characterize the global chemo-dynamical evolution of dSphs.

\begin{deluxetable*}{lcccccl}
\tablewidth{0pt}
\tabletypesize{\footnotesize}
\tablecaption{Summary of Galactic Archaeology Survey
\label{tab:GA_survey}}
\startdata \hline\hline 
Survey        & Mode &  Mag. Range & Exposures & No. Fields  & Survey Time & Comments \\
              &      &   (mag)     & (sec)     &             &  (nights)   &          \\
\hline\hline
{\it Bright time:} & &             &                 &             &             &          \\
The Milky Way & MR+LR(blue) & $V<19$ & 2700 & 208 &  20   &  Thick disk: $b=30$, $0<l<270$    \\
The Milky Way & MR+LR(blue) & $V<19$ & 2700 &  46 &   5   &  Thick disk: $b=-30$, $60<l<120$  \\
\hline
{\it Grey time:}   & &             &                 &             &             &           \\
dSph          & MR+LR(blue) & $V<21$ & 7200 &   28  &  7  &  Fnx, Scl, LeoI, UMi \& Dra  \\
The Milky Way & MR+LR(blue) & $V<21$ & 7200 &   40  & 10  &  Halo: $47<b<73$, $l=90$ \& 270   \\
The Milky Way & LR   &  $V<22$      &  7200 &   48  & 12  &  Outer disk: $l=180$       \\
The Milky Way & LR   &  $V<22$      &  7200 &   24  &  6  &  `Field of Streams'       \\
\hline
{\it Dark time:}  &  &             &                  &            &             &           \\
dSph          & MR+LR(blue) & $V<21$ & 7200 &   8  &   2   &  Sextans                \\
M31 halo      & LR   & $21.5<V<22.5$ & 18000 & 50  &  31   &  HSC sample             \\
dIrr          & LR   & $V<22.5$      & 18000 &  4  &   3   &  NGC6822, IC10 \& WLM   \\
\hline
Total         &      &               &       & 456  &  96   &                  
\enddata
\end{deluxetable*}

% 3.3
\subsection{GA survey plan}

Now that the MR mode will allow us to 
address GA science subjects
beyond what is achievable with the LR mode as described above,
we summarize our survey plan in Table \ref{tab:GA_survey}.
The primary targets are stars in the Milky Way and dSphs using both the MR and LR
modes and bright red giant stars in a wide area of M31's halo with the LR mode.

\subsubsection{The Milky Way and dSphs}

As detailed in Section \ref{sec:alphameas}, the MR mode will enable 
us to obtain the ratios [X$/$Fe] with X$=${Mg, Si, Ca, Ti} with precisions of 
of $\Delta$[X$/$Fe]$\sim 0.2$~dex for spectra with S$/$N of about 60 per resolution element.
Primary Milky Way targets in our dedicated large survey 
using this higher-resolution mode will be bright stars with $V < 19$ mag,
which are observable even in bright time; exposure times of 2700 sec are needed
to obtain sufficiently accurate [X$/$Fe] ratios. The MR mode will be further dedicated
to the survey of fainter Milky Way stars, in addition to  member stars of dSphs with $V < 21$ mag
in grey/dark time, for which we require exposure times of 7200 sec.
Thus, this mode will offer us
an efficient usage of allocated PFS nights including some bright/grey time.

The LR PFS survey of radial velocities combined with Gaia astrometry data will
provide six-dimensional phase-space information for Galactic stars, with which
it is possible to identify individual progenitors as building blocks of the Milky
Way, in the form of kinematically distinct clumps in phase space
\citep[][]{HelmiZeeuw:00,Gomezetal:10}.
With [$\alpha$/Fe] ratios of these stars obtained from our MR spectroscopy,
it will be possible to set important limits on the star formation and chemical
evolution histories in each building block of the Milky Way halo. Furthermore,
the [$\alpha$/Fe] ratios are excellent discriminators of the thick disk from
the thin disk, with which one can set important limits on the formation scenario
of the thick disk by investigating its rotational and orbital properties
as a function of [Fe/H] 
\citep[][]{Leeetal:11}.
We will also quantify the  velocity distribution of the thick disk as a function of spatial coordinates and the asymmetry/symmetry of its dynamical
properties,  providing constraints on the role of dynamical heating by CDM subhalos and/or visible satellites in thick disk formation
\cite[e.g.,][]{HayashiChiba:06}.
To address these issues, we focus on three 
regions sensitive to azimuthal streaming velocity (disk rotation) at the
halo/disk interface:
$(l,b)=(90^\circ,\pm 30^\circ)$ and $(270^\circ, +30^\circ)$,
where
each block includes a sufficient number of targets
(e.g., as many as 2000 stars per PFS FoV, including more than 500 F/G turnoff stars
with $V < 20$ mag)
to derive the spatial dependence of the velocity distribution.
Full spatial motions of these stars, in synergy with Gaia, and their [$\alpha$/Fe] ratios
from the MR mode of PFS will allow us to gain
new insights into the origin of 
the thick disk as well as the star formation history of halo progenitors.
In addition, taking advantage of the wide FoV of PFS, we will also target other Milky Way
samples and obtain the distribution of [$\alpha$/Fe] ratios
for selected bright stream stars ($V < 19$ mag), so that their star-formation
history is constrained and compared to other regions. This survey requires in total $\sim 300$ PFS
pointings ($\sim 390$ sq. degree), for each of which 2700 sec of exposure is required.

\begin{figure}
\begin{centering}
%\leavevmode\epsfxsize=7.cm \epsfbox{monoceros.eps}
\includegraphics[width=7.cm]{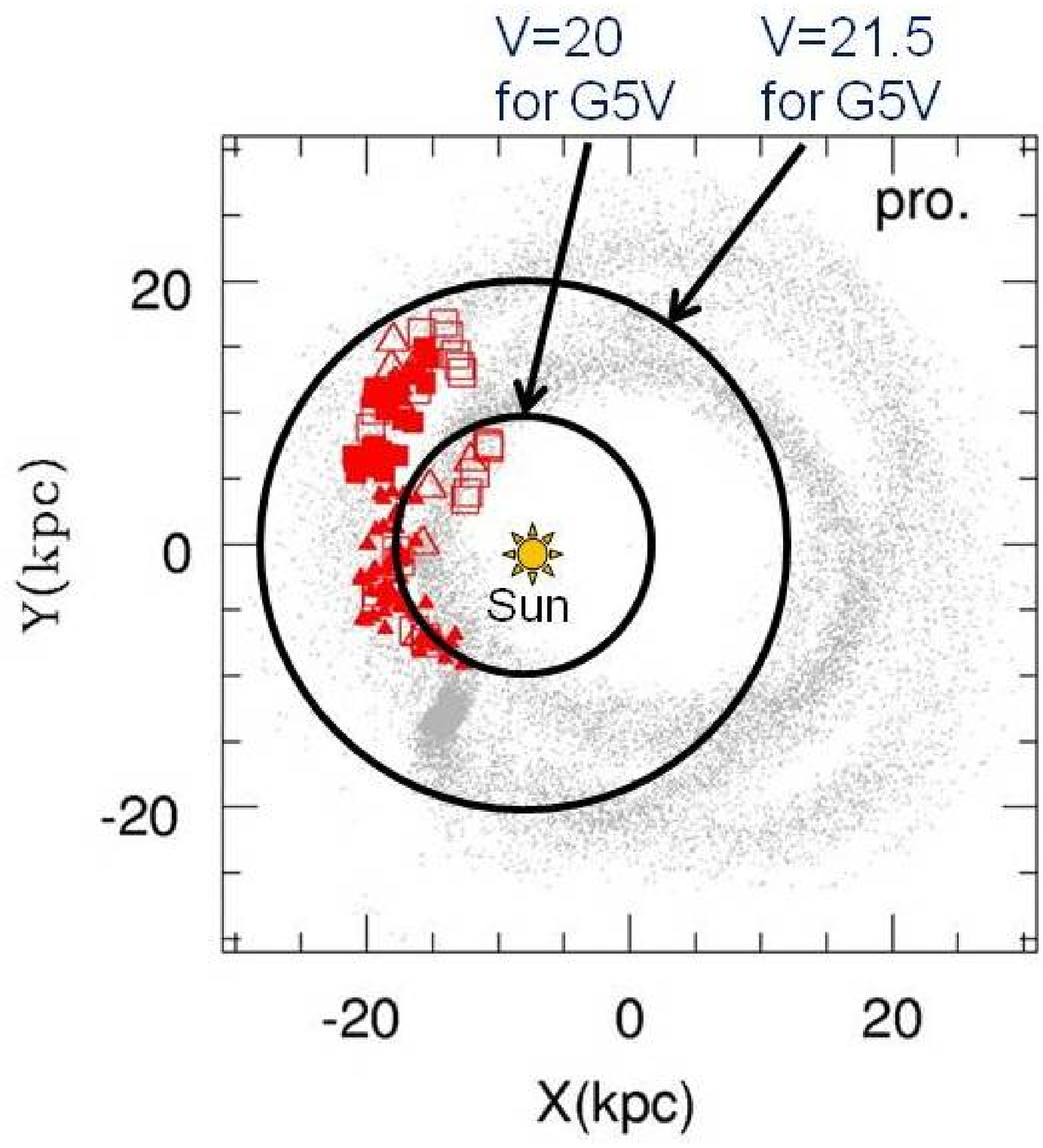}
\caption{
Spatial distribution of simulated tidal debris of a merging dwarf galaxy
associated with the Monoceros stream 
\citep[adapted from][]{Penarrubiaetal:05}.
%(adapted from Pe\~narrubia et al. 2005).
The Sun is placed at $(X, Y, Z) = (-8, 0, 0)$~kpc. Red triangles and squares
denote positions of previously detected stream stars. Note that $V = 20$ mag limit
for G5V stars is largely incomplete in probing these tidal debris
compared with $V = 22$ mag limit.
}
\label{fig:monoceros}
\end{centering}
\end{figure}

In addition to these survey regions using the MR mode, we plan to observe Milky Way stars
as faint as $V = 22$ mag, with the LR mode, in the regions of the outer disk near $l = 180^\circ$
and in the `Field of Streams', to clarify the chemo-dynamical nature and origin of these
substructures. In particular, outer disk regions including the so-called Monoceros stream
are expected to contain abundant tidal debris from a merging dwarf galaxy   
\citep[e.g.][]{Penarrubiaetal:05}
which can be probed by
observing G5V stars as faint as $V = 21.5-22$ mag with Subaru/PFS; 
a $V = 20$ mag limit, practical for 4m telescope projects, would be largely incomplete in probing these merger remnants
(see Fig.~\ref{fig:monoceros}).

\begin{figure}
\begin{centering}
%\leavevmode\epsfxsize=11.cm \epsfbox{Survey_Fields.eps}
\includegraphics[width=8.cm]{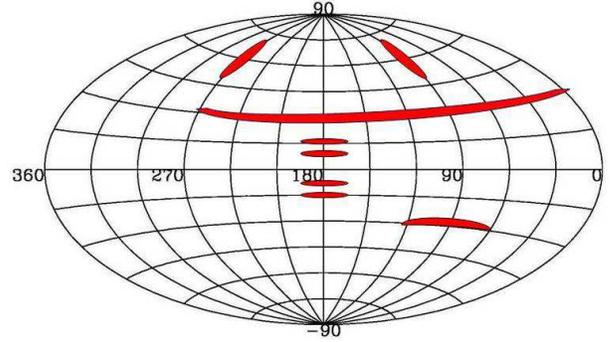}
\caption{
Proposed survey fields in the thick disk, stellar halo, and outer disk regions
of the Milky Way in the galactic coordinate system.
}
\label{fig:Survey_Fields}
\end{centering}
\end{figure}

Our proposed survey regions 
are 
summarized in Fig.~\ref{fig:Survey_Fields}.
We note the number of sample stars required in each survey mode, i.e.
for thick disk, stellar halo and outer disk, follows the justification of the required sample
size described in Section~\ref{sec:GA_action}.

As described in the previous section, the MR mode in Subaru/PFS will be essential
to determine the
 global chemo-dynamical state of Galactic dSphs across their radial extent.
We will use the MR mode for most of these dSphs in grey time, where we will observe
member stars as faint as $V = 21$ mag with 7200 sec of exposure for each PFS
pointing. 
We require four PFS pointings for those dSph with large tidal radii
(Fornax, Sculptor and UMi: see Fig.~\ref{fig:dSph_obs})
and one PFS pointing for each of LeoI and Draco.
Since repeat velocity measurements are necessary for 
identifying
binary effects,
we require at least 28 PFS pointings in total for these dSphs. Sextans has 
a large tidal radius ($\sim 160$ arcmin), for which we need many PFS pointings. 
To efficiently execute this observing run, we will request
dark time in spring for this particular target.
For a sample of Local Group dIrr galaxies,
we will target RGB stars down to $V
= 22.5$ mag (in the summer/fall season) using the LR mode and
in order to obtain S$/$N as large as 20 at $\lambda = 5,000$ \AA,
we require dark time and 5~hr exposures per pointing for these stars .

The Gaia intermediate data release is
expected by the time the PFS GA survey commences. Gaia and ground-based imaging surveys, such as
Pan-Starrs, VST, and Skymapper, are likely to find many streams and
ultra-faint dwarf galaxies. The MR mode of PFS will uniquely provide the resolution and field of
view necessary for follow-up spectroscopy of these new-found structures.

\begin{figure*}[htpd]
\begin{centering}
%\leavevmode\epsfxsize=11.cm \epsfbox{M31_PFSobs.eps}
\centerline{\includegraphics[width=11.cm]{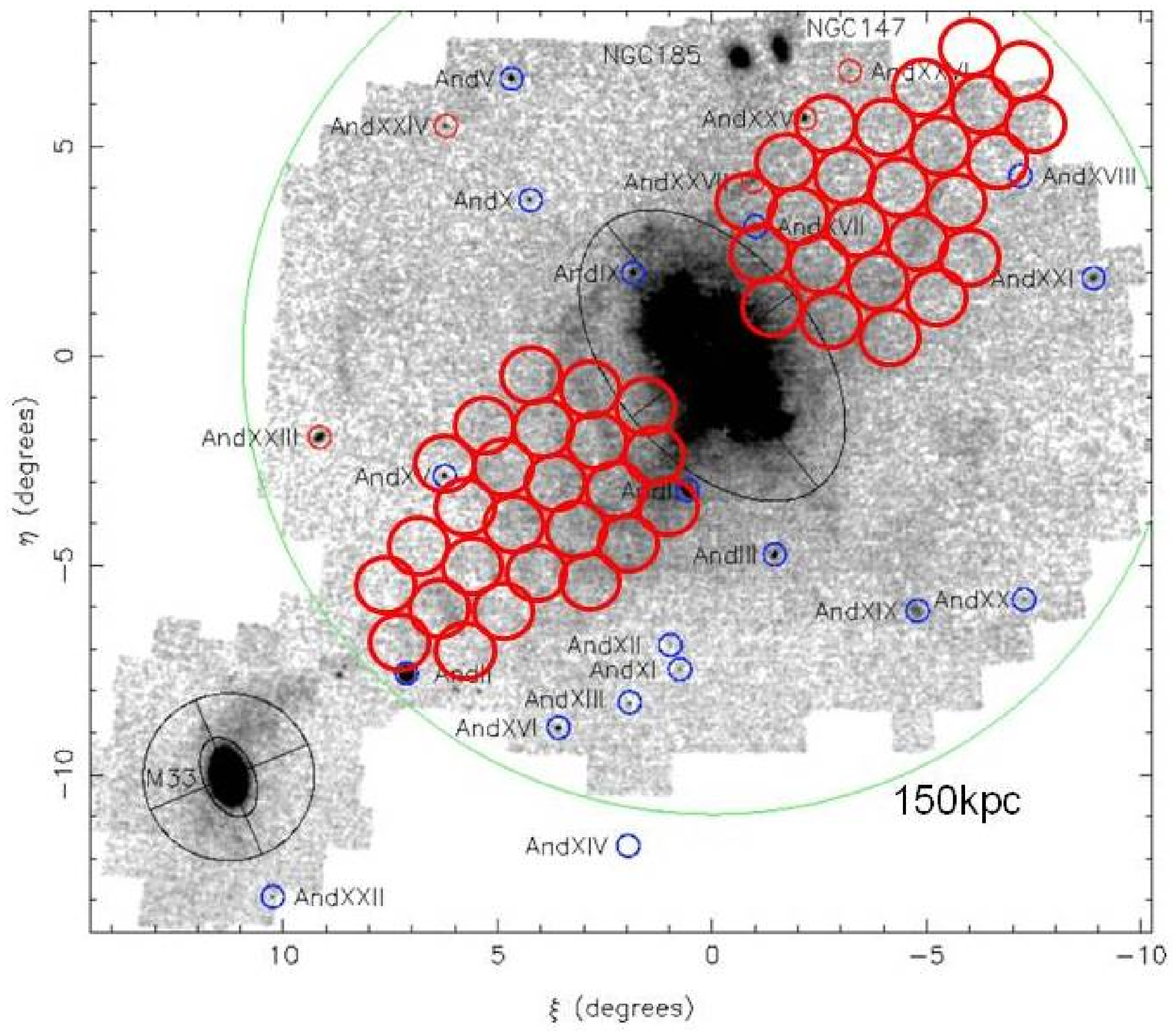}}
\caption{
Proposed PFS pointings along the minor axis of M31's halo ($\sim 50$ pointings),
in which several stellar streams are included. The map of RGB stars is taken
from 
\cite{Richardsonetal:11}.
%Richardson et al. (2011).
}
\label{fig:m31_HSC}
\end{centering}
\end{figure*}

%%%
\subsubsection{The M31 halo}

The primary targets in the M31 survey are bright red giants with $21.5 < V < 22.5$ mag, i.e.,
stars around the tip of the RGB with $I \simeq$ 20.5 selected along the minor axis of
the galaxy (Fig. \ref{fig:m31_HSC}). Pre-imaging observations and selection of the candidate halo giants
will be provided by an HSC imaging survey and we will utilize a newly developed narrow-band filter,
NB515, to assist in removing foreground Galactic
dwarfs. A survey along the minor axis will enable us to derive the chemodynamical
properties of the general M31 halo as well as to study in detail several stellar streams,
including a cold stream-like feature in the north-west part of the halo, part
of which corresponds to Stream~F discovered by previous Subaru observations
\citep[][]{Tanakaetal:10}.
The distribution of stars along these M31 streams will be especially useful in 
constraining the number of orbiting dark matter subhalos, which are predicted to induce  detectable  density 
variations along the stream through dynamical disturbances
\citep[][]{Carlberg:11}.
A careful subtraction of foreground Milky Way stars through spectroscopic observations will 
be crucial for constraining the number of orbiting CDM subhalos in Andromeda and 
comparing with theoretical predictions. 

PFS will measure radial velocities and metal abundances for M31 halo stars to set important 
constraints on their intrinsic chemo-dynamical properties. Spectroscopic 
metallicities derived for individual stars will represent a major advance over those derived photometrically or by stacking
spectra. To estimate the number of RGB stars in the M31 halo per PFS field, we follow
the recent observations with Keck/DEIMOS by \citet{Gilbertetal:12},
in which the pre-selection of candidate RGB targets using the DDO51
filter was made.
At a projected distance of 80~kpc from the M31 center, we expect 
$\sim 2600$ objects per PFS field (including significant foreground/background
contamination even after NB515 pre-selection), among which
about 100 secure RGB stars of the M31 halo can be extracted.
We propose to observe $\sim 66$ sq. degrees ($\sim 50$ pointings)
as shown in Fig.~\ref{fig:m31_HSC}.
In order to obtain spectra with S/N$\simeq 20$ per resolution element
at $\lambda = 5000$ \AA,  5 hr exposures are required at  $V = 22.5$ mag. 
Thus, $\simeq$30 nights are required for this component of the survey.

In summary, for all the GA survey targeting the Milky Way, dSphs and the M31 halo,
we require a total of 96 nights.

\begin{figure*}[htpd]
\begin{centering}
%\leavevmode\epsfxsize=13.cm \epsfbox{MR_spec.eps}
\centerline{\includegraphics[width=13.cm]{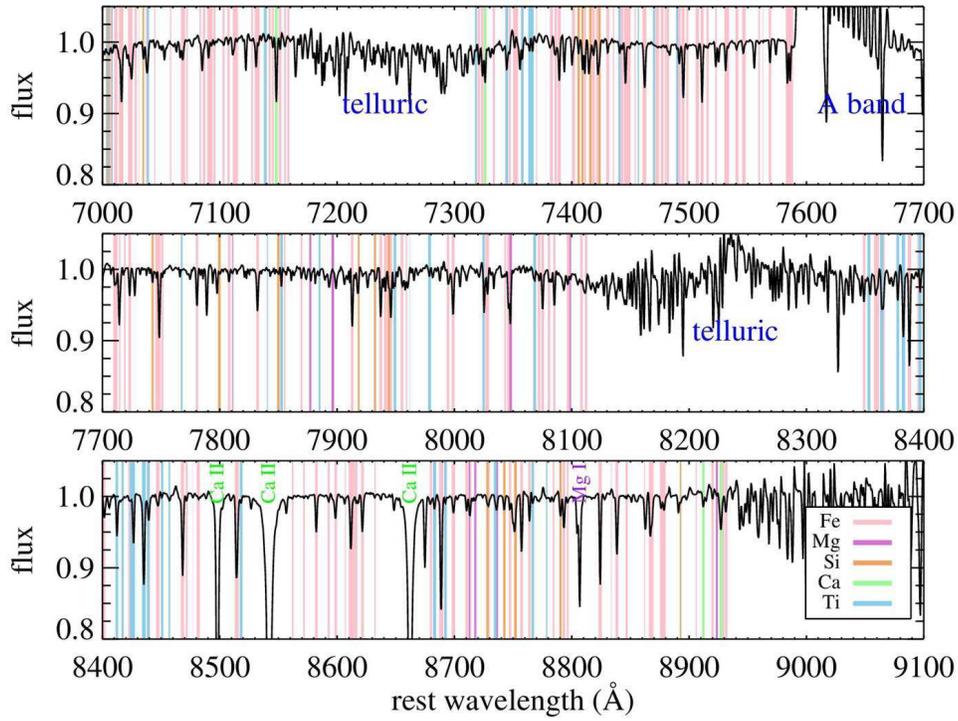}}
\caption{
Spectrum of a red giant star in the Sculptor dSph, with $\Delta\lambda = 1.5$ \AA. This has been  obtained by slightly degrading a Keck/DEIMOS spectrum with $\Delta\lambda = 
1.2$ \AA. This spectral region contains many weak metal lines as well as significant 
Ca II triplet lines.
}
\label{fig:MR_spec}
\end{centering}
\end{figure*}

% 1.4
\subsection{Estimation of stellar atmospheric parameters from MR spectra}
\label{sec:alphameas}

This medium-resolution survey at $R=5,000$ will yield stellar spectra similar to those 
provided by the Keck/DEIMOS instrument, with which we have sufficient experience and 
well-calibrated techniques to measure multi-element $\alpha$ abundances and quantify
their uncertainties
\citep[][]{Kirbyetal:08,Kirbyetal:10}.
MR spectra with 1.3 \AA \ to 1.5 \AA \ FWHM in the red spectral region (7,000 \AA \
$\le \lambda \le$ 9,000 \AA, see Fig.~\ref{fig:MR_spec}) yield
multiple abundance dimensions, such as [Fe/H] and 
[$\alpha$/Fe], as well as other stellar atmospheric parameters, as clearly demonstrated 
by E.~Kirby and his collaborators using Keck/DEIMOS data 
\citep[][]{Kirbyetal:08,Kirbyetal:09}.
The method makes use of a large grid of stellar synthetic spectra parametrized by 
effective temperature ($T_{\rm eff}$), surface gravity 
($\log~g$), metallicity ([Fe/H]), and $\alpha$-element abundance 
[$\alpha$/Fe]). The atmospheric parameters and metal 
abundances are then derived with the $\chi^2$ fitting to an observed PFS spectrum.
This spectral modeling method 
for the determination of metallicities is advantageous compared to an empirical 
spectrophotometric one, such as the Ca~II infrared triplet method, where the latter 
is affected by the assumed ratio of [Ca/Fe] and limited to the calibrated range of metallicity.
Furthermore, this method enables us to obtain individual 
abundances of $\alpha$-elements in addition to Fe, namely Mg, Si, Ca, and Ti, which 
characterize the yields of massive stars that exploded as Type II SNe.

\begin{figure}[htpd]
\begin{centering}
%\leavevmode\epsfxsize=7.cm \epsfbox{LR_Mg.eps}
%\leavevmode\epsfxsize=7.cm \epsfbox{MR_Mg.eps}
\includegraphics[width=7.cm]{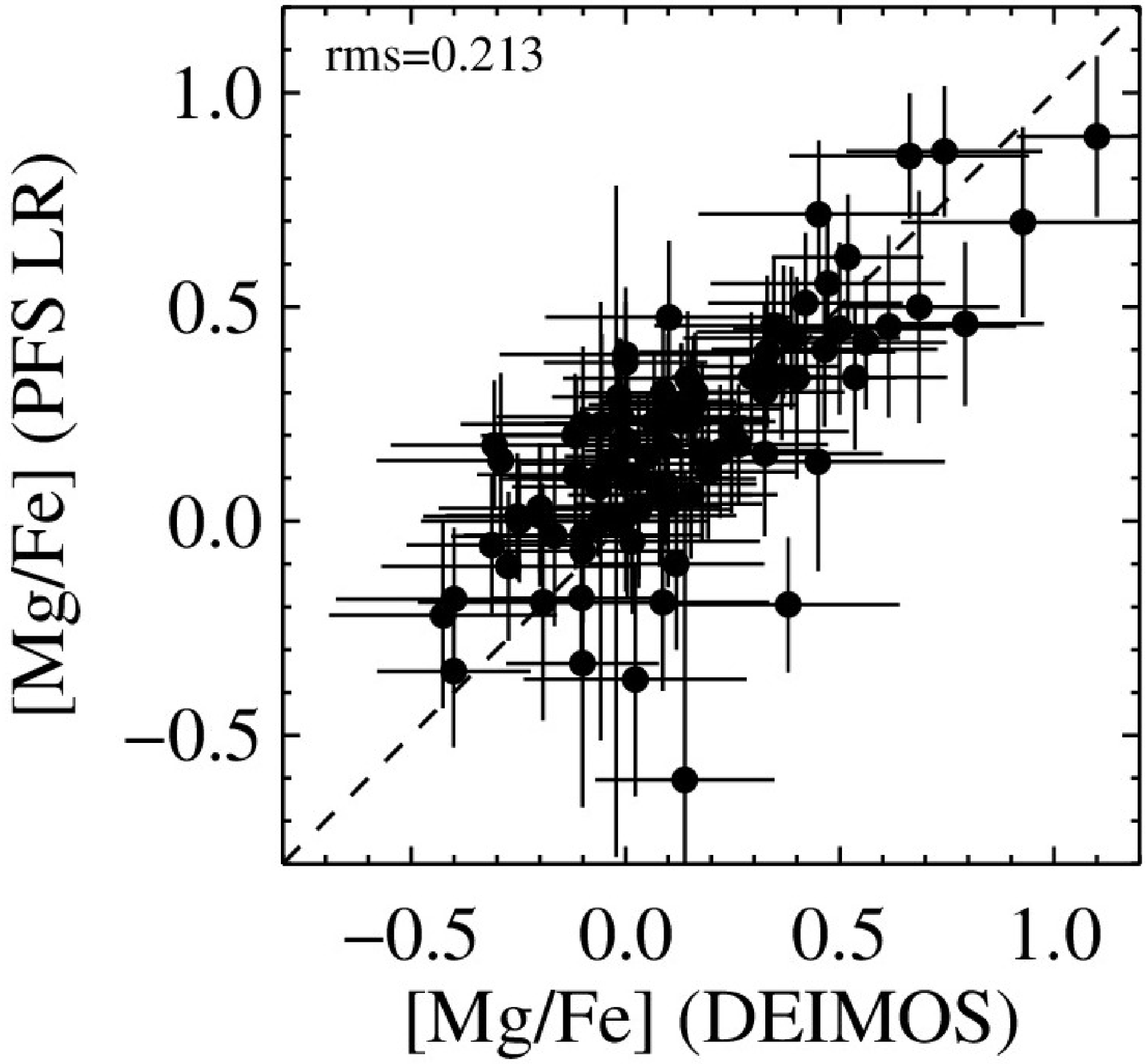}
\includegraphics[width=7.cm]{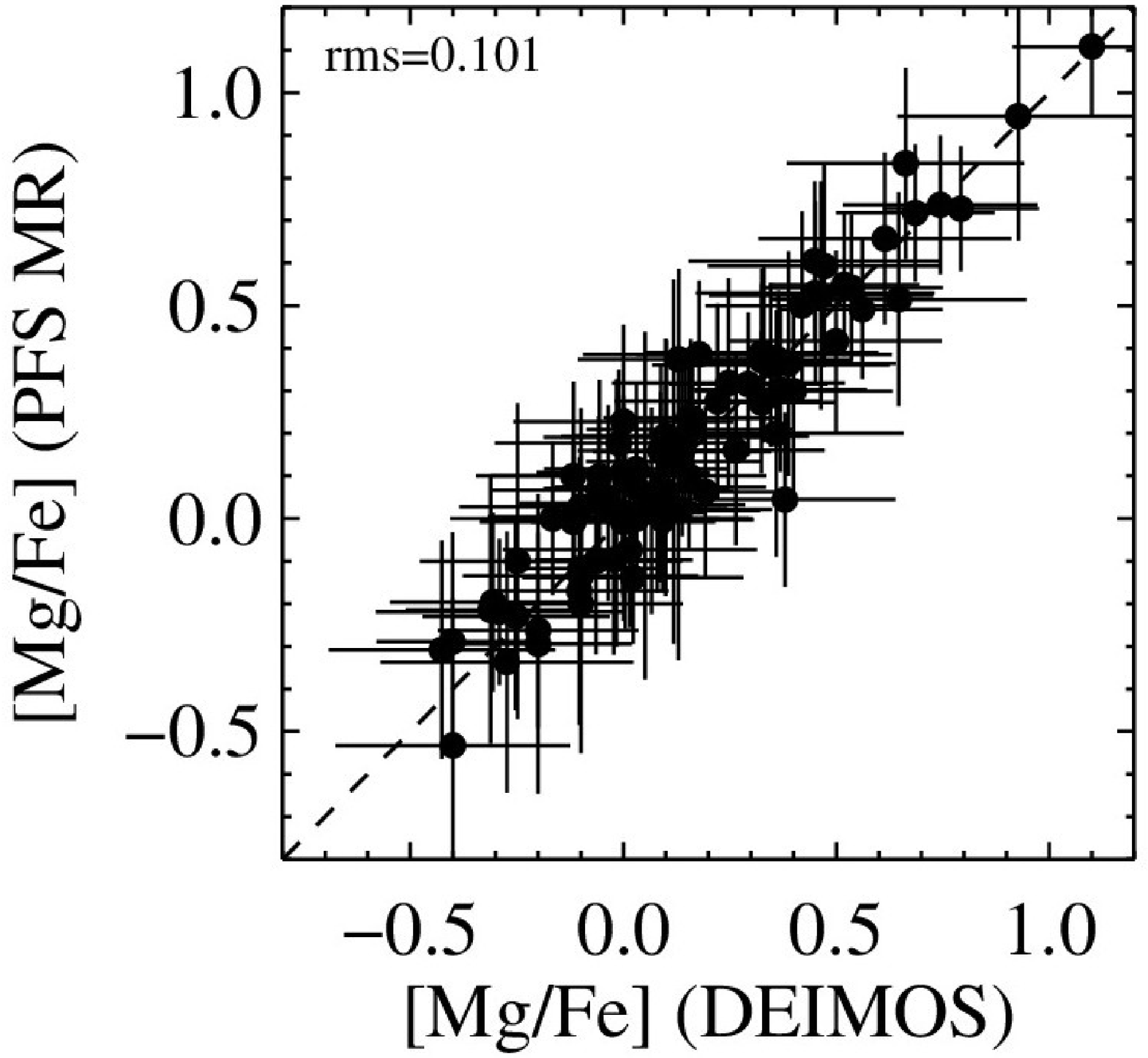}
\caption{
Left: Comparison of [Mg/Fe] ratios derived from PFS-LR spectra ($\Delta\lambda = 
3$ \AA) to those of the original DEIMOS data ($\Delta\lambda = 1.2$ \AA; where the 
former spectra are made by degrading the resolution of the latter ones), for 400 red 
giant stars in the Sculptor dSph.. Right: The same comparison but for PFS-MR spectra 
($\Delta\lambda = 1.4$ \AA). These panels indicate that the expected errors in [Mg/Fe] derived 
from PFS-LR spectra ($\sim 0.3$ dex) are too large to distinguish field halo stars 
with [Mg/Fe]$\sim 0.3$ dex from disk stars and/or recently accreted stars with 
[Mg/Fe]$\le 0.0$.  On the other hand, PFS-MR provides the [Mg/Fe] measurement
precision required for this determination.
}
\label{fig:Mg_comp}
\end{centering}
\end{figure}

Fig.~\ref{fig:Mg_comp} shows the feasibility of deriving [Fe/H] and [Mg/Fe] 
from PFS-MR spectra of $\Delta \lambda = 1.4$ \AA, in comparison with 
$\Delta \lambda = 3$ \AA (LR mode).
It is clear from these figures that the uncertainty in [Mg/Fe]
derived from PFS-MR spectra should be of order  $\sim 0.2$ dex
(taking into account an error of [Mg/Fe]$\sim 0.15$ dex in DEIMOS) and is
thus small enough to distinguish field halo stars 
with [Mg/Fe]$\sim 0.3$ dex from disk stars and/or recently accreted stars with 
[Mg/Fe]$\le 0.0$. Our simulations using DEIMOS data also suggest that this
precision of multi-element $\alpha$ abundances is possible with $\sim 0.75$ hour
exposures for bright stars of $V < 19$ in bright time and $\sim 2$ hour
exposures for stars as faint as $V = 21$ in grey/dark time.
\begin{deluxetable*}{cl}
\tablewidth{0pt}
\tabletypesize{\footnotesize}
\tablecaption{Summary of Galactic Archaeology Requirements
\label{tab:GA_requirements}}
\startdata \hline\hline 
Mode    & Requirements \& Comments \\
\hline
LR               & For the Milky Way stars ($V < 22$) and M31 halo ($21.5 < V < 22.5$) \\
($R=$2000-3000)  & Velocity precision of 5--10 km~s$^{-1}$ \\
                 & [Fe/H] to $\sim 0.2$ dex, no other elements measurable       \\
                 & $\lambda=$3800 \AA \ to 1 $\mu$m incl. Ca~II HK, Ca~I, Mgb/MgH, CaT \\
                 & $\sim 2200$ fibers, $\sim 10^6$ stars                          \\
\hline
MR               & For the Milky Way stars ($V < 19$ in bright and $V < 21$ in grey/dark time) \\
($R=$5000)       & Velocity precision of 3 km~s$^{-1}$  \\
                 & [Fe/H] to $0.15$ dex, [X/Fe] (X=Mg, Si, Ca, Ti) to $\sim 0.2$ dex  \\
                 & $\lambda=$7100 to 8850 \AA \ incl. CaT and $\alpha$-element lines  \\
                 & $\sim 1000$ fibers, $\sim 10^5$ stars                        
\enddata
\end{deluxetable*}

% 1.5
\subsection{Instrument requirements for MR GA survey}

For the MR mode, a spectral resolution at least as high as
$\Delta\lambda=1.5$ \AA\ and S/N$=60$ per resolution
element over the wavelength range 7,100 \AA \ to 8,850 \AA\ will  allow
us to determine [X/Fe], with X$=${Mg, Si, Ca, Ti}, to a precision of
$\sim 0.2$ dex.  The resolution of the LR mode ($\Delta\lambda=3$ \AA,
$R \sim 3,000$) yields $\Delta$[X/Fe]$\sim 0.3$ dex, insufficient to
distinguish field halo stars, with [X/Fe]$\sim 0.3$, from disk and
tidal stream stars with [X/Fe]$\sim 0$ (see Fig.~\ref{fig:Mg_comp}).
A MR capability for PFS, such as described here, is necessary to
achieve the science goals proposed in this document.

We also note that in order to determine radial velocities with accuracy of $\sim 3$
km~s$^{-1}$, a resolving power of $R = 5,000$ 
(i.e. that of the planned MR mode) is required with a
wide wavelength coverage, in particular for the red region where highest S$/$N ratios
are expected.

The Galactic Archaeology requirements are summarized in 
Table~\ref{tab:GA_requirements}.

In addition to the dedicated GA survey described here, there are 
a number of other important questions that may be addressed with both LR and MR modes
of PFS. These include: (1) chemo-dynamical properties of the Galactic bulge/bar 
and the transition(s) between the bulge, inner disk and halo components in the Milky Way,
(2) kinematics of the outer parts of halo globular clusters 
in the Milky Way, and (3) identification and analysis of globular clusters
in M31. These studies of these issues
can be carried out through PI-led programs.

\section{PFS Galaxy Evolution Survey}
\label{chap:galaxy}

{\it Summary:
 The goal of the PFS Galaxy Evolution Survey is to follow the growth of
the full panoply of modern-day galaxies from cosmic dawn to the
present.  While we know some of the basic physical processes that
drive galaxy evolution (dark matter halo merging, gas accretion, star
formation and associated energy release, galaxy merging, black hole
accretion and associated energy release), how and when they operate,
and their relative importance, remain unknown.  We will use the
unprecedented wavelength coverage of PFS, in particular from $1$ to
 $1.26
\mu$m, to explore the redshift desert between $1.4 < z < 2.2$
when the star formation rate density and black hole growth were at
their peak.  Using \lya\ emission, we will trace the growth of
galaxies and black holes all the way to the epoch of reionization, $2
< z < 7$. Thanks to deep broad- and narrow-band imaging enabled by the
HSC survey, we will be able to study these young galaxies with
unprecedented statistics using PFS.  We propose a 100 night survey
covering 16 deg$^2$ with these main components:
(i) A color-selected galaxy survey of 500,000 galaxies using 785,000 fiber-hours 
over 16 deg$^2$ from $z = 1$ to $z=2$ to a limiting magnitude of $J_{\rm AB} = 23.4$ mag,
  with a $z < 1$ component limited to $J_{\rm AB} = 21$ mag, and a
  magnitude-limited component covering 2.6 deg$^2$;
(ii)
A survey of
$140,000$ bright dropout galaxies and \lya\ emitters 
over $2 < z < 7$, using 247,000 fiber-hours; 
(iii) A survey of 
$\sim 50,000$ color-selected quasars from $3 < z < 7$, using a total 
of 1600 fiber hours in this survey and 21,000 fiber hours in the BAO survey.
}

The most important requirements for our science are the following:

\begin{enumerate}
\item
  The multiplexing capability of PFS is a critical component of our
  survey design, allowing us to observe hundreds of thousands of
  objects.

\item
  The sensitivity in the NIR is most crucial to our redshift success.
  We assume that in 3 hrs of integration we can measure continuum
  redshifts from the 4000\AA\ break down to $J_{\rm AB} = 23.4$ AB
  mag.  We find that this is achievable with the default instrument
  configuration, at an effective $R\approx300$ in the NIR arm, 
  if the scattered light going down each fiber is no more than a few percent of the
  sky brightness; the nominal requirement is 0.5\%.

\item 
  We assume that in a 3 hr integration we can achieve $>5 \sigma$ line
  detections of $\sim 10^{-17}$ erg~s$^{-1}$~cm$^{-2}$ at effective 
resolution $R=1300$
  for the full spectral band in the blue and red arms and 75\% of the NIR arm, 
  including systematic errors in sky subtraction (see \S 4.3).
\end{enumerate}

\subsection{Extragalaxy Science Objectives}

\begin{figure*}[t]
\begin{center}
%\vbox{ 
%\vskip -5mm
%\hskip +5mm
%\psfig{file=dmdz_clusters.ps,width=0.25\textwidth,keepaspectratio=true,angle=-90}
%}
\centerline{\includegraphics[width=0.25\textwidth,angle=-90]{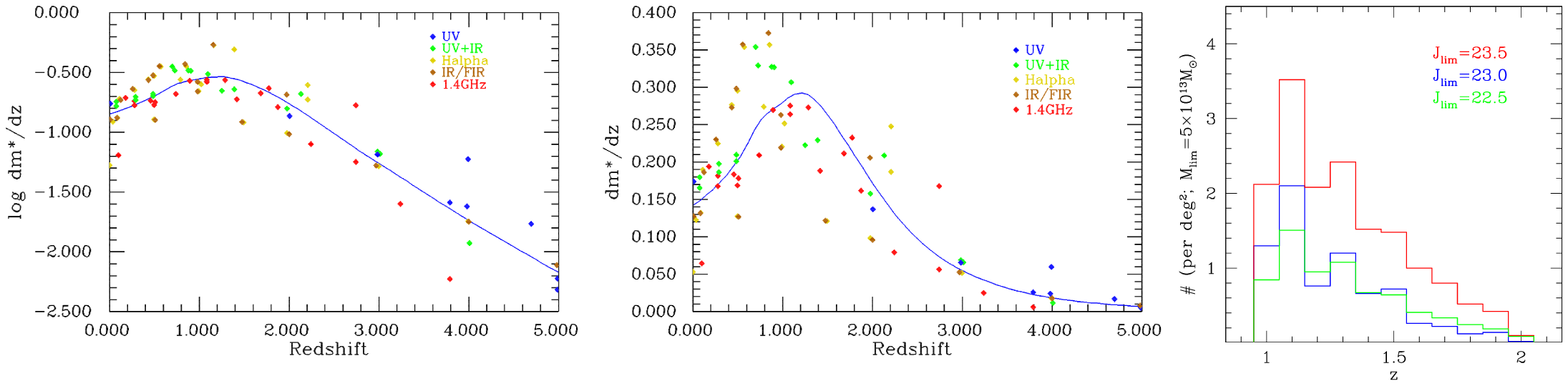}}
%\centerline{\includegraphics[width=0.98\textwidth,angle=0]{dmdz_clusters_sm.eps}}
\vskip -0mm
%\figcaption[]{
\caption{
%\small
{\it Left}:
The differential build-up of stellar mass as a function of redshift using 
both a logarithmic (left) and linear (middle) 
stretch.  The data are from \citet{behroozietal2012}, and 
$dm^\ast/dz$ is in units of [$M_\odot/$yr/Mpc$^3$]. 
{\it Right}: Expected
number of clusters with $M>5 \times 10^{13}$~\msun\ per  deg$^2$ survey
volume with the stated depths.  The estimates use the ``lightcone''
simulations of \citet{sehgaletal2010} and populate halos with galaxies using the
halo occupation distribution model from \citet{tinkerwetzel2010}.
\label{fig:sfcluster}}
\end{center}
\end{figure*}

The Sloan Digital Sky Survey has provided us with a detailed
understanding of the optical properties and large-scale distribution
of present-day galaxies.  Large redshift surveys such as zCOSMOS
\citep{lillyetal2009}, DEEP2 \citep{cooperetal2006}, and VVDS
\citep{lefevreetal2005} have studied areas of $\leq 2$ square degrees
at $z \sim 1$; at this epoch galaxies had similar morphologies,
luminosity ranges, and environmental characteristics as they do today.
However, the bulk of the stellar and black hole mass was assembled at 
$1 < z < 2$ (Fig.~\ref{fig:sfcluster}), when the universe
was a very different place.  We want to understand how galaxies and
gas are distributed relative to dark matter halos at these epochs, how
each of these components interact (via merging, cooling flows, star
formation, energy feedback, and reionization), and whether star
formation proceeded in a fundamentally different way at that crucial 
epoch of mass assembly. To date
there are spectroscopic samples of only a few thousand blue,
star-forming galaxies with $z > 1.4$, covering $<1$ deg$^2$
\citep[e.g.,][]{steideletal2010}.  Particularly difficult to study is
the so-called redshift desert $1.4 < z < 2.2$, since no strong and
reliable redshift indicators fall in the optical bandpass.  Even with
the current generation of multi-object near-infrared spectrographs, it
will not be possible to survey the solid angles proposed here, which allow us
to tie together galaxy evolution with the large-scale environment.
The PFS spectrograph, with a wavelength coverage of 0.38-1.26 $\mu$m,
is specifically designed to explore the redshift desert over volumes
comparable to the SDSS at low redshift.  {\it We propose the first
  large-area spectroscopic survey of roughly half a million galaxies
  in the redshift range $1 < z < 2$, when the bulk of stellar mass was
  assembled.}

While the bulk of the stellar mass in galaxies was assembled between $1 < z < 2$,
the star formation rate density likely peaked at even earlier times
(Fig.~\ref{fig:sfcluster}), and so to get a full picture of galaxy
evolution we must study the earliest cosmic epochs as well. 
The wide wavelength coverage of PFS allows us to observe
galaxies from $z\sim 2$ to the redshift frontier of $z\sim 7$ 
(Fig.~\ref{fig:sfcluster}) within {\it the same survey design}.  It is
prohibitive to perform a rest-frame optical continuum selection at
$z>2$.  Instead we will target high-redshift Lyman break galaxies
(LBGs) and \lya\ emitters (LAEs) at $2 \lesssim z \lesssim 7$.  We
will obtain unprecedented numbers of spectra of both populations,
particularly at $z > 5$, to probe the epoch of reionization. With
these samples we will chart the evolution of star formation rate,
metallicity, galaxy density, and large-scale structure as a function
of redshift.  We will probe the epoch when the star formation of early
galaxies changed the ionization state of the intergalactic medium
(IGM) in what is called cosmic reionization.  The large area of the
PFS spectrograph allows us to survey wide enough areas on the sky to
study the topology of ionized bubbles in the IGM at this epoch ($\sim
10^6$~Mpc$^3$) as well as to conduct a census of young galaxies at $2
\lesssim z \lesssim 7$.  We will have a sample two orders of magnitude
larger than those that came before.

%%%%%%%%%%%%%%%%%%%%%%%%%%%%%%%%%%%%%%%%%%%%%%%%%%%%%%%%%%%%%%%%%%%%
%%BoundingBox: 
\vskip 0mm
%%%%%%%%%%%%%%%%%%%%%%%%%%%%%%%%%%%%%%%%%%%%%%%%%%%%%%%%%%%%%%%%%%%%%

It is now recognized that all massive galaxies contain supermassive
black holes (SMBHs) in their centers, with masses ranging from
$10^5-10^{10}$~\msun.  The evolution of black hole growth with cosmic
time resembles that of the star-formation activity.  The steep
drop-off of both at $z<2$ points to a link between supermassive black
hole growth and galaxy formation. Moreover, the masses of SMBHs are
correlated with those of their host bulges in the present-day
Universe, suggesting that they evolved together \citep[e.g.,][]{
  kormendyrichstone1995,magorrianetal1998,marconihunt2003}.  A full
understanding of galaxy evolution requires study of the accretion
history of SMBHs.  We will target color-selected quasars
that are $\sim$ three magnitudes fainter than the SDSS to a 
redshift of $z=7$ and
study their clustering properties, the evolution of BH mass density
with cosmic time, and evolution in the metal content and ionization
state of the inter-galactic medium.

\begin{figure*}[t]
\begin{center}
%\vbox{ 
%\vskip 5mm
%%\hskip +5mm
%\psfig{file=survey_param.eps,width=0.35\textwidth,keepaspectratio=true,angle=-90}
%}
\centerline{\includegraphics[width=0.45\textwidth,angle=-90]{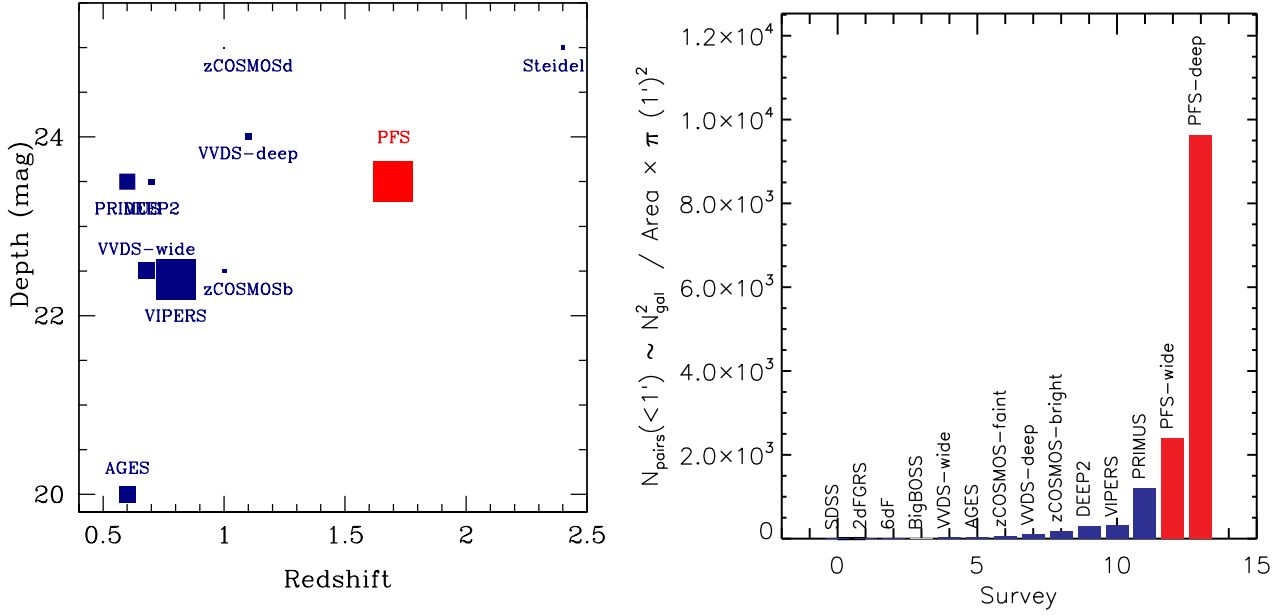}}
\vskip -0mm
\caption{ 
%\small
{\it Left}:
Depth versus redshift for existing and planned large redshift surveys with $z < 2.5$. 
The symbol size represents survey area.  Each survey is placed roughly 
at the median redshift of the survey.  PFS occupies a unique position in this 
parameter space, as it is the {\it only} large-area survey capable of filling in the 
redshift regime between $1.5 < z < 2$.
{\it Right}: 
This figure shows the number of spectroscopic pairs separated by
less than one arcminute. It highlights the power of PFS to (i) study
small-scale clustering on the group scale and (ii) probe the gas
distribution in galaxy halos using absorption line probes. Fiber-based 
spectroscopic surveys doing only one pass over the sky suffer from 
the fiber collision limitation and cannot access small-scale
 pairs. ``PFS-wide'' 
here is the BAO survey described in \S 2, while ``PFS-deep'' is the survey here.
%}
\label{fig:surveyparam}}
\end{center}
\end{figure*}

{\it We propose to study galaxy formation and cosmic reionization from
  $z=1$ to $7$ seamlessly. We will conduct the first spectroscopic census of young
  galaxies at $2\lesssim z\lesssim 7$ up to the reionization epoch,
  based on 60,000 spectra of LBGs and LAEs in a wide 
range of environments.  No existing
  or planned galaxy evolution surveys in the next ten years can
  compete with our combination of depth, redshift coverage, and area (Fig.~\ref{fig:surveyparam}).}

The wide wavelength coverage and large area of the PFS will enable an
enormously powerful galaxy evolution survey.  We envision a $16$
deg$^2$ survey over $\sim 100$ nights, or a million fiber-hours.  All
observations will be taken with twenty-minute exposures; given the
rapid reconfiguration of PFS, we thus have tremendous flexibility in
observing various targets for total integration times ranging from
twenty minutes to six hours or more.  At $1 < z < 2$, using color
selection, we will survey to a depth of $J_{\rm AB} \approx 23.4$ mag
with three-hour integrations, which will allow us to study a fair
sample of galaxies down to a stellar mass of $\sim 10^{10}$~\msun\ at
$z \approx 2$, with a stellar-mass matched $z < 1$ sample down to
$J=21$ mag observed with twenty-minute integrations. Finally, to
understand our color selection, we will include a 2.6 deg$^2$
magnitude-limited survey to $J_{\rm AB} \approx 23.4$ mag. To extend
our study of typical galaxies to yet higher redshift, at $2\lesssim
z\lesssim 7$ we will target star-forming galaxies.  We expect to
observe 60,000 Lyman break and Lyman alpha emitting galaxies, an order
of magnitude increase over existing spectroscopic samples.
Taken together, the survey will give us enormous leverage to link the 
evolution of galaxies from cosmic dawn to the peak epoch of stellar mass growth.

We will focus on three imaging surveys for target selection. We will
target {16 deg$^2$} of the HSC Deep survey that contains existing
deep NIR imaging down to $J_{\rm AB}= 23.4$ mag.  HSC Ultradeep covers
{3.5 deg$^2$} and we will perform deeper spectroscopy in this
area.  Finally, HSC Wide covers {1400 deg$^2$}, over which we will
search for rare high-redshift quasars ($z > 3$).

%\subsection{Scientific Goals of the PFS Galaxy Evolution Survey}

%%%%%%%%%%%%%%%%%%%%%%%%%%%%%%%%%%%%%%%%%%%%%%%%%%%%%%%%%%%%%%%%%%%%
%%BoundingBox: 
\vskip 0mm
%%%%%%%%%%%%%%%%%%%%%%%%%%%%%%%%%%%%%%%%%%%%%%%%%%%%%%%%%%%%%%%%%%%%%

\begin{figure}[t]
\begin{center}
%\vbox{ 
%\vskip -4mm
%\hskip +20mm
%\psfig{file=line_z_new.eps,width=0.45\textwidth,keepaspectratio=true,angle=90}
%}
\centerline{\includegraphics[width=0.45\textwidth,angle=90]{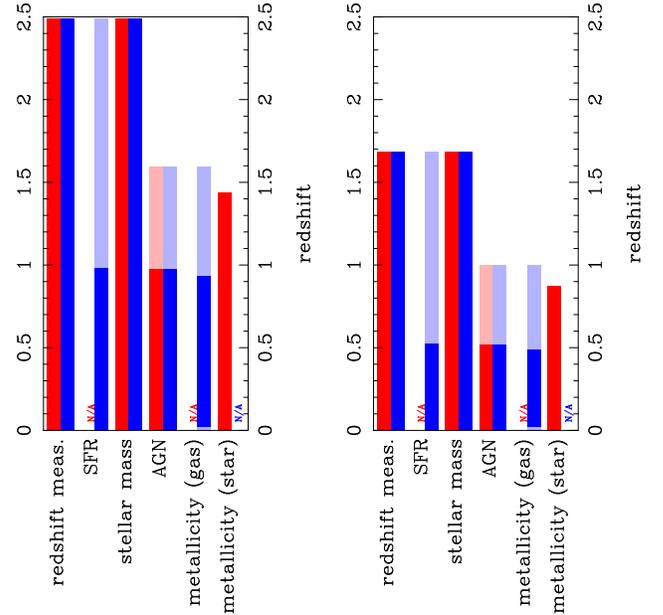}}
\vskip -0mm
%\figcaption[]{
\caption{
%  \small
  Redshift range over which various spectroscopic properties of
  galaxies can be tracked assuming a flux-limited survey down 
  to $J_{\rm AB} = 23.4$ mag. 
  {\it Left}: The capabilities
  of PFS with the full wavelength coverage (optical+NIR).
  {\it Right}: The same, but with only the blue+red coverage. The dark and
  light areas show the redshift range where a primary and secondary
  set of spectral features will be available for galaxies with masses
  above a characteristic value ($M*$). Passive and star forming galaxies
  are marked in red and blue, respectively.
\label{fig:linez}}
\end{center}
\end{figure}

\begin{figure*}[t]
\begin{center}
%{\centering \leavevmode
%\epsfxsize=.50\textwidth \epsfbox{lumifun_full_diff_nb921_evolution.eps} \hfil
%\epsfxsize=.50\textwidth \epsfbox{line_profile_evolution.eps}}
%\centerline{
\includegraphics[width=0.43\textwidth]{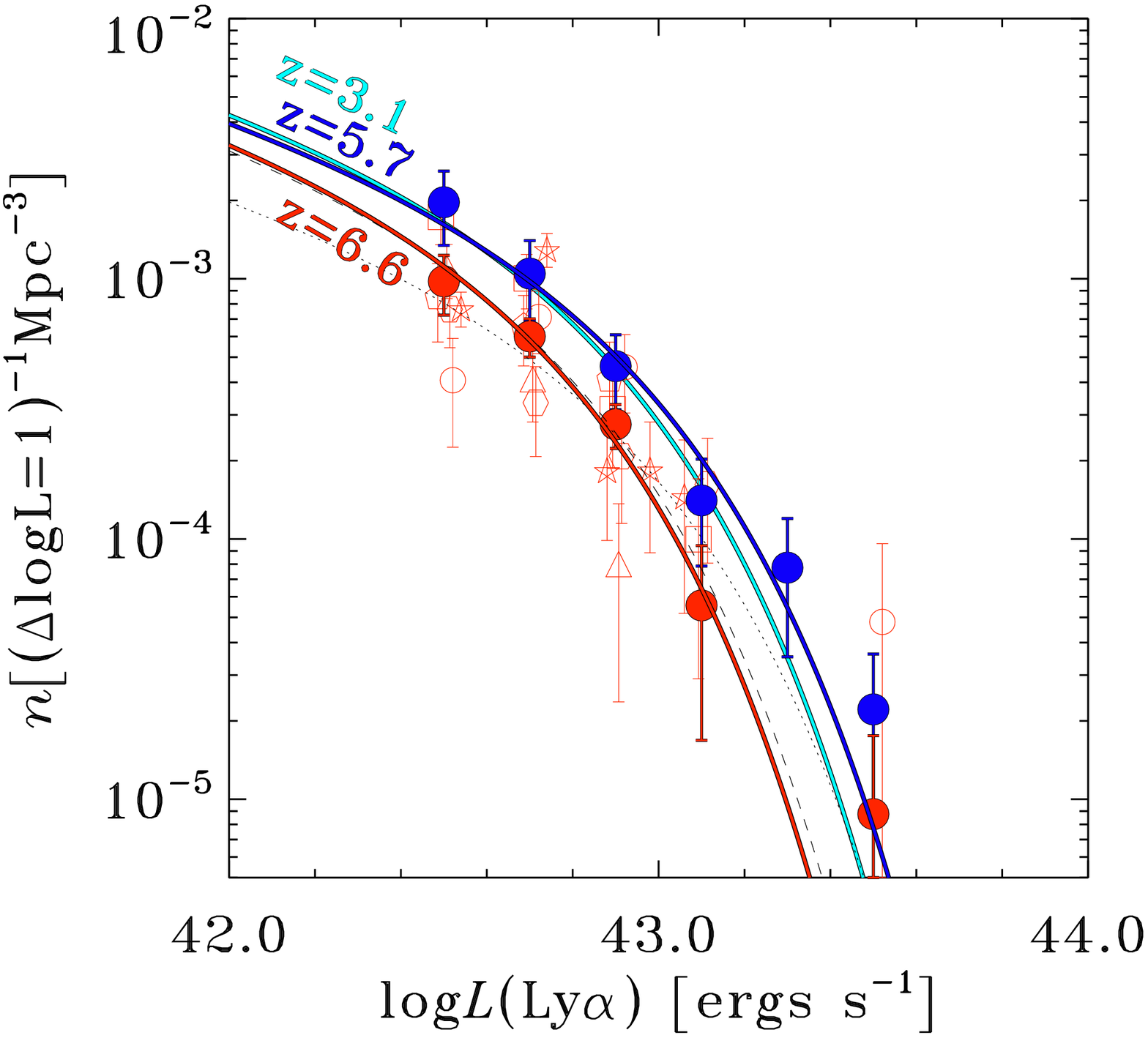}
\includegraphics[width=0.43\textwidth]{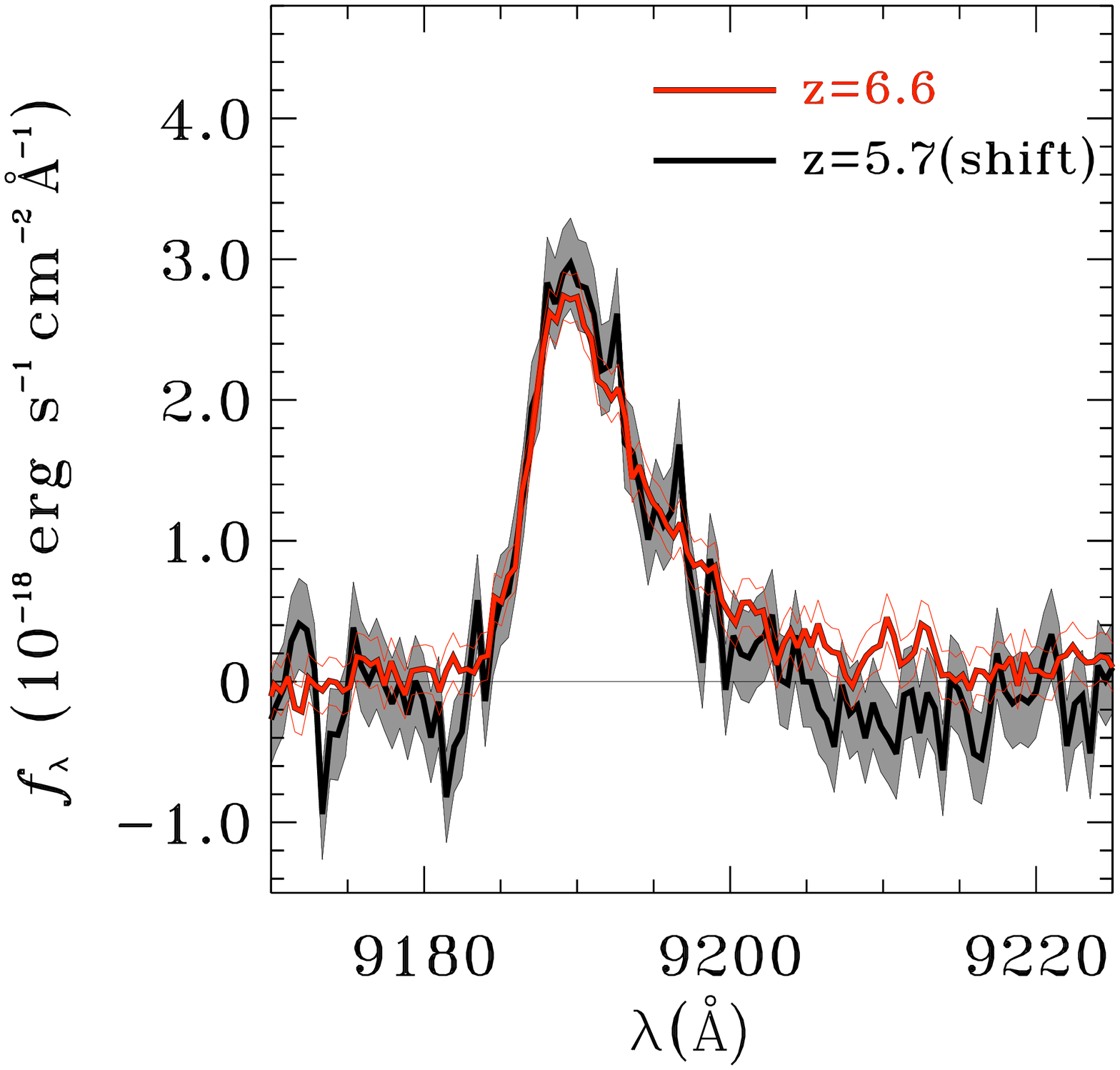}
\caption{
%\small
{\it Left}:
Evolution of \lya\ luminosity functions from $z=3.1$ to $z=6.6$
\citep{ouchietal2010}.  Red filled circles are the best estimates for
$z=6.6$ LAEs, and the red solid line is the best-fit Schechter function of
$z=6.6$ LAEs.  Blue filled circles and solid line are data points and
the best-fit Schechter function, respectively, of the $z=5.7$ LAEs.  For 
reference, the cyan solid line is the best-fit Schechter function of the $z=3.1$ LAEs
\citep{ouchietal2008}.  We will use evolution in the \lya\ luminosity 
function to search for the signature of reionization as absorption from 
neutral hydrogen systematically suppresses \lya\ emission.
{\it Right}: Evolution of \lya\ line
profiles in existing data \citep{ouchietal2010}.  Red and black thick lines represent
composite spectra of LAEs at $z=6.6$ and $5.7$, respectively.  For
comparison of line shapes, the composite spectrum of the $z=5.7$ LAEs
is redshifted to $z=6.6$, and scaled by an arbitrary factor.  One
sigma errors of the composite spectra are shown with thin red lines
for $z=6.6$ LAEs and gray shading for $z=5.7$ LAEs.  
We hope to do seven times better in any deg$^2$ patch 
of the sky by stacking the much larger sample of objects in the PFS survey.
\label{fig:LyaLF_LyaLineProfile}}
\end{center}
\end{figure*}

The primary science goals of the PFS galaxy evolution survey are the following:

\begin{enumerate}

\item {\bf The Build-up of Stellar Mass Density:}
  We will derive stellar masses and star formation rates from the
  spectra (Fig.~\ref{fig:linez}).  It has been shown that stellar
  population modeling works well even at our low resolution
  \cite[e.g.,][]{krieketal2006,krieketal2011,panteretal2007,
    chenetal2012}.  We will measure star formation rates both from the
  modeling and from \oii\ emission out to $z \sim 2.5$, which we will
  calibrate using \halpha\ out to $z \approx 1$. At yet higher redshift, we
  can use the UV continuum as a star formation rate indicator 
(Fig.~\ref{fig:linez}). We will trace the stellar mass function and star
  formation history of galaxies as a function of color and galaxy
  density at $1 < z < 2$.

\item {\bf The Growth of Structure:} We will measure the spatial
  correlation functions of galaxies on both small and large scales
  (i.e., one and two halo terms; Fig.~\ref{fig:surveyparam} {\it
    right}). Spectroscopic redshifts give significantly better
  clustering measurements than do photometric redshifts, and will
  allow us to perform halo occupation modeling as a function of
  redshift. On large scales, we will measure the galaxy bias as a
  function of galaxy properties and redshift, which is a key
  constraint on galaxy formation models.  On smaller scales, we can
  study the properties of galaxies in cluster and group environments,
  and see how galaxy formation is related to environment and dark
  matter halo mass.  We will use a combination of weak lensing from the
  Hyper Suprime-Cam survey and clustering from the PFS survey 
  at low redshifts, and clustering alone at higher redshifts to
  tie the average properties of the galaxy population to their
  underlying dark matter halos, and the group and cluster halos in
  which they live. We will also study the
  cluster/proto-cluster population as a function of redshift.  Over
  the survey volume, we expect to find $\sim 325$ clusters with mass (at that epoch)
  $M > 5 \times 10^{13}$~\msun\ (including 
$\sim$65 massive clusters with $M
  > 10^{14}$~\msun) at $1 < z < 2$ (Fig.~\ref{fig:sfcluster}), an
  unprecedented sample when compared to the very few found in current
  deep pencil-beam surveys \citep{sehgaletal2010}. At $2 \lesssim z
  \lesssim 6$ we expect to find $\sim 100$ proto-clusters using LBGs
  and LAEs as tracers.

\item{\bf Gas Inflow and Outflow:}
  We will trace the interplay between gas accretion and feedback using
  both direct and indirect means.  We will trace the mass-metallicity
  relation using strong emission-line diagnostics for $z < 1.6$ and
  with stellar photospheric lines in the UV at $z > 2$ by stacking
  spectra \citep{andoetal2007}.  We will search for outflow or inflow
  using interstellar absorption lines such as Mg {\small II}, S
  {\small IV}, C {\small IV}, etc in stacked spectra.  We will also
  stack background galaxy and quasar spectra as a function of impact
  parameter around foreground galaxies to map out the gas kinematics
  in the outer halos of star forming galaxies
  \cite[e.g.,][]{steideletal2010,tumlinsonetal2010,bordoloietal2011}.
  Using the galaxies themselves as a backlight, we will search for the
  signature of cold-gas inflow and test the cold accretion hypothesis 
\cite[e.g,][]{dekelbirnboim2006}
  with a sample $\sim$20 times larger than the one of Steidel et al.\

\item{\bf The Build-up of Supermassive Black Holes:}
  We will measure the quasar luminosity function from $3 < z < 7$ to
  $\sim 3$ mag deeper than the SDSS.  A wide-area search of 
  $>1000$ deg$^2$ is needed to find rare luminous $z > 6$ quasars,
  while we will study fainter populations over the 16 deg$^2$
  galaxy survey. Using quasar emission lines we will study the
  evolution of the BH mass function, and the accretion history of
  these black holes. We will be able to study the clustering
  properties of the quasar population using cross-correlation with the
  dropout galaxies in our sample, which will reveal the typical dark
  halo mass of the quasar population as a function of redshift and
  luminosity \cite[e.g.,][]{adelbergersteidel2005}, and relate the relative 
  growth of galaxies and BHs. Finally, we will
  uncover obscured and low-luminosity active galaxies using
  emission-line diagnostics in the $1 < z < 2$ survey.

\item {\bf The Epoch of Reionization, Ionized Bubbles:} For the first
  time, we will constrain the topology of ionized bubbles at $z\sim 7$
  with a complete spectroscopic sample of LAEs.  Current surveys
  \cite[e.g.,][]{ouchietal2010} have shown the promise of using the
  mean \lya\ profile shape as a tracer of neutrality, but the small
  samples offer no information on large-scale spatial variations
  (Fig.~\ref{fig:LyaLF_LyaLineProfile}). We will exploit the $\sim
  9,000$ LAEs from our PFS survey to search for variations in \lya\
  profiles over scales of 10 Mpc, comparable to the expected size of
  typical ionized bubbles.  We will calibrate the intrinsic \lya\
  line shape using a critical sample of LAEs at $z \sim 2.2-2.4$, for
  which we will have robust systemic redshifts from \oii.

\item {\bf The Epoch of Reionization, Neutral Fraction:} We will
  utilize the evolution of the \lya\ luminosity function 
(Fig.~\ref{fig:LyaLF_LyaLineProfile}) \cite[e.g.,][]{ouchietal2010} and
  the complementary measure of the fraction of galaxies with \lya\
  emission, with unprecedented large samples of LAEs and LBGs,
  respectively \cite[][]{starketal2011}.  This dual approach will allow us to
  calibrate possible intrinsic evolution in the \lya\ luminosity
  function arising from changes in the comoving star
  formation density. Linking emission line demographics to LBG
  colors will address possible effects of dust extinction. The
  resulting decline in the visibility of \lya\ over $6<z<7$ will
  then be interpreted in terms of a change in the neutral fraction
  using appropriate models \cite[e.g.,][]{mcquinnetal2007}. We will obtain
  complementary constraints on the neutral fraction using the
  Gunn-Peterson troughs of $z \sim 6-7$ quasars.

\end{enumerate}

%%%%%%%%%%%%%%%%%%%%%%%%%%%%%%%%%%%%%%%%%%%%%%%%%%%%%%%%%%%%%%%%%%%%
%%BoundingBox: 
%\vskip -5mm
%%%%%%%%%%%%%%%%%%%%%%%%%%%%%%%%%%%%%%%%%%%%%%%%%%%%%%%%%%%%%%%%%%%%%

%%%%%%%%%%%%%%%%%%%%%%%%%%%%%%%%%%%%%%%%%%%%%%%%%%%%%%%%%%%
%%%%%%%%%%%%%%%%%%%%%%%%%%%%%%%%%%%%%%%%%%%%%%%%%%%%%%%%%%%

We wish to strongly emphasize the power of the wide wavelength
coverage of PFS.  The NIR arm allows us to measure the \oii\ line (and
thereby reliable redshifts) out to $z=2.4$ (Fig.~\ref{fig:linez}).  The NIR arm thus opens the opportunity to study the
crucial, and largely unexplored, epoch of $1 < z < 2$.  It allows us
to trace galaxy populations from low to high redshifts with no gaps. 
There will be no redshift desert with this spectrograph.
As illustrated in Fig.~\ref{fig:linez}, we will have access to many
important emission and absorption features over an unprecedented range
of redshifts, which will enable new and important science.  We will
cross-calibrate star-formation and metallicity indicators such as
\oii\ and \halpha; both can be observed to $z \approx 1$.  We can also
study the physics of \lya\ emission via its profile shape 
by obtaining accurate redshifts from the 
\oii\ line for thousands of LBGs with $2.1 < z < 2.4$.

\subsection{Simulated Data}

In this section we will show that our main science goals are
achievable.  Our exposure times will be 20 min, after which we can 
plan to leave our brightest targets, thanks to the rapid fiber repositioning time. 
For the faintest $1 < z < 2$ targets ($J_{\rm AB} = 23-23.4$ mag) we will spend 
three hours per galaxy.  The integration times for higher redshift sources 
will range from one to sixteen hours.
We have generated realistic simulated spectra to determine what types
of measurements are feasible as a function of continuum and line
luminosity.  Specifically, we take the instrument specifications and
reasonable assumptions for the properties of the sky emission and our
ability to model it.  We demonstrate that with our chosen exposure
times, we can accomplish our science goals of (1) measuring emission line
fluxes and line-widths and thus redshifts and (2) measuring
absorption-line redshifts. Finally, we explore 
a variety of assumptions about sky brightness and scattered light from 
the point of view of survey design.

Our model for the throughput of the spectrograph includes
the quantum efficiency of the chips, light-losses at the fiber due to
seeing, losses in the fiber and fiber coupling, reflectivity of the
Subaru primary mirror and transmission of the corrector, throughput of
the camera and collimator (including vignetting) and dichroics as a
function of wavelength, the grating transmission, and the atmospheric
transparency. The spectral simulator models the two-dimensional 
distribution of source and sky photons on the chip, then collapses 
along the spatial direction to produce simulated one-dimensional spectra 
and the corresponding error arrays.  The noise calculation includes shot 
noise from the source, sky emission lines, and sky continuum, as well 
as readnoise, dark current, and thermal background.

The most important, and most challenging, aspect of our simulations
involves modeling the sky, which dominates the noise in the red and
NIR arms.  Our treatment is the same as that outlined in \S 2 and
assumes that we can measure the overall sky spectrum at a given time
at the $0.5\%$ level.  This requirement is theoretically possible,
given Poisson errors, but technically challenging in practice.  For
instance, \citep{boltonetal2012} achieve $\sim 1\%$ sky subtraction in
the red in BOSS.  Relative to our system, the BOSS fibers are short and move
very little, and BOSS takes calibration frames at every pointing. On
the other hand, the BOSS spectrographs move with the telescope, 
and thus suffer flexure
and temperature variations that do not impact the fixed and
temperature-controlled PFS spectrographs.  Furthermore, the native PFS
spectral resolution is higher, which will facilitate sky subtraction.
Accordingly, an additional Gaussian random deviate with an amplitude
that is 0.5\% of the sky flux, is added to the error budget of each
pixel, with bright sky lines bleeding over into adjacent
pixels. Finally, we assume that 0.5\% of the total flux in the sky is
spread across the detector as a smooth background and include the
relevant shot-noise.  We revisit these assumptions later in the
section, and show that our survey is robust even if our assumptions
for each are overly optimistic.

\begin{figure*}[t]
\begin{center}
%\vbox{ 
\vspace*{-0em}
\includegraphics[width=13cm, angle=0]{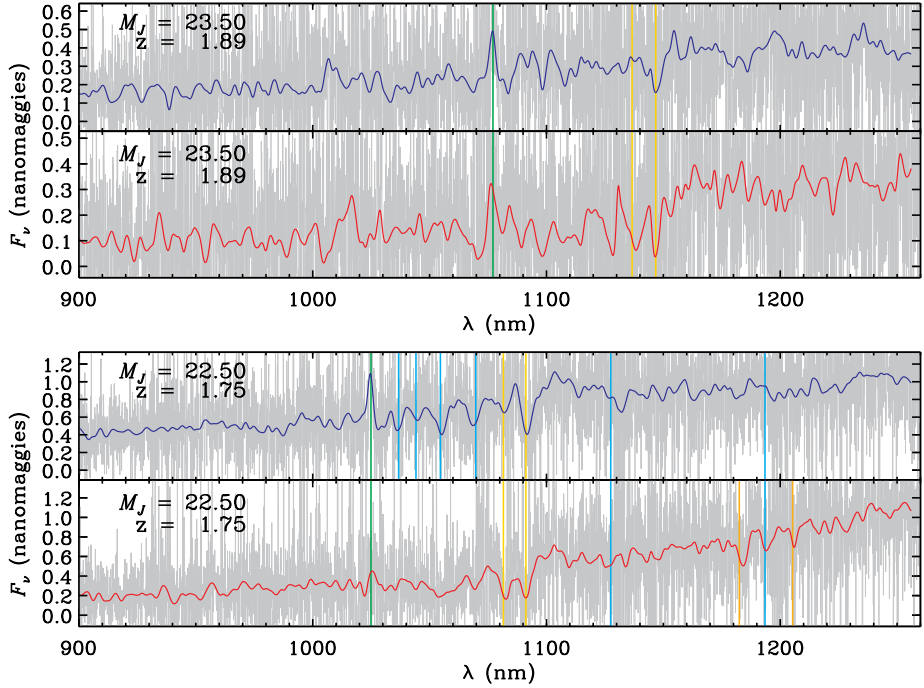}
\caption{
%\scriptsize
%\small
Simulated spectra as observed in three-hour integrations (see text) demonstrating the power of PFS.
The spectra are plotted in units of ``nanomaggies'', a unit of flux 
used by the SDSS, defined such that one maggie is the flux 
from an AB$=0$ object, 3631 Janskys. Thus 1 nanomaggie is equivalent to 
an apparent AB magnitude of 22.5, $3631 \times 10^{-9}$ Jy or 
$3.631 \times 10^{-29}$ erg~s$^{-1}$~cm$^{-2}$~Hz$^{-1}$.
The grey lines show the observed spectra, while the red and blue overplotted spectra
are binned to a resolution of $R = 400, 300, 300$ in the blue, red, 
and NIR arms respectively.   The green 
line indicates the position of the [O {\tiny II}] doublet, the blue lines indicate 
the higher-order Balmer lines, the gold lines show the 
Ca H+K lines, and the orange lines indicate the G-band and Fe4383. 
The spectral templates are partly based on 
stellar population synthesis models 
of NEWFIRM Medium Band Survey galaxy photometry, which provides 
some constraint on the emission line equivalent widths.  In the 4000~\AA\ break region, 
we splice in a higher-resolution composite spectrum from the DEEP2 survey \citep{schiavonetal2006}.
We show total integration times of three hours.
From top to bottom, we show the blue and red galaxy pair with  
$J_{\rm AB} = 23.5$ mag at $z=1.89$ and the same pair with 
$J_{\rm AB} = 22.5$ mag at $z=1.75$.  In this case, we will have very rich spectral 
diagnostics to determine the physical properties of the galaxies.}
\label{fig:simspectragal}
\end{center}
\end{figure*}

\begin{figure*}[t]
\begin{center}
\includegraphics[width=13cm,angle=0]{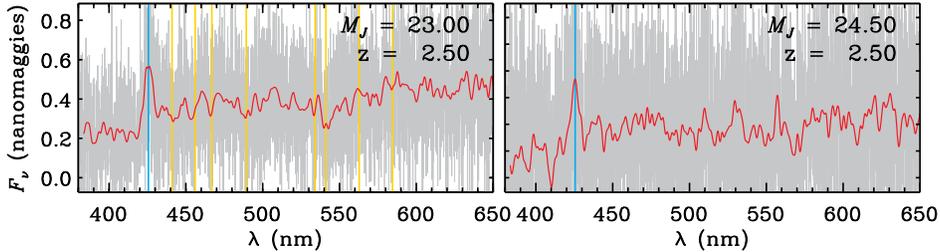}
\caption{
%\small 
Star-forming galaxies selected as a $U-$band dropouts with $J_{\rm AB} = 23, 24.5$ 
mag, as observed in a three-hour total integration with PFS.  As above, the grey lines are the observed PFS spectra, while the red overplotted spectra
are at effective resolution of $R = 400, 300, 300$ in the blue, red, 
and NIR arms respectively. The cyan line denotes 
Ly $\alpha$.  The gold lines demarcate strong interstellar features
Si{\tiny II}$\lambda 1260$, O{\tiny I}+S{\tiny II}$\lambda 1303$, C{\tiny II}$\lambda 1334$, 
Si{\tiny IV}$\lambda 1398$, Si{\tiny II}$\lambda 1526$, C{\tiny IV}$\lambda 1546$, 
Fe{\tiny II}$\lambda 1608$, and Al{\tiny II}$\lambda 1670$.
The spectral energy distribution is based on a spectral synthesis model from
    \citet{reddyetal2012}, with emission line EWs based on \citet{shapleyetal2003}. 
For the brighter targets (left) we will detect ISM absorption 
lines in individual spectra, while we will stack fainter targets (right).}
\label{fig:simspectralbg}
\end{center}
\end{figure*}

\begin{figure*}[t]
\begin{center}
\includegraphics[width=13cm,angle=0]{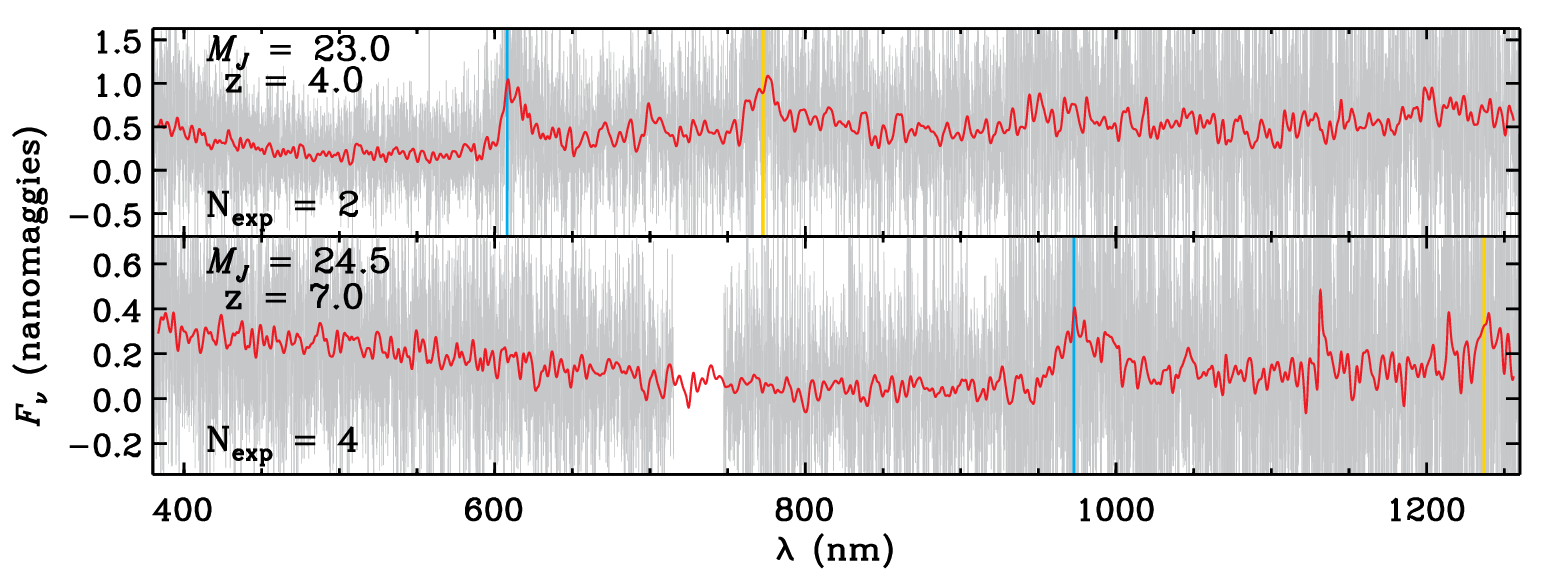}
\caption{
%\small 
Two QSO spectra. As above, the grey lines are the observed PFS spectra, while the red 
spectra are at effective resolution of $R = 400, 300, 300$ in the blue, red, 
and NIR arms respectively. The top one assumes a 30 min total integration time, 
as we will have in the BAO survey described in Section~\ref{chap:cosmology}, 
while the bottom one 
assumes a full hour of total integration time as is proposed for the QSOs 
observed in the galaxy-evolution survey described in this Section.  
In both, the cyan line marks the position 
of Ly $\alpha$ while the gold line denotes C{\small IV}.}
\label{fig:simspectraqso}
\end{center}
\end{figure*}

The spectra shown here are inverse variance weighted and smoothed to
$R=400,300,300$ for the blue, red, and NIR arms respectively 
(Figs.~\ref{fig:simspectragal}, \ref{fig:simspectralbg}, and
\ref{fig:simspectraqso}) for viewing purposes only.  When we examine
\oii\ emission lines, we instead use $R \approx 1200$ to properly
resolve emission lines with $\sigma = 100$~\kms. A large fraction of
the spectrum is eaten by night sky lines; these low resolution spectra
give a convenient representation of their information content.  For
reference, the unsmoothed spectra are shown as well. All of our
assumptions are summarized in Table \ref{tab:gal_req}.

%%%%%%%%%%%%%%%%%%%%%%%%%%%%%%%%%%%%%%%%%%%%%%%%%%%%%%%%%%%%%%%%%%%%
%%BoundingBox: 

%\vskip -5mm
%%%%%%%%%%%%%%%%%%%%%%%%%%%%%%%%%%%%%%%%%%%%%%%%%%%%%%%%%%%%%%%%%%%%%

In the following, we show simulation results for four different templates: a
blue and red galaxy at both $z \approx 1.7$ and at our redshift limit of 
$z \approx 1.9$, a Lyman-break galaxy (LBG), and a QSO.
The galaxy templates are stellar population model fits to NEWFIRM
Medium-Band Survey photometry for galaxies spanning the
color-magnitude diagram.  These fits, from the code EAZY
\citep{brammeretal2008} include emission lines.  Since the region
surrounding the 4000\AA\ break and the \oii\ line is of particular
importance, we splice in spectra with considerably higher spectral
resolution between 3700-4500\AA.  We splice a stack of DEEP2 galaxies
from \citet{schiavonetal2006} into the red galaxy template.  For the
blue galaxy splice, we use a model star-forming galaxy with
emission-line EWs based on \citet{krieketal2011}. These are shown in 
Fig.~\ref{fig:simspectragal} at our detection limit of $J=23.4$ mag 
at $z \approx 2$, observed in a 3hr integration with PFS.

The LBG template is based on the spectral energy distribution from
\citet{reddyetal2012}.  We add emission lines to match the EWs in
\citet{shapleyetal2003}; the objects shown in Fig.~\ref{fig:simspectralbg} in the bottom panel are representative of
objects with high \lya\ EW in their sample. From these simulations it
is clear that we will be able to detect both \lya\ down to our detection
limits and interstellar absorption features in our brighter targets. 
Finally the QSO spectra are generated from the composite
spectrum from \citet{vandenberketal2001}.

\subsubsection{Continuum Redshifts}

One of our greatest challenges comes in measuring absorption-line
redshifts for $z \approx 2$ galaxies at our flux limit.  In practice,
we expect that many of our targets will have some \oii\ emission,
since most red galaxies have low EW emission lines
\citep[e.g.,][]{hoetal1997,yanetal2006}.  Nevertheless, there will
certainly be passive red galaxies for which the 4000\AA\ break will be
the primary source of redshift information. It will be measured in the
NIR arm.  While we do not yet have full redshift measurement software
in place for our spectral simulations, we use the $S/N$ in the 4000\AA\
break as a proxy for our ability to measure a reliable continuum-based
redshift.  We assume that with a $S/N>5$ in the break itself we will
be able to measure a redshift \citep[e.g.,][]{tonrydavis1979}.  Note that this 
is a crude proxy for a redshift measurement: in principle we want a code that 
can uniquely identify the 4000\AA\ feature.  On the other hand, we expect 
the task will be eased by the presence of some \oii\ emission in most of our real sources.

\begin{figure*}[t]
\begin{center}
%\vbox{ 
\vspace*{-0em}
\includegraphics[width=13cm,angle=0]{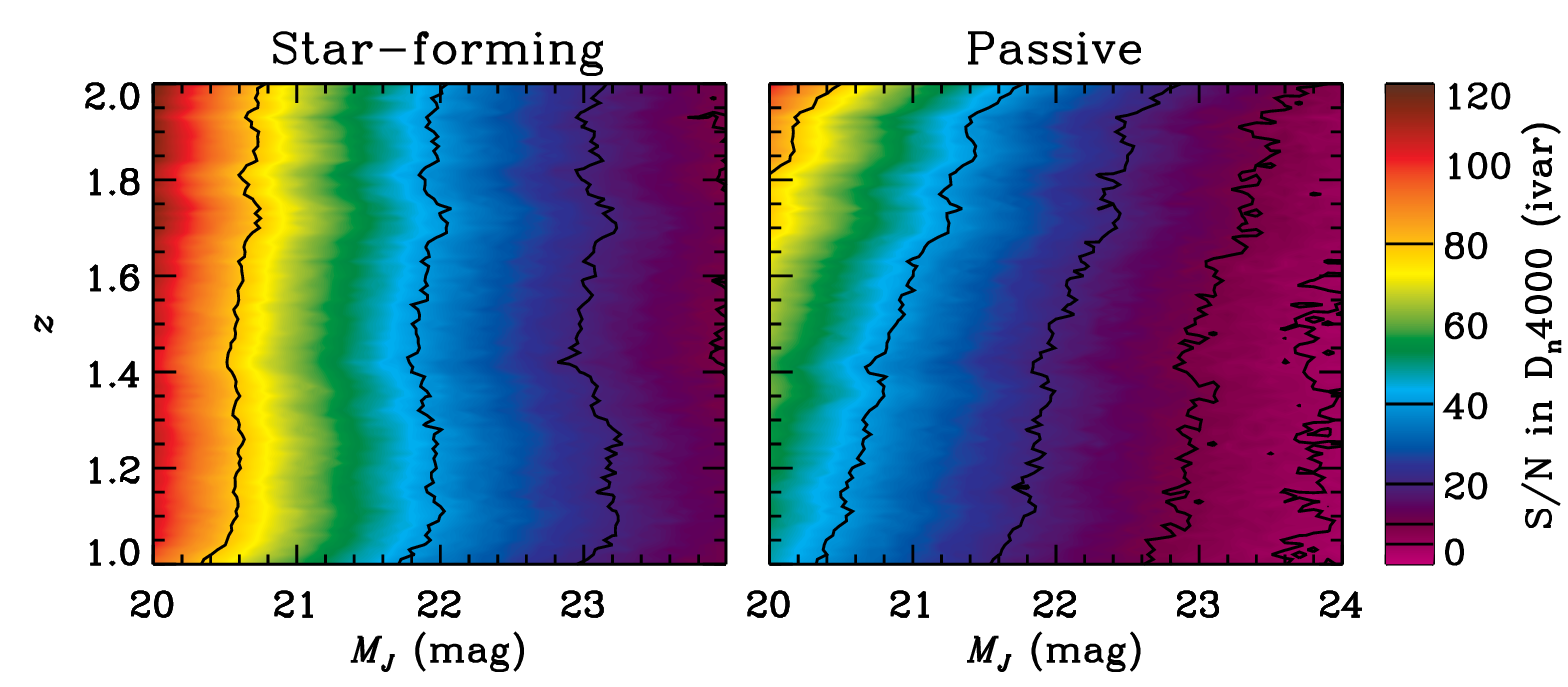}
\caption{
{\it Top}: The $S/N$ ratio in the 4000~\AA\ index as a function of $J_{\rm AB}-$band magnitude 
and redshift.  The color scale indicates $S/N$, as indicated by the scale bar, while the contours 
denote $S/N=5, 10, 20, 40, 80$.  Here we assume a 
fiducial scattered light fraction of 
0.5\%, and that we can subtract the sky to within 0.5\% as well.
}
\label{fig:heatmap}
\end{center}
\end{figure*}

Fig.~\ref{fig:heatmap} (top) demonstrates the expected $S/N$ in the 4000\AA\ break as
a function of apparent $J-$band magnitude and redshift, for both a
typical red and blue galaxy.  These figures reveal a
number of important things.  First, even at our flux limit, we expect
to achieve redshift success ($S/N>5$ in the 4000\AA\ break) for both
red and blue galaxies, independent of their break strength.  Second,
because we are selecting galaxies in the $J-$band, the blue galaxies
are considerably brighter at the 4000\AA\ break than the red galaxies
for the majority of the redshift space.  Thus, although their break
strengths are weaker, there is still higher $S/N$ in the continuum for
these templates.  Third, because the red galaxies have a relatively steep spectrum, 
their break strength actually increases as the break shifts into the $J-$band 
at higher redshift at a fixed apparent magnitude.

We have made a few critical assumptions in modeling the background,
particularly in the NIR, which we will now relax.  We assumed that the
scattered light fraction is only 0.5\% of the sky photons (as is the
specification for PFS), and that we can subtract the sky to within
0.5\% as well.  We investigate our redshift success if either the
scattered light fraction is larger or we are unable to perform such
reliable sky subtraction.

First, we assume that we can only achieve 2\% sky subtraction, rather
than the specified 0.5\%.  In practice, this assumption has little
impact on the $S/N$ calculations for a broad feature like the 4000\AA\
break, because we inverse-variance weight the pixels and mask strong
sky features anyway.

Changing the scattered light fraction does make a difference. Rather than taking 
0.5\% of the power in the sky and spreading it uniformly over the entire chip, 
we take 5\% of the power in the sky.  Since we increase the number of 
photons in the continuum with this test, we directly degrade our $S/N$. 
To calculate the level of degradation when we increase the scattered 
light fraction, we first note that over the $J-$band, 
the continuum photons (both from true sky continuum and from thermal 
background) contribute $\sim 6\%$ of all background photons. The 
rest come from sky lines.   We also note that $\sim 60\%$ of the scattered 
photons fall on science pixels.  For a scattered light fraction of 5\%, 
we will increase the background continuum level by 60\% of 5\% or 3\%, or 
a factor of 1.5. This gives a $\sqrt{1.5} \approx 20\%$ reduction in $S/N$ due to the increased 
continuum level.  With this degradation, we can still expect to achieve 
redshift success based on our simulations.  However, our
ability to measure more detailed spectral properties of the galaxies
would suffer.  Thus, our proposed program will be feasible even if the
assumptions about scattered light are overly optimistic. 

%%%%%%%%%%%%%%%%%%%%%%%%%%%%%%%%%%%%%%%%%%%%%%%%%%%%%%%%%%%%%%%%%%%%
%%BoundingBox: 
%%%%%%%%%%%%%%%%%%%%%%%%%%%%%%%%%%%%%%%%%%%%%%%%%%%%%%%%%%%%%%%%%%%%%

\subsubsection{Galaxy Physical Parameters}

Our spectra will facilitate more than just a redshift survey, as
Fig.~\ref{fig:simspectragal} demonstrates.  With the typical spectra
at $z \approx 1.5-1.7$ shown here, we can achieve a $S/N>5$ in
broad-band spectral indices (Fig.~\ref{fig:heatmap}), allowing us to
measure the continuum shape and thus constrain stellar populations.
For a $z \approx 1.5$ galaxy, we will even be able to detect
indicators of the stellar abundances from stellar absorption features
(Fig.~\ref{fig:simspectragal}).  Finally, we will have access to
emission features such as \oii\ and higher-order Balmer lines
(including H$\beta$ to $z=1.6$) that will allow us to constrain the
star formation rate and the internal extinction (see next section).
These simulations motivate our three-hour total integration times.

\subsubsection{Emission Line Fluxes and Widths}

To test our ability to detect and characterize emission-line objects,
we generate a picket-fence of emission lines across the entire
waveband, simiarly to the discussion in Section~\ref{cosmology:sec:sensitivity}. 
We generate \oii\ models as a pair of Gaussians at
$\lambda, \lambda~3726.2,3728.9$, with a ratio of $2:1$.  The pair of
lines are assumed to have the same intrinsic width, and we generate
two models with $\sigma_{\rm gas} = 100 \, {\rm and} \,
200$~km~s$^{-1}$ respectively. We also generate \lya\ models as skewed Gaussians.

Many of our science cases depend on our ability to measure not only
the line centroid of emission lines but also the line widths and fluxes.
For \oii\ lines with a combined flux of $F =
10^{-17}$~erg~s$^{-1}$~cm$^{-2}$, we find that we can recover the
total flux within $<20\%$, and the line-width to within $<40\%$. 
This is true over the full extent of the blue and red arms, and for $\sim 75\%$ 
of the NIR arm (the remainder are cases where the \oii\ overlaps directly 
with sky emission lines).
We recall that this flux is typical of \oii\ emitters at $z \approx 2$
\citep{takahashietal2007}.  We repeat the experiment for the \lya\
emission lines with $F = 0.8 \times 10^{-17}$~erg~s$^{-1}$~cm$^{-2}$,
which is again typical of LAEs at $5 < z < 7$.  We
recover \lya\ fluxes and linewidths at the $10\%$ level in the blue arm.
In the red arm we can only recover linewidths at the $20\%$ level.  However,
these limits are for individual objects; we will do better using stacking.  Thus, we will do a
factor of $\sim 20$ better at measuring the composite \lya\ profile of
high redshift emitters.  Finally, we attempt linewidth measurements
for our QSO spectra.  Even with half-hour integration times at our magnitude
limit, we expect to measure the C{\small IV} linewidth with $40\%$
precision.  The $S/N$ of the spectra will not compromise 
our goal of measuring spectral shapes and estimating 
SMBH masses from these spectra.  While 
 there are well-known uncertainties 
in calculating reliable SMBH masses using broad UV lines, 
we will be in a good position to compare the SMBH masses derived from different 
emission lines observed in the same spectra \citep[e.g.,][]{shenetal2008}. 

\subsection{The Survey Fields}

Over the coming years, the Hyper Suprime-Cam on Subaru will perform
an ambitious imaging survey in $grizy$.  The survey will have three
components with different depths.  That of most interest to us is the
HSC Deep layer, which will reach a $5 \, \sigma$ depth of $i \approx
27.2$ mag.  There is also narrow-band imaging planned that will be of
crucial importance in selecting \lya\ emitters at high redshift.
$J-$band imaging provides the cleanest method for selecting a sample
of redshift $1 < z < 2$ galaxies for study with PFS.  From the PFS
spectroscopy, our strongest continuum redshift constraint comes from
the $4000$~\AA\ break, which falls in the NIR arm (the $J-$band) for
$z \gtrsim 1.5$, and thus purely from $S/N$ considerations, we select
targets in the $J-$band for the $z < 2$ survey.  Selection down to
$J_{\rm AB} = 23.4$ mag will survey galaxies with $M^* \gtrsim
10^{10}$~\msun\ (\S 4.6).

To select a complete sample, we should have imaging $\sim 1$ mag
deeper than the magnitude limit \cite[e.g.,][]{newmanetal2012}.  For
$z < 2$ selection, we want $J_{\rm AB}=24.5$ mag photometry over 16
deg$^2$.  For the high-redshift star-forming galaxies, the ideal
photometry would have $J_{\rm AB}=26$, $H_{\rm AB}=25.6$, $K_{\rm
  AB}=25$ over 16 deg$^2$, and a bit deeper over the HSC 3.5 deg$^2$
Ultradeep area.  The fields we have chosen ({\it XMM}-LSS, ELAIS-N1,
and DEEP2-3) all have infrared coverage now to $J_{\rm AB} = 23.4$
mag.  Ongoing and planned imaging including UKIDSS-DXS, VISTA-VIDEO,
and COSMOS cover 6.5 deg$^2$ to $J_{\rm AB}=24.5$ mag.  Thus, at the
absolute minimum, we need an additional 10.5 deg$^2$ to $J_{\rm AB}=24.5$
mag and $K_{\rm AB}=23.5$ mag.  The most efficient way to achieve this
goal at the present time is VISTA - assuming 9h per band per deg$^2$,
we would need $\sim$190 hrs or thirty VISTA nights.  This represents a
large, but not prohibitive, investment.  To achieve the full depths of
$J_{\rm AB}=26$, $H_{\rm AB}=25.6$, $K_{\rm AB}=25$ over 16 deg$^2$
and the deeper limits in the Ultradeep area would require a 1500 hr
program, which is more difficult.

Our fields have been selected to maximize overlap with multiwavelength
data sets, including X-ray data to study accretion activity and
mid-infrared data to probe obscured star formation and accretion.  In
particular, we are excited about the prospect to use {\it Spitzer} and
{\it Herschel} data for measuring complementary star formation rates
to those derived from emission lines or continuum fitting.  Between
the SERVS, SWIRE, and COSMOS surveys, there is {\it
  Spitzer} data to at least 4 $\mu$Jy at 3.6 $\mu$m in the same 9
deg$^2$ with the NIR coverage. There is $U-$band data with 
AB=$26$ mag depth from CFHT that we will use in the selection of 
high redshift star forming galaxies. We also have good overlap with the
Hermes fields \cite[][]{hermesetal2012}.  By the same token, our
survey fields will provide natural targets for ALMA and/or CCAT.

\subsection{Survey Design and Target Selection: Continuum-selected $z < 2$ Survey}

%%%%%%%%%%%%%%%%%%%%%%%%%%%%%%%%%%%%%%%%%%%%%%%%%%%%%%%%%%%%%%%%%%%%
%%BoundingBox: 
\begin{figure*}[t]
\begin{center}
%\vbox{ 
%\vskip -10mm
%\hskip +10mm
%\psfig{file=mass_lim.eps,width=0.4\textwidth,keepaspectratio=true,angle=-90}
%}
\includegraphics[width=0.4\textwidth,angle=-90]{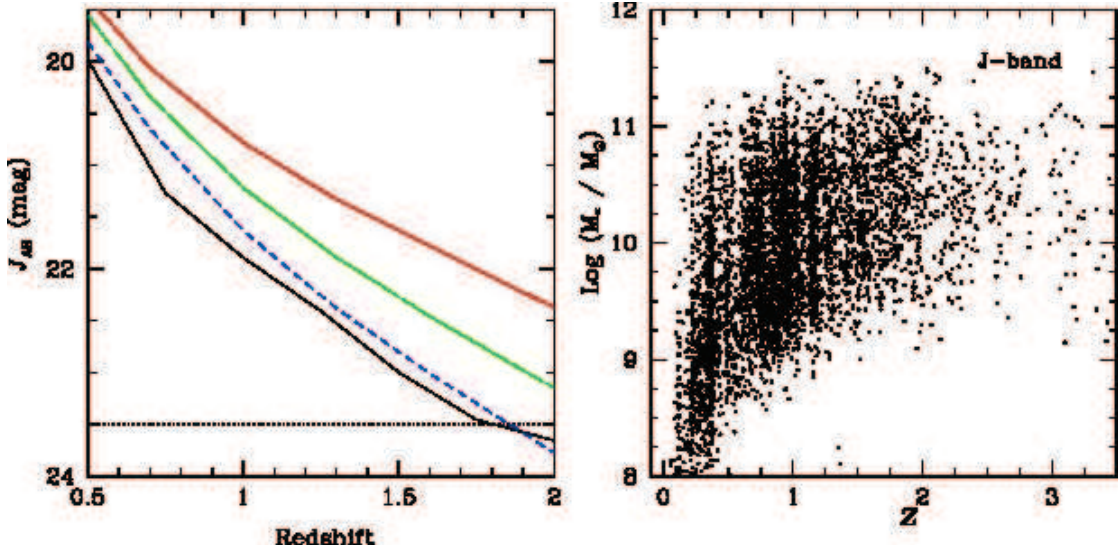}
\vskip -0mm
%\figcaption[]{
\caption{
%\small
{\it Left}: Illustrative models of galaxies with different 
evolutionary histories based on \citet{maraston2005} 
stellar population models. 
All models have $\tau e^{-\tau}$ star formation histories, and are 
normalized to be $L^*$ galaxies today.  In red is a red galaxy progenitor 
with a final mass of $M^* = 6 \times 10^{10}$~\msun.  
In green is a more typical star-forming galaxy with a final 
mass of $M^* =3.4 \times 10^{10}$~\msun with an extended history, while the blue 
line is a more extreme star forming system with a final mass of 
$M^* =2.3 \times 10^{10}$~\msun.  For reference, in black is the 
mean apparent $J-$band magnitude of a $3 \times 10^{10}$~\msun\ galaxy 
at each epoch from the $K-$band selected NEWFIRM medium-band survey.
{\it Right}: 
Expected stellar mass distribution for our $1<z<2$ survey, based on the NEWFIRM 
Medium Band Survey \citep{whitakeretal2011}.  We apply a $J-$band magnitude limit 
of 23.4 mag, but note that the total number of points in the figure 
corresponds to a 0.2 deg$^2$ survey; we will spectroscopically 
observe 125 times as many objects. 
Without a redshift selection, a large fraction of the galaxies in a magnitude-limited 
sample have $z < 1$.  The banding in the figure is due to large scale structure; our 
much larger volume will average over this cosmic variance.
\label{fig:mass_lim}}
\end{center}
\end{figure*}
%%%%%%%%%%%%%%%%%%%%%%%%%%%%%%%%%%%%%%%%%%%%%%%%%%%%%%%%%%%%%%%%%%%%%

%Throughout the document, 
As we discuss the time needed for each component, 
we will use the natural unit of fiber-hours (we will also convert to night-equivalents, 
but note that these are misleading since we plan to observe all types of targets at 
all times for efficiency). In a 100-night survey, assuming 2000 science fibers and 6 hr
nights, we have 1.2 million fiber hours.  

Our primary goal for the $z < 2$ survey is to study the evolution of 
the $\sim L^*$ galaxy population, which leads to the following choices:

\begin{enumerate}

\item 
  {\bf Magnitude limit, with color selection}: 
  Based on our spectral simulations, we can expect to reach a
  $S/N>5$ in the 4000~\AA\ break in 3 hr
  exposures at our limiting magnitude of $J_{\rm AB} = 23.4$ mag (Fig.~\ref{fig:heatmap}).
  With this magnitude limit, we expect to probe masses as low as
  $10^{10}$~\msun\ at $z \approx 2$ based on the NEWFIRM Medium-Band
  redshift survey (Fig. \ref{fig:mass_lim}), although we are only
  complete for stellar masses $\gtrsim 10^{11}$~\msun.  We will describe 
  various color selections below. In order to
  observe the same population at low redshift, we plan to observe all
  objects to a limiting magnitude of $J_{\rm AB}=21$ mag.  Since we
  will spend only 20 min per galaxy at these bright limits, we get
  these 6000 galaxies per deg$^2$ essentially for free, and they
  provide a low-redshift comparison sample (e.g., for comparing
  \halpha\ and \oii-derived star formation rates).

\item
{\bf Magnitude-limited component}:
We propose to perform a purely magnitude-limited survey to $J_{\rm AB}
= 23.4$ mag, with the same exposure times as outlined above, in 2 PFS
pointings. With two independent pointings, we will sample different
large-scale structures, and given that we expect $\sim 10$ clusters
with $M>10^{14}$~\msun\ and $z > 1$ in this volume, we expect to sample a range of
environments in these pointings. Ideally, this component would be
completed early in the survey.  This small ({82,000 fiber hours} or
$\sim 8$ night) investment, includes observations of an additional
24,000 galaxies per deg$^2$ over the nominal color-selected survey, that
will allow us to tune and understand our color selection in detail in
preparation for the main survey. Furthermore, we can investigate rare
populations that our color selection may exclude.  Finally, the bias
of the galaxies used for the cosmology survey may be a strong function
of redshift, as it involves a color cut which may select
astrophysically distinct populations of galaxies at different
redshifts.  The bias, and the relationship of the selected galaxies to
the full galaxy population, needs to be understood in order to use the
sample for cosmological inferences, especially when combining
clustering measurements at different redshifts.  (Note that the BAO
scale itself is actually fairly insensitive to these details, as the
bias is scale-independent on these very large scales).

\item 
  {\bf Area}: The area should be large enough to minimize cosmic variance. In 16
  deg$^2$ we cover a volume a few tenths that of the SDSS main galaxy sample in a
  redshift bin at $z \approx 2$, which should sample a fair volume of
  the universe.  To quantify the level of cosmic variance, in Table \ref{tab:gal_cosmicvar}
  we show the expected error bars on the number density of galaxies in
  mass and redshift bins derived using the method of
  \citet{mosteretal2011}.  Thus over the survey area of 16 deg$^2$ our analysis
  will not be cosmic variance limited.

\end{enumerate}

%\begin{table}[t]
%\begin{center}
%%\caption{\bf Impact of Cosmic Variance}
%{Impact of Cosmic Variance}\\
%%%\tablenum{2}
%\begin{tabular}{lllll}\hline\hline
\begin{deluxetable}{lllll}
\tablewidth{0pt}
\tabletypesize{\footnotesize}
\tablecaption{ Impact of Cosmic Variance
\label{tab:gal_cosmicvar}}
\startdata \hline\hline 
      9.1  &  0.02  &  0.02  &  0.02 &   0.03 \\
      9.6  &  0.02  &  0.02 &    0.03 &   0.03 \\
      10.1  &  0.02  &  0.02 &    0.03 &    0.03 \\
      10.6 &    0.02  &  0.03 &    0.03 &    0.04 \\
      11.1 &     0.03 &   0.03 &    0.04 &    0.05 \\
      11.6  &  0.04  &  0.05 &    0.06 &    0.07
%\\ \hline\hline
\enddata
%\end{tabular}
%\caption{
\tablecomments{
The logarithmic error bars on the number density in each
logarithmic stellar mass bin (log $M_{\odot}$; column 1) are shown for four
redshift ranges: $1 < z < 1.4$ (Col. 2), $1.4 < z < 1.8$ (Col. 3), 
$1.8 < z < 2.2$ (Col. 4), $2.2< z < 2.6$ (Col. 5), assuming a 16 deg$^2$ area.  
The error bars are derived using the method of Moster et al. (2010).
%\label{tab:gal_cosmicvar}
}
\end{deluxetable}
%\end{center}
%\end{table}

\subsubsection{Color Selection}

%\begin{table}[t]
%\begin{center}
%%\caption{\bf All Targets}
%{All Targets}\\
%%\tablenum{5}
%\label{tab:sample}
%\begin{tabular}{llllllll}\hline\hline
\begin{deluxetable*}{llllllll}
\tablewidth{0pt}
\tabletypesize{\footnotesize}
\tablecaption{All Targets
\label{tab:sample}
}
\startdata \hline\hline 
Mag.\ & $z$ & Selection & N deg$^{-2}$& Sampling & Int Time & Area & Fiber Hours\\
(1) & (2) & (3) & (4) & (5) & (6) & (7) & (8) \\
\hline
$J_{\rm AB}<21$     & $z < 1$ & photoz & 6k & 0.8 & 0.3 & 16 & 23k\\
$J_{\rm AB}= 21-22$ & $1 < z < 2$ &  photoz & 1.5k  & 0.8 & 0.3 & 16 & 5.8k\\
$J_{\rm AB}=22-23$ & $1 < z < 2$ &  photoz & 8.4k  &  0.8 & 2 & 16 & 215k\\
$J_{\rm AB}=23-23.4$ & $1 < z < 2$  &  photoz & 12k & 0.8  & 3 & 16 & 460k\\
\hline
$J_{\rm AB}<23.4$     & $z < 2$ & all & 47k & 0.8 & 0.3-3 & 2.6 & 82k \\
%$J_{\rm AB}<23.4$     & $z < 2$ & all & 47k & 0.3-3 & 82k \\
\hline
$NB387=25$ & $z=2.2$  &  LAE & 300 & 1 & 2 & 16 & 9.6k\\
$NB816=25$ & $z=5.7$  &  LAE & 230  & 1 & 5 & 16 & 18.4k\\
$NB921=25$ & $z=6.6$  &  LAE & 240  & 1 & 5 & 16 & 19.2k\\
$NB101=26$ & $z=7.3$  &  LAE & 14  & 1 & 6 & 3.5 & 0.3k \\
\hline
$i=24$ & $2 < z < 3$ & LBG & 4.6k & 0.4 & 3 & 16 & 88k \\
$i=24$ & $3 < z < 4$ & LBG & 840 & 1 & 3 & 16 & 40k\\
$i=25$ & $4 < z < 6$ & LBG & 100 & 0.1 & 6 & 16 & 4.8k\\
%$i=24.4$ & $z \sim 5.0$ & LBG & 21 & 6 & 25 \\
%$i=24.4$ & $z \sim 5.9$ & LBG & 0.5 & 6 & 25 \\
\hline
$z=26$ & $2 < z < 4$ & LBG & 8k & 0.1 & 16 & 3.5 & 45k\\
$z=26$ & $4 < z < 6$ & LBG & 4k & 0.1 & 16 & 3.5 & 22k\\
%$i=24$ & $z \sim 3.8$ & LBG & 18k & 3 & 3.5 \\
%$i=26$ & $z \sim 5.0$ & LBG & 3.2k & 6 & 3.5 \\
%$i=26$ & $z \sim 5.9$ & LBG & 227 & 6 & 3.5 \\
\hline
$J<24$ & $3 < z < 7$ & QSO & 30 & 1 & 0.5 & 1400 & 21k \\
$J<25$ & $3 < z < 7$ & QSO & 100 & 1 & 1 & 16 & 1.6k
%\\ \hline\hline
\enddata
\tablecomments{
%\end{tabular}
%\caption{
Col. (1): $J-$band magnitude (AB) or narrow-band magnitude.
Col. (2): Redshift range.
Col. (3): Selection method.
Col. (4): Number of galaxies per deg$^2$.  For the continuum-selected 
sources, we have adopted the mean of the number densities from the 
COSMOS and NEWFIRM catalogs. The LBGs are grouped in redshift with 
$\delta z \approx 1$.  Note that the 1400 deg$^2$ QSO search will be part of 
the BAO survey.
Col. (5): Fraction of galaxies that we will sample.
Col. (6): Integration time (hrs).
Col. (7): Survey area (square degrees).
Col. (8): Fiber hours.
}
\end{deluxetable*}
%\end{center}
%\end{table}

A major strength of the SDSS was the simple magnitude-limited
selection with an easy-to-model selection function. We would like to
replicate that ideal to the extent possible, but require a color selection if 
we want to spend more than half of our time observing galaxies with $1 < z < 2$.  
We have explored various color-selection techniques to maximize the
time spent in the redshift range of interest and outline their
strengths and weaknesses here.  Note that in addition to the
main survey of 16 deg$^2$, we require a purely magnitude-limited
sample over  2 PFS pointings to be observed early in the
survey, which will allow us to calibrate our selection methods. A
similar strategy was adopted by DEEP2 \cite[][]{davisetal2003}. In Table \ref{tab:sample}
we show the number of targets by magnitude, number density, and
exposure time.

\begin{figure*}[t]
\begin{center}
%\vbox{ 
%\vskip -1mm
%\hskip +5mm
%\psfig{file=color_select.eps,width=0.4\textwidth,keepaspectratio=true,angle=-90}
%}
%\centerline{\includegraphics[width=0.4\textwidth,angle=-90]{color_select.eps}}
\includegraphics[width=0.45\textwidth,angle=0]{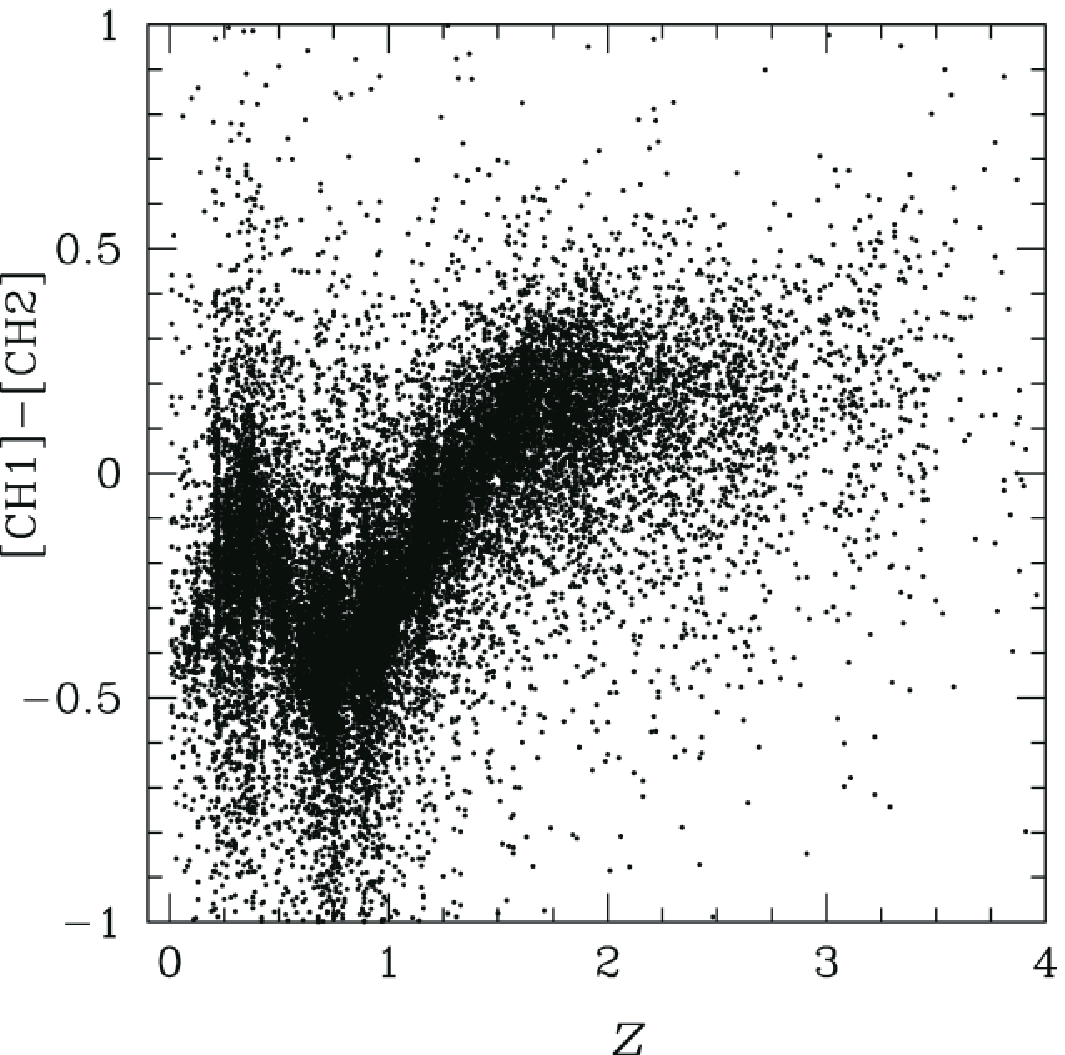}
\includegraphics[width=0.5\textwidth,height=0.42\textwidth,angle=0]{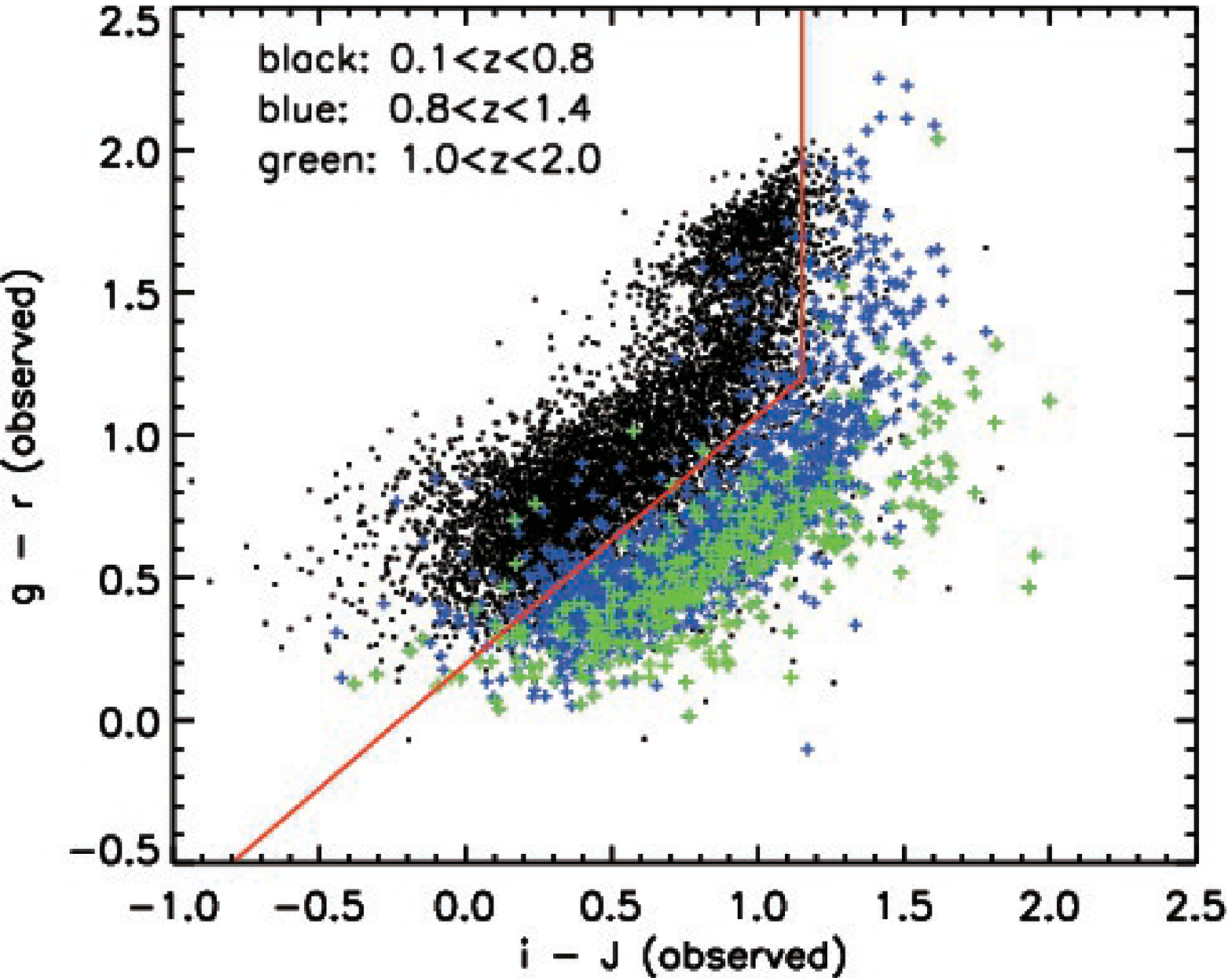}
\vskip -0mm
%\figcaption[]{
\caption{
%\small
{\it Left}:
The dependence of galaxy \spitzer/IRAC CH1-CH2 ($3.6-4.8 \mu$m) color on redshift.  
Galaxies are from the NEWFIRM Medium Band Survey, and IRAC 
data have 1 $\mu$Jy depth.
The structure is due to the $1.6 \mu$m peak in galaxy spectra 
moving through the CH1 band.
{\it Right}: The $griJ$ color selection, where galaxies are from the COSMOS survey. 
We select galaxies with $i-J > 1.15$ or $(g-r) < 0.87(i - J) + 0.196$;
%g-r < (0.5 + 0.359 [i-J + 0.8])$; 
the 
selection is highly complete ($\sim 90\%$) while the contamination is only 
$\sim 20\%$ for $z < 0.8$.
\label{fig:color_select}}
\end{center}
\end{figure*}

We outline three potential color selections here:
\begin{enumerate}

\item {\bf Photometric redshifts}: As a straw-man, using photometric
  redshifts is a very appealing strategy, as it utilizes all of the
  exquisite HSC imaging.  We are testing the
  capability of HSC Deep to measure reliable photometric redshifts.
  Based on simulations run by M. Tanaka of photometric redshifts from 
  HSC Deep data, the completeness, or the fraction of objects with
  $z_{\rm spec} >1$ that correctly have $z_{\rm phot}>1$, is 93\%.  The
  contamination, which is the fraction of objects at $z_{\rm spec}<1$ that
  have $z_{\rm phot}>1$, is 5\%.  Our ability to measure reliable 
  photometric redshifts should improve dramatically in the coming years, as 
  we gain access to real HSC data, and take advantage of ongoing redshift 
  surveys in our field (e.g., VIPERS).  A major concern with using the photometric 
  redshift is systematic error (for instance, in the template set).   
  We will be measuring photometric redshifts for a population without 
  good spectroscopic coverage, and therefore there is room for 
  significant spectral differences between the templates and the targets.
  Much more work is needed to determine whether photometric redshift selection 
  can provide an unbiased sample selection.

\item {\bf IRAC CH1-CH2 Selection}: Galaxy spectral energy
  distributions peak at $\lambda = 1.6 \, \mu$m, redward of
  which is the Raleigh-Jeans tail.  As this feature
  moves through the IRAC bandpasses (particularly CH1~$=3.6 \,\micron$
  and CH2$=4.8 \,\micron$), the CH1-CH2 color goes through a distinct
  minimum at $z \approx 0.8$ and then rises monotonically towards
  higher redshift (Fig.~\ref{fig:color_select}~{\it left}).  
  A simple cut of CH1-CH2~$>-0.1$ effectively selects $z > 1.3$ 
  galaxies \cite[e.g.,][]{papovich2008}.  Using the NEWFIRM Medium 
  Band Survey \citep{whitakeretal2011},
  we find that the completeness, or the fraction of objects with
  $z > 1$ that are selected by this method, is $64\%$, while the
  contamination, or fraction of selected objects that have $z < 1$, is
  30\%.  The numbers are not as impressive as the photometric
  redshifts, but the selection is easily modeled and relatively free
  of systematics.  There is \spitzer\ data of sufficient 
  depth over $\sim$ half of the 16 deg$^2$ field.

\item {\bf {\it griJ} Selection}: Inspired by the DEEP2 color
  selection, we have designed an optical/NIR color selection that is
  effective at removing $z < 0.8$ interlopers. The color region we
  select is defined by $i-J > 1.15$ or $(g-r) < 0.87(i - J) + 0.196$
%g-r < (0.5 + 0.359 [i-J +
%  0.8])$ 
and $J_{\rm AB} < 23.4$ mag (Fig.~\ref{fig:color_select}~{\it right}).  Again based on the NEWFIRM
  Medium Band Survey, the completeness (fraction of galaxies with $z >
  1$ that we select) in this case is 93\%.  The contamination
  (galaxies in the final sample with $z < 1$) is 42\%.  However, if we
  look at galaxies in the final sample with $z < 0.8$, the
  contamination is only 24\%.  Therefore, this simple color selection,
  with all bands to be observed, is very promising if we decide to
  target $z > 0.8$ galaxies.  The total number of objects per deg$^2$
  would increase by 37\%.  These numbers come from the NEWFIRM catalog
  but change little when we use the COSMOS photometric redshift
  catalog instead. We view this color selection as a very promising
  option.  In what follows, the numbers assume a perfect cut at $z=1$,
  but the appropriate correction would only be an increase of 37\% for
  the number density of continuum-selected galaxies.
\end{enumerate}

All of these selections assume $J-$band imaging to the desired depth
of $J_{\rm AB}=$24.5 mag for the entire 16 deg$^2$ field.  Should we
fail to acquire the additional 7 deg$^2$ of $J-$band imaging, we can
also use a pure optical-band color selection: $1.405 (g-r) + 0.505
(r-z) - 2.150 (i-y) < 0$, which has high completeness (93\%) and 29\%
contamination for $z > 0.8$, based on COSMOS data.  In this way, we
can mitigate incompleteness due to including shallow $J-$band data.

%%%%%%%%%%%%%%%%%%%%%%%%%%%%%%%%%%%%%%%%%%%%%%%%%%%%%%%%%%%%%%%%%%%%
%%BoundingBox: 
%\vskip -5mm
%%%%%%%%%%%%%%%%%%%%%%%%%%%%%%%%%%%%%%%%%%%%%%%%%%%%%%%%%%%%%%%%%%%%%

We hope to additionally mitigate contamination through real-time
redshift determinations. The main contaminant in a continuum-limited
sample is moderate-redshift, intrinsically faint galaxies that are
likely strongly star forming and thus have strong emission lines.  We
should be able to identify their redshifts with a single 20 min
exposure, if we can carry out real-time analysis of the spectra.
Thus, we can move to another target in $<1$ hr, given the fact that
the fibers are individually deployable.  In addition, we will sharpen
our target selection by using the redshift information from the
cosmology survey, and the VIPERS spectroscopy, which overlaps with our
fields.  Finally, with our magnitude-limited component, we will work to understand 
our selection bias.

%%%%%%%%%%%%%%%%%%%%%%%%%%%%%%%%%%%%%%%%%%%%%%%%%%%%%%%%%%%%%%%%%%%%
%%BoundingBox: 
\begin{figure*}[t]
\begin{center}
%\vbox{ 
%\vskip 0mm
%\hskip +0mm
%\psfig{file=color.eps,width=0.3\textwidth,keepaspectratio=true,angle=-90}
%}
\includegraphics[width=0.3\textwidth,angle=-90]{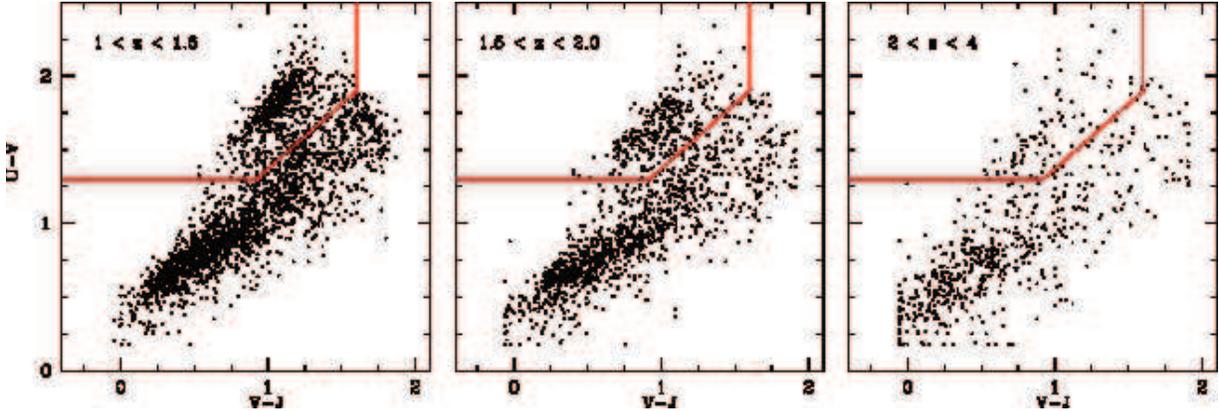}
\vskip -0mm
%\figcaption[]{
\caption{
%\small
Rest-frame color-color plots of the NEWFIRM Medium 
Band Survey catalog, in the redshift bins indicated and 
with the $J-$band cut of 23.4 mag.  In red are the color 
cuts used to define the red sequence by 
\citet{williamsetal2009}.  It is clear that we will sample the 
red sequence all the way to $z \approx 2$.
\label{fig:color}}
\end{center}
\end{figure*}
%\vskip -5mm
%%%%%%%%%%%%%%%%%%%%%%%%%%%%%%%%%%%%%%%%%%%%%%%%%%%%%%%%%%%%%%%%%%%%%

%{\it Target Sampling}
\subsubsection{Target Sampling}

We must also make a scientific choice about sampling.  If we sparsely
sample the galaxy population, we can cover a wider area more quickly,
and still sample the average properties of the population.  However, we will 
lose information on small scale clustering (Fig. \ref{fig:surveyparam}).
Specifically, the PFS survey will allow accurate measurements of the evolution of
the abundance and clustering of galaxies as a function of stellar mass
and of the galaxy-galaxy lensing signal (utilizing the imaging from
the Hyper Suprime-Cam survey), from $1 \le z \le 2$. These results can
be used to constrain the evolving relation between the stellar mass of
galaxies and the mass of the halos in which they reside. To
investigate how well we need to sample galaxies in our survey in order
to measure this evolution, we have generated mock catalogs using the
semi-analytical model of \citet{deluciaetal2006}, which is based upon
the dark matter halo merger trees drawn from the Millennium simulation
\citep{Springel2005}. We have considered four redshift bins of $\Delta
z=0.3$ and used bins in stellar mass of $\Delta$log $M_*=0.3$ each. 

\begin{figure*}[t]
\begin{center}
%\vbox{ 
%\vskip 0mm
%\hskip +0mm
%\psfig{file=figsampling.eps,width=1.\textwidth,keepaspectratio=true,angle=0}
%}
\centerline{\includegraphics[width=.8\textwidth]{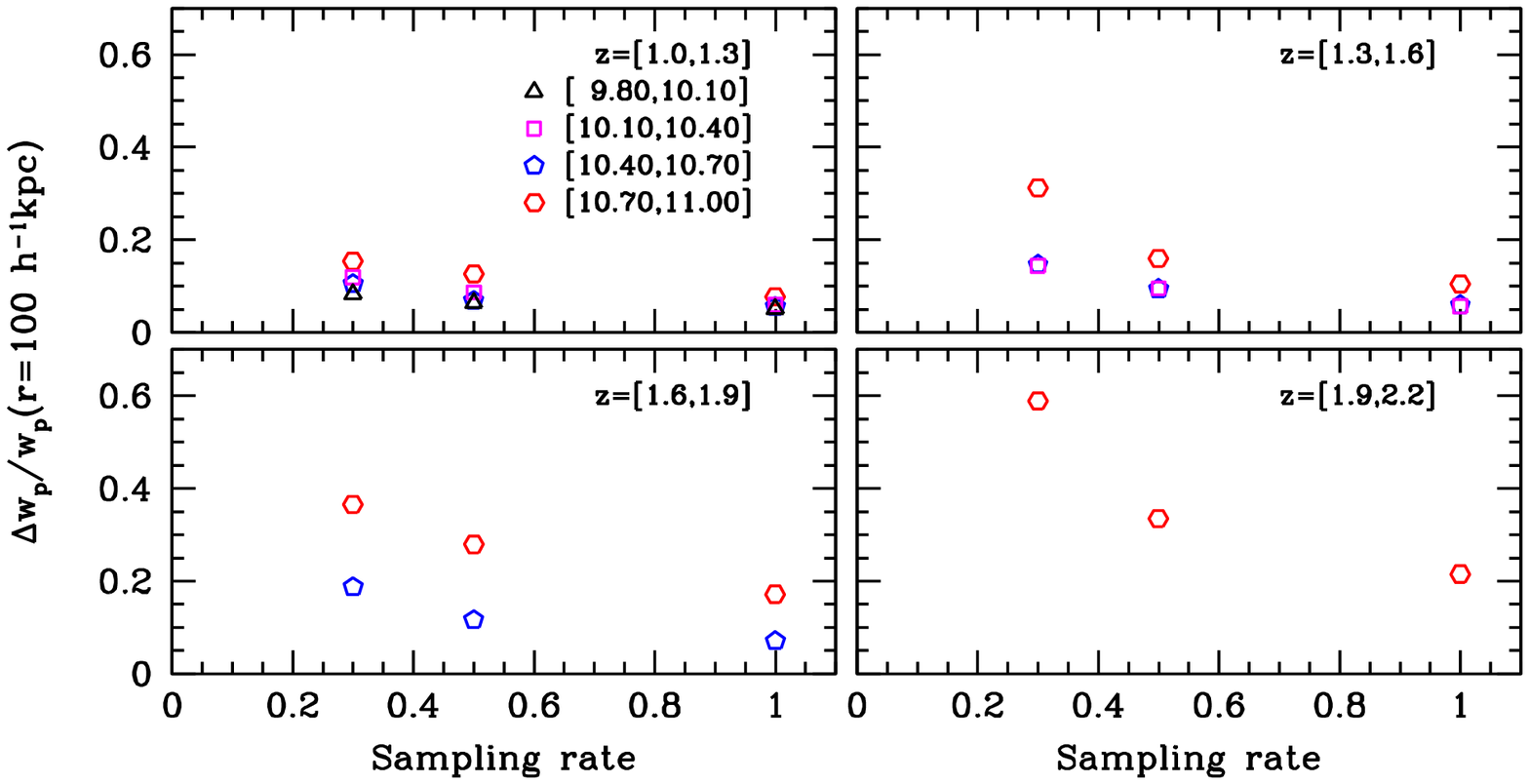}}
\vskip -0mm
%\figcaption[]{
\caption{
%\small
We plot the fractional error in the projected correlation function $w_p$ at 100 kpc scales, 
as measured from semi-analytic models to match the PFS selection.  The measurements are 
made in four redshift bins, with different symbols corresponding to different stellar 
mass bins, as indicated. We see that the uncertainties improve substantially as sampling 
approaches $\sim 70-80 \%$, after which we reach diminishing returns.
\label{fig:clustering}}
\end{center}
\end{figure*}

Fig.~\ref{fig:clustering} shows the fractional error in the
projected correlation function $w_p$ on 100 kpc scales for sampling
rates of 30, 50, and 100\%.  On the smallest scales the errors are
dominated by the number of pairs of galaxies expected at a given
separation (thus benefiting from increased sampling), while on the
largest scales, the errors are dominated by sample variance.  We can see 
that reaching sampling rates $\sim 80\%$ is important for the errors, while 
for higher levels of completeness we get diminishing returns.  In practice it 
is challenging to achieve higher levels of completeness anyway, so we adopt 
80\% sampling.

In a similar manner we would also like to simulate the galaxy-galaxy
lensing signal expected around the PFS galaxy sample, using the
redshift distribution of sources from the HSC imaging survey. We will
model the galaxy clustering results, the stellar mass function
measurements, and the galaxy-galaxy lensing results in the framework
of a halo occupation distribution model \cite[see e.g.,
][]{leauthaudetal2012}.  The joint analysis of
such data with realistic errors can be used to judge the sampling
requirement for the galaxy evolution component of the PFS survey.

%%%%%%%%%%%%%%%%%%%%%%%%%%%%%%%%%%%%%%%%%%%%%%%%%%%%%%%%%%%%%%%%%%%%
%%BoundingBox: 
%\vskip -5mm
%%%%%%%%%%%%%%%%%%%%%%%%%%%%%%%%%%%%%%%%%%%%%%%%%%%%%%%%%%%%%%%%%%%%%

\subsubsection{Summary}

{Taking 80\% sampling, we plan for a total of 69 nights, or
  704,000 fiber-hours, for this component of the survey, assuming 3 hrs
  per galaxy in the faintest half-magnitude bin.}  We additionally include 
a magnitude-limited component over 2.6 deg$^2$ that adds {82,000 fiber
  hours} ($\sim 8$ nights). In Fig.~\ref{fig:mass_lim}, we show that we will be able to observe galaxies
with masses $M> \sim 10^{10}$~\msun\ at our redshift limit, allowing
us to study the progenitors of $L^*$ galaxies today.  Thanks to the
$J-$band selection, we will be sensitive to the red sequence even at
$z \approx 2$, as shown in Fig.~\ref{fig:color}.  The PFS survey is
compared in redshift, depth, and area to other existing and upcoming
surveys in Fig.~\ref{fig:surveyparam}.  Our survey is designed to
explore a hitherto unknown epoch in cosmic history, at the time when
the bulk of the stellar mass in galaxies was assembled.

\subsubsection{Uniqueness of the PFS Survey}

As we plan for the PFS survey, many powerful NIR multi-object
spectrographs are now coming online (e.g., MOSFIRE on Keck).  When the
PFS survey begins, large samples of $1 < z < 2$ galaxies will already
be observed.  What will make PFS unique?  There are two aspects.  The
first aspect is volume.  Thanks to the large field-of-view and
tremendous multiplexing capability of PFS, we will achieve both high
sampling and a large volume, allowing us to study the importance of
environment from small groups up to Mpc scales and allowing for large
samples of foreground/background pairs for absorption studies. The
second is our wide wavelength coverage, which allows us to study both
the rest-frame optical and rest-frame UV properties of these galaxies
at one time.  It is conceivable that earlier surveys will have
observed a few thousand galaxies with such wide wavelength coverage
when the PFS survey begins \cite[e.g.,][]{Steidel2003}; we will study
half a million galaxies.  The wide area of our survey, combined with
the measurements of weak lensing and clustering, will allow us to link
the evolution of galaxies with the evolution of dark matter
halos. Furthermore, the large numbers will allow us to study
spectroscopic properties of galaxies in fine bins in stellar mass,
redshift, star formation rate, and environment over the full range of
our survey.  
%In addition, PFS will also provide excellent targets for
%follow-up with ALMA and {\it JWST}.

\subsection{Survey Design and Target Selection: 
Galaxies at Cosmic Dawn}

At higher redshifts, we wish to study the evolution in the star-forming population 
out to $z \approx 7$, and use this population as a tracer of cosmic reionization. 
While $L^*-$progenitors are very faint at these redshifts, we will find them via the
\lya\ emitter population.  We will study more massive star-forming galaxy populations 
selected via their continuum colors.

\subsubsection{Survey Design and Target Selection}

The characteristics of the instrument that are important for the
high-redshift component include: (1) the high multiplexing of 2000 science
fibers (with 400 fibers devoted to sky); (2) the wide field-of-view (1.3 deg$^2$); (3) sensitivity
between 3800 - 12600${\rm \AA}$, (4) quick redeployment of fibers
between exposures, (5) accurate and stable positioning (over a period
of hours), and (6) stable fiber throughput.

In particular, for the high-redshift science case, the wide-area
survey enabled by the PFS instrument allows us to cover both a high
dynamic range in galaxy properties and cover a wide range of
environments.  Over our uniform 16 deg$^2$ survey, we expect to find
60,000 $z > 2$ galaxies, with those at $2<z<4$ brighter than {$i_{\rm AB}<24$} and
those at $z>4$ reaching fainter magnitudes {$i_{\rm AB}<25$}.  The large
area will also allow us to observe multiple ionized bubbles in the IGM at
$z \approx 7$, probed by \lya\ emitters. The size of the
ionized bubbles may be as large as $\sim 1$ degree at the end of the
reionization epoch \citep{mcquinnetal2007}.

\begin{figure*}[t]
\begin{center}
%\vbox{ 
%\vskip 0mm
%\hskip +0mm
%\psfig{file=photvspecz_highz.eps,width=0.3\textwidth,keepaspectratio=true,angle=-90}
%}
\centerline{\includegraphics[width=0.45\textwidth,angle=-90]{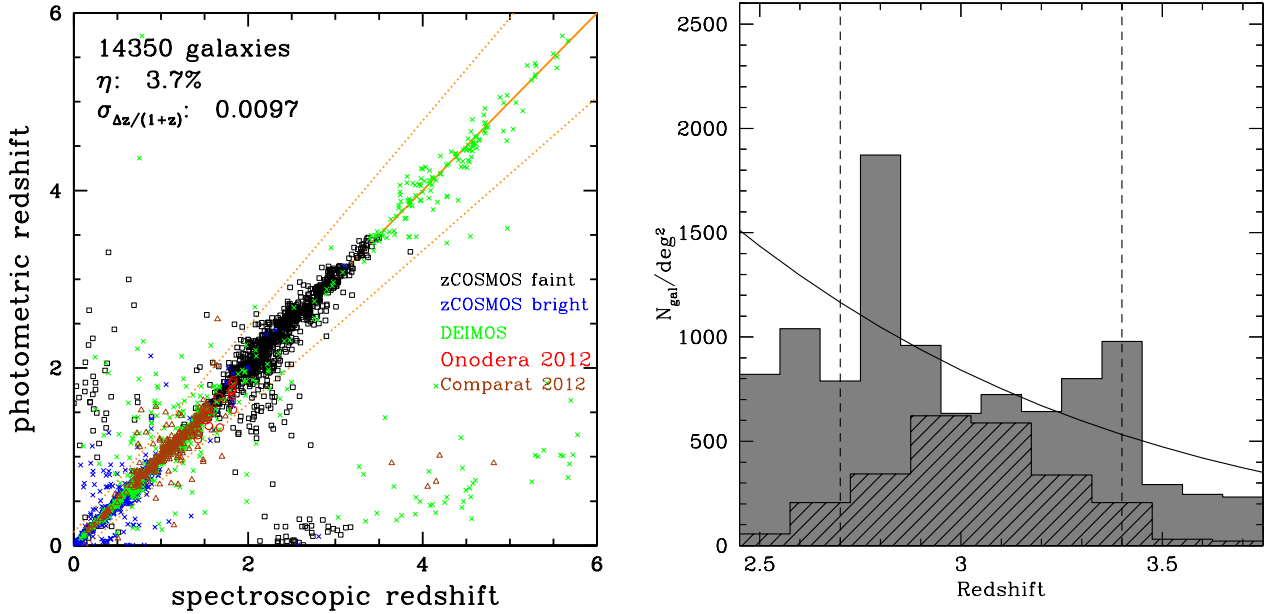}}
\vskip -0mm
%\figcaption[]{
\caption{
%\small
{\it Left}: A comparison between photometric and spectroscopic 
redshifts from O. Ilbert et al.\ in preparation from
redshift measurements (zCOSMOS + DEIMOS programs) in the COSMOS field.
With enough photometric coverage, photometric redshifts provide a reasonable selection 
method.
{\it Right}: 
We contrast the gbell-shapedh spectroscopic redshift $N(z)$ that results from a 
classic three-color LBG selection (hatched) with the result from the VVDS spectroscopic 
redshift sample magnitude-selected with $i_{\rm AB} < 24.75$ (shaded)
 \citep[see also][]{lefevreetal2013}. 
The solid curve represents the best fit to the VVDS spectroscopic $N(z)$
equivalent to a photometric redshift distribution.
\label{fig:photspecz}}
\end{center}
\end{figure*}

The majority of the high-z sample will be selected via photometric
redshift selection using $ugrizyJHK$+IRAC when available ($M_{\rm
  gal}>10^{10}$ $M_{\odot}$). Photometric redshift selection will
deliver a smooth distribution in $N(z)$ (Fig. \ref{fig:photspecz}),
but may miss some targets.  Therefore, we will also target galaxies
that are selected via standard three-color techniques but not selected
via their photometric redshift.  For the sake of simplicity we will
refer to all continuum-selected targets as Lyman break galaxies (LBG hereafter).

%%%%%%%%%%%%%%%%%%%%%%%%%%%%%%%%%%%%%%%%%%%%%%%%%%%%%%%%%%%%%%%%%%%%
%%BoundingBox: 
%\vskip -5mm
%%%%%%%%%%%%%%%%%%%%%%%%%%%%%%%%%%%%%%%%%%%%%%%%%%%%%%%%%%%%%%%%%%%%%

The second component to the high-z galaxy survey is the spectroscopic
followup of \lya\ emitters (LAE) selected from the HSC Deep and
Ultradeep narrow-band imaging.  The narrow-band filters in the
Ultradeep survey are $NB387; z=2.2$, $NB816; z=5.7$, $NB921; z=6.6$,
and $NB101; z=7.3$.  In the Deep survey, we will have imaging in
$NB387, NB816,$ and $NB921$.  This technique provides the most certain way 
to identify the low-mass galaxies that likely grow into $L^*$ galaxies today 
\cite[][]{shimasakuetal2004}.

Here, we provide specific details regarding the target selection that
varies as a function of redshift.  The exposure times are chosen to 
achieve a $S/N$ of $\sim 3$ per resolution element, assuming $R \sim 400$ 
in the red continuum of the typical galaxy 
in each subsample, as justified below.

\begin{enumerate}

\item  At $2 < z < 4$, we will target the bright end of the LBG luminosity
  function down to $i_{\rm AB} \approx 24$ uniformly over the full 16
  deg$^2$ survey area. We will observe sources brighter than $L^*$,
  with masses $M^*> 10^{10}$~\msun, typically $M^* \approx
  10^{11}$~\msun\ \cite[see, e.g.,][]{cucciatietal2012}.  Existing
  $u-$band imaging from the CFHT archive reaches 26 AB mag, which is
  deep enough to detect the strong drop in flux shortward of \lya\ for
  low-z dropouts. With these data, we can identify high-z candidates
  down to $i_{\rm AB} \approx 24$. If we want to study galaxies in $\delta z=0.5$ 
redshift intervals with $\sim 1$ mag (or ten $10^{0.2}$ \msun) bins, 
as well as in bins of galaxy type and environment with at least $\sim 200$ 
galaxies per bin, then we require $\sim 8,000$ targets.  

Over 16 deg$^2$, we do not expect to be cosmic variance limited. At 
$\langle z \rangle =2.5$ the cosmic variance uncertainty in the 
stellar mass function will be 0.02 dex, 
while at $\langle z \rangle =3.5$ it will be 0.02 dex and 0.08 dex 
for log $M^* = 10.25/$~\msun\ and log $M^* = 11.25/$~\msun, 
respectively, according to \citet{mosteretal2011}. Accounting for the decreasing 
source density per redshift bin towards higher redshifts (see Table 4.2), 
due to both a flux-limited selection and an intrinsic decrease in the number 
density of the bright population, we plan to target $\sim 100\%$ of the sources at 
$3 < z < 4$ ($i_{\rm AB}=24.3$) and $\sim 40\%$ of the sources ($i_{\rm AB}=24$) 
at $2 < z < 3$. Over the full 16 deg$^2$, the total sample will contain 
about 47,000 galaxies between $2 < z < 4$. Three-hour integration 
times are required to ensure continuum detection, as described below. 
The total cost is 130,000 fiber hours or effectively 11 nights of observing time. 
  
\item Galaxies at $4 < z <6$ with $i<25$ mag will be targeted over the
  16 deg$^2$ survey field with $10$\% sampling. The intrinsic evolution in
  \lya\ luminosity from $4 < z < 6$ will serve as an important
  benchmark for evaluating the impact of reionization on the \lya\
  luminosity function at $z > 5$.  The required integration time for
  this faint sample is 6 hours, again to ensure continuum detection.  We will 
  target a total of 1,600 galaxies in {4,800 fiber hours} ($< 1$ night).

\item
  We will target all LAEs with $z=5.7, 6.6$ to $NB=25$ mag over the
  full 16 deg$^2$ area.  These galaxies are very faint: we will devote
  5 hrs of integration time per galaxy. Their number density is 600
  per PFS FOV for both redshifts combined, so we need {38,000
    fiber hours} (3 nights) for this component of the
  survey. Crucially, we will also study the intrinsic line shapes of
  LAEs at $z=2.2$, for which we can obtain a systemic redshift from
  the \oii\ line.  It is important to quantify the intrinsic
  variability in \lya\ line shape to distinguish from changes caused
  by neutral absorption at the epoch of reionization. We will observe
  these with 2hr integrations down to a limiting \lya\ flux of $2
  \times 10^{-16}$~erg~s$^{-1}$~cm$^{-2}$, corresponding to an \oii\
  flux of $\sim 8 \times 10^{-18}$~erg~s$^{-1}$~cm$^{-2}$.  The
  expected space density is $\sim 390$ per PFS FOV.  We will have a
  final sample of 4800 objects at a sampling rate close to 100\%.  The
  cost of this component will be {9,600 fiber hours} ($<1$ night).

\item
We will also target LBGs over the same redshift range ($2<z<6$) to a
  fainter limiting magnitude of $i_{\rm AB} = 26$ mag (20k galaxies,
  using 10\% sampling) within the HSC Ultradeep area (3.5 deg$^2$).
  These require very deep exposures of 16 hours to spectroscopically
  identify the faintest galaxies yet observed.  Based on previous work
  with much smaller samples, we expect a \lya\ fraction of $\sim 30\%$
  at $z=3$ and at the magnitude limit, rising to $40-50\%$ at $z=6$
  \citep{starketal2011}.  Thus, crudely speaking, we
  can expect a $30-40\%$ redshift success.  This sample can be used to
  constrain the low end of the mass function at high redshift and to
  probe the circumgalactic medium of foreground galaxies as a function
  of impact parameter and galaxy type.  This ultradeep spectroscopic
  component will take {67,000 fiber hours} (6 nights) to target
  $\sim$4,200 galaxies (1,500 redshifts) using a 10\% sampling at 
  $i_{\rm AB}=24-26$ mag over 3.5 deg$^2$.
\end{enumerate}

Combining large samples of bright continuum-selected galaxies at intermediate 
redshifts with LAEs at the highest redshifts will allow us to trace the evolution 
in number-density and clustering of star-forming galaxies all the way to $z=7$, 
as well as to probe cosmic reionization.  Our wedding-cake structure will allow us to 
study both the large-scale distribution of these galaxies, as well as their detailed 
spectral properties for smaller, deeper samples.

\subsubsection{Spectral Measurements}

Here we summarize the main demands on the spectra for the high-z galaxy survey 
for each subsample, and justify that with the depths we plan, we will be able to 
achieve these goals.

\begin{enumerate}

\item
  For the continuum-selected samples, our first goal is to accurately
  measure a spectroscopic redshift.  Our exposure times and depths are
  chosen to detect the continuum at the 3$\sigma$ level.  Based upon
  similar surveys with the VLT, most spectra will have detectable
  spectroscopic features that provide a redshift
  \citep{lefevreetal2005,lillyetal2007}.  In most cases, the presence
  of \lya\ in emission or absorption is the dominant feature. We will
  also search for HeII$\lambda$1640 in emission that may indicate the
  presence of population III stars. The expected equivalent widths for
  HeII range from $1-20$\AA, and have been detected in 10-15\% of
  drop-out samples \cite[e.g.,][]{schaerer2008}. 
In stacks, we will use the strength
  and width of the line to look for signatures of winds (from
  Wolf-Rayet stars) and possibly signs of accretion/gravitational
  cooling.

\item 
  We will search for absorption lines due to the interstellar medium,
  such as O{\small I}, C{\small II}, Si{\small IV}+O{\small IV}, Si{\small
    II} and C{\small IV} are present (see Fig. \ref{fig:simspectralbg}).
  These can provide information on the stellar populations and
  metallicities of high redshift galaxies especially for the rare,
  bright ($i<23$) examples that exist in small numbers in current surveys
  \cite[e.g.,][]{shapleyetal2003}.  We note that for the subsample of
  objects at $z \approx 2$, for which we have both [O {\tiny II}] and
  \lya, we will also know the true rest frame of the galaxy very well.
  For the typical objects, using stacks of $\sim 100$ spectra, we will
  measure line profiles and velocity offsets to indicate the presence
  of inflows or outflows \cite[e.g.,][]{steideletal2010}.

\item  
  Line fluxes of \lya\ emitters will reach a depth, at a $S/N$
  level of 5, of $F = 0.8 \times 10^{-17}$~erg~s$^{-1}$~cm$^{-2}$,
  which is typical of LAEs at $5 < z < 7$ depending on the line
  profile.  We can recover fluxes and linewidths at the $10\%$ level in the
  blue arm and at the $20\%$ level in the red arm.

\item Broad emission lines (FWHM$>1200$ km s$^{-1}$) and the presence of CIV/CIII 
in emission will be used as a signature of AGN activity.

\end{enumerate}

\subsubsection{Target Selection: Quasars down the luminosity function}

Like the LBG samples described above, we will pursue a two-tiered
quasar selection.  We will perform a quasar survey within the BAO
redshift survey over 1400 deg$^2$ and a deeper survey within the 16
deg$^2$ area described here.  Our target strategy is summarized in
Table \ref{tab:sample}.

The PFS-AGN wide survey is based on the wide layer of the HSC
multi-band imaging legacy survey (5-band optical images of $g, r, i,
z,$ and $y$ over 1400 deg$^2$, 
combined with available wide NIR surveys by VISTA and
UKIRT). We focus on the color-selected quasar
candidates at $3 < z < 7$ using color cuts similar to those used by
the SDSS \citep{richardsetal2002}.  Within the BAO survey, we will
target a few tens of quasar candidates per PFS pointing to a magnitude
limit of $J \leq 24$ with half-hour integration times, to achieve the
number-density of quasars shown in Table \ref{tab:sample}.  The
primary goals include identifying $6 < z < 7$ candidates for deeper
spectroscopy to study reionization, measuring the quasar luminosity
and mass function over $3.5 < z < 7$, and measuring the quasar auto-
and cross-correlation functions.  The total number of targets is
small, and it will be integrated into the wide-field cosmology survey.
As shown above, we can achieve these goals in 0.5 hr integration times
even at our magnitude limit (Fig.~\ref{fig:simspectraqso}).
We note that the PFS cosmology survey requires 2 pointings per field,
each of which has 0.5 hr integration time, giving us time to observe quasars.

The PFS-AGN deep survey is based on the deep layer of the HSC legacy
survey.  In addition to color selection, we will also target variable
objects and X-ray and mid-infrared selected objects.  We will go a
magnitude deeper than the PFS-AGN wide survey, and expect to target
100 active galaxies per deg$^2$.  The goals include finding $\sim 10$
faint $z \approx 7$ quasars, studying X-ray and variability selected
active galaxies at $1 < z < 4$, cross-correlating absorption systems
in the quasar spectra with the foreground galaxy population, and
studying obscured populations in the galaxy survey that are identified 
via narrow emission lines.  If we spend an hour per target for the majority 
of these targets, then it will take very few fibers from the main galaxy 
evolution survey.

\subsubsection{Summary of the Survey}

%\begin{table}[t]
%{\small \begin{center}
%{Survey Parameters}\\
%\label{tab:gal_survey}
%\begin{tabular}{ll}\hline\hline
\begin{deluxetable*}{ll}
\tablewidth{0pt}
\tabletypesize{\footnotesize}
\tablecaption{Survey Parameters
\label{tab:gal_survey}
}
\startdata \hline\hline 
$1 < z < 2$  &   \\
Selection (optical)     & HSC Deep: $i<27.2$ mag\\
Selection (NIR)     & {\it XMM}-LSS, ELAIS-N1, and DEEP2-3: $J_{\rm AB}<23.4$ mag \\
Exposure time at flux limit & 3 hrs \\
Area & 16 deg$^{2}$ \\
Number of nights & 69 nights\\
\hline
Flux-limited survey & \\
Selection (NIR)     & {\it XMM}-LSS, ELAIS-N1, and DEEP2-3: $J_{\rm AB}<23.4$ mag \\
Exposure time at flux limit & 3 hrs \\
Area & 2 PFS pointings \\
Number of nights & 8 \\
\hline 
High-redshift galaxies & \\
Selection (optical)     & HSC Deep: $i<27.2$ mag\\
Selection (narrow-band) & HSC Deep: $NB < 25$ mag \\
Exposure time at flux limit & 3-16 hrs \\
Area & 16 deg$^{2}$ \\
Number of nights & 23 \\
\hline
Total number of nights & 100 
%\\
% \hline\hline
\enddata
\end{deluxetable*}
%\end{tabular}
%\end{center}}
%\end{table}

In Table \ref{tab:sample} we show all the sub-samples that we plan to
target, their expected number density per square degree, the required
exposure times, and the proposed survey area.  Throughout, when we
quote the required number of nights, we are assuming 2000 science
fibers per PFS pointing and no overhead in changing targets.  We
quote the number of required clear nights (that is, no weather factor
has been applied).  We outline a two-tiered survey, with our primary
galaxy survey covering {16 deg$^2$} and a deeper component
covering {3.5 deg$^2$}.  To summarize, we will use the following
selections to trace the evolution of galaxies from cosmic dawn to the 
peak epoch of stellar mass build-up:
\begin{itemize}

\item
  We will perform a continuum-selected survey of $1 < z < 2$ galaxies
  with $J_{\rm AB} < 23.4$ mag over 16 deg$^2$ and $0 < z < 1$
  galaxies with $J_{\rm AB} < 21$ mag. We will target $\sim 28,000$
  galaxies per deg$^2$, requiring multiple pointings per field.  
  The primary survey will be color-selected,
  using a combination of HSC and NIR imaging.  We will also
  perform a purely magnitude-limited survey over two PFS pointings early in
  the survey to calibrate our selection strategy.

\item
  At $2 < z < 7$, we trace the galaxy population using bright
  dropouts down to a limiting magnitude of {$i<24$} mag over the
  full 16 deg$^2$.  We expect to target $430$ galaxies per deg$^2$.
  We will survey much fainter LBGs to $i<26$ mag over 3.5 deg$^2$ to
  study intervening absorption systems.

\item
  We focus on LAEs at $z=5.5,6.6,7.3$, in order to use the \lya\
  emission as a probe of reionization. Additionally we will study LAEs
  at $z \sim 2.3$, where we can use the \oii\ emission line to measure
  the systemic redshift.

\end{itemize}

Adopting these survey areas and target densities, the survey requires
100 clear nights. The survey parameters are summarized in Table \ref{tab:gal_survey}.

\subsection{Requirements}

We are assuming the nominal sensitivities and resolution of PFS.  The
feasibility of our survey relies most heavily on the sensitivity of
PFS in the red and NIR arms, from which we will derive redshifts and
key diagnostic information.  Our numbers are summarized in Table
\ref{tab:gal_req}. Our most important assumptions are:

\begin{enumerate}

\item
  We assume 2000 science fibers per PFS pointing (with 400 fibers
     devoted to sky).  We also assume that it is 
  possible to reposition fibers rapidly, so that we move off of bright 
  targets in a short time.  This large multiplexing 
  capability is crucial for our science case, otherwise we cannot achieve 
  the large areas and high sampling rates that make our survey unique.

\item 
  We assume that we can measure continuum-based redshifts at our 
  magnitude limit of $J_{\rm AB} = 23.4$ mag.  Based on our spectral 
  simulations, we find that in a 3 hr integration we will achieve a $S/N$ of at least five in 
  the 4000~\AA\ break in the NIR arm to measure continuum-based redshifts 
  \citep{tonrydavis1979}.  We suggest real-time 
  spectral analysis to identify low redshift interlopers after the initial 
  20 min integration.

\item
  We expect to reach a 5$\, \sigma$ line-flux limit of at least
  $10^{-17}$ erg s$^{-1}$~cm$^{-1}$ over $>75\%$ of the NIR arm.  We
  estimate that this limit is reached in the NIR arm in 3 hrs of
  integration on the \oii\ line. We are assuming intrinsic line widths
  in the galaxies of $\sigma_{\rm gas} \approx 100$~\kms.  The $S/N$
  will degrade by $\sim 30\%$ for a galaxy with twice the dispersion,
  as the light is spread over roughly twice as many pixels. This gets
  us to $L_{\rm [OII]}\sim 3 \times 10^{41}$ erg s$^{-1}$ at $z
  \approx 2$, which is close to the \oii\ luminosity of a typical $L^*$
  galaxy at this epoch \cite[e.g.,][]{takahashietal2007}.

\item
  In the blue and red arms, our detection limit is deeper. 
  We find that we can detect \lya\ with $7 \times 10^{-18}$ erg
  s$^{-1}$~cm$^{-2}$ in the blue arm with $R \approx 1300$ 
  at the 5$\sigma$ level in 6 hrs of 
  integration.  Again we assume an intrinsic gas dispersion of 100~\kms, 
  which is typical of measured LAEs \citep{ouchietal2010}.  We are thus detecting 
  sources with $L > 1.5 \times 10^{42}$~erg~s$^{-1}$, which is a factor of 
  five fainter than the typical LAE at these redshifts \citep{ciardulloetal2012}.
  In the red arm, our limits are slightly better.  In 6 hrs our 5$\sigma$ 
  detection limit is $5 \times 10^{-18}$ erg s$^{-1}$~cm$^{-2}$.  
  Here our limits of $8.5 \times 10^{42}$~erg~s$^{-1}$ at $z=5$ and 
  $2 \times 10^{43}$~erg~s$^{-1}$ at $z=6$ are around or slightly exceed $L^*_{\lya}$ at 
  these epochs \citep{ouchietal2010}.  Our limits are comparable to those 
  achieved by \citet{ouchietal2010} for an individual object, and we will 
  increase the sample size by two orders of magnitude.

\item
  In order to achieve the above continuum and emission-line limits above, 
  we place strong requirements on our software, specifically in regard to 
  sky subtraction in the red and NIR arms.  We assume that we are able to 
  determine the sky level at a given time to $0.5\%$.  This is 
  supported by the Poisson noise in the lines themselves, and experience with BOSS 
  \citep{boltonetal2012}. Also,
  we assume that $0.5\%$ of the light hitting the detector is 
  distributed across the entire chip as a smooth background.  We show that 
  our primary science goals will succeed if the background subtraction is poorer and 
  scattered light fraction higher than the specifications.  However, our more challenging 
  goals of measurement continuum properties of $z \approx 1.5$ galaxies do require the 
  three-hour integration times proposed here.

\end{enumerate}

%%%%%%%%%%%%%%%%%%%%%%%%%%%%%%%%%%%%%%%%%%%%%%%%%%%%%%
\begin{figure*}[t]
{\centering 
\includegraphics[width=0.45\textwidth,angle=0]{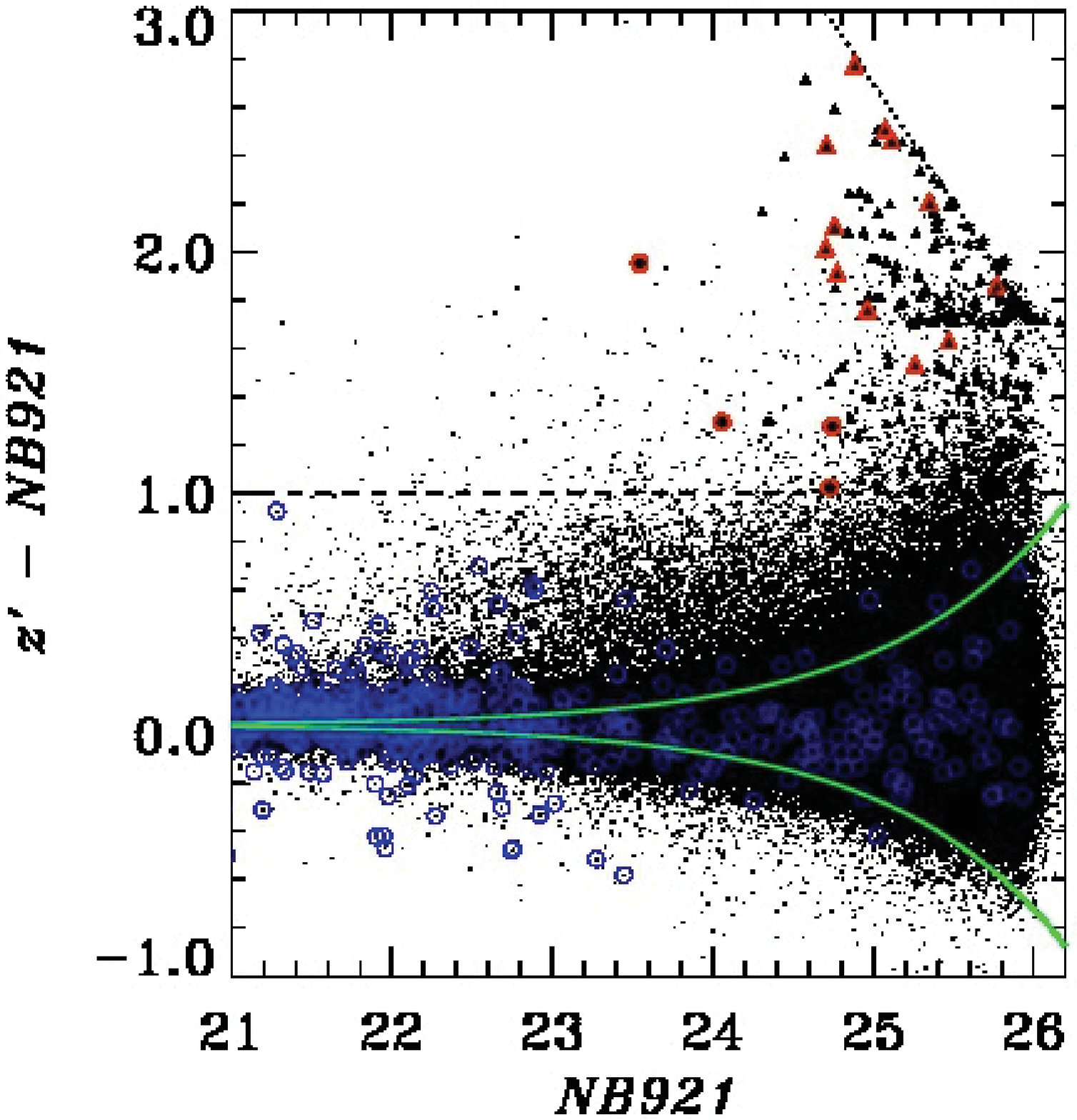}}
{\centering 
\includegraphics[width=0.45\textwidth,angle=0]{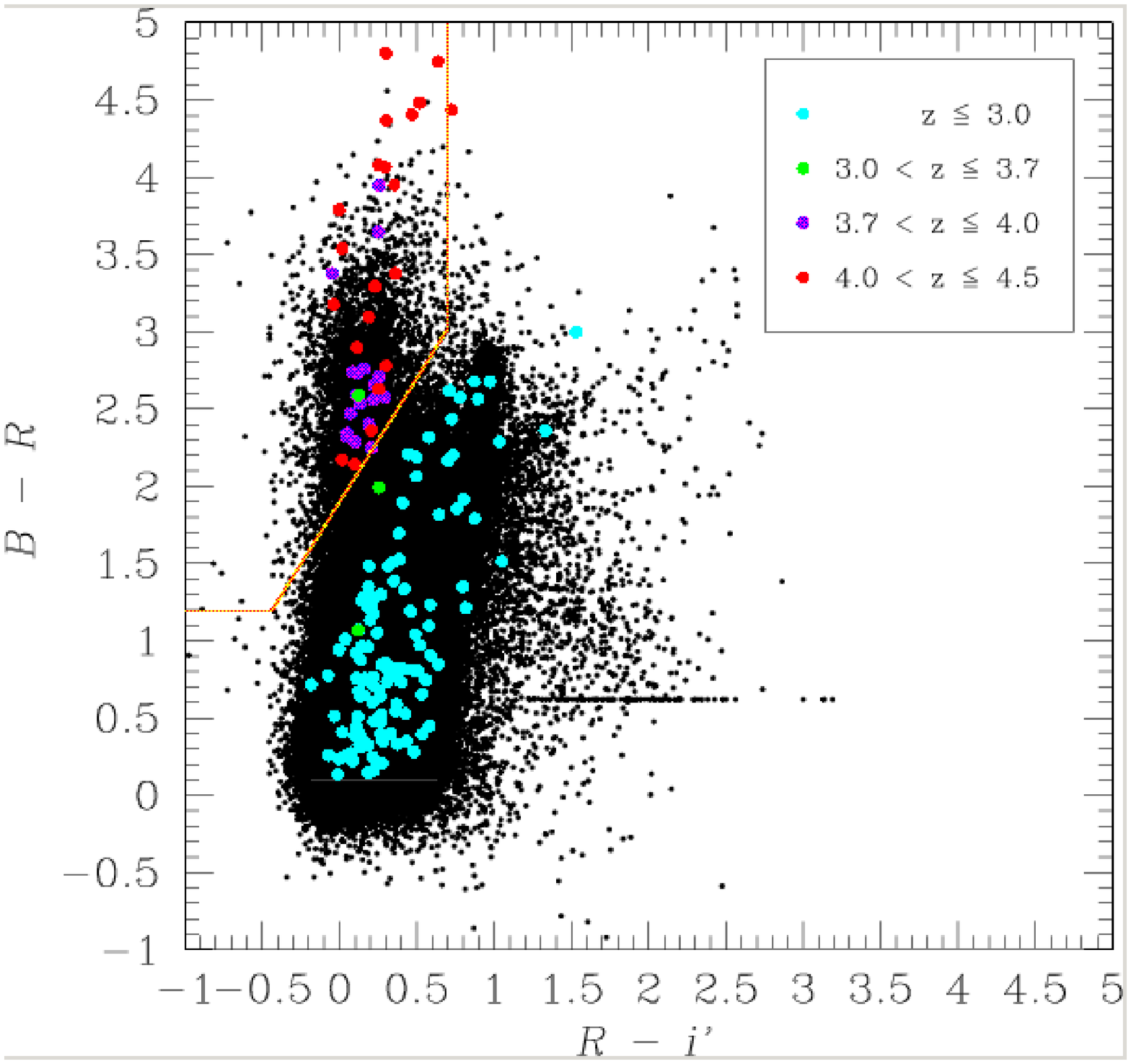}}
\caption{ 
%\small 
{\it Left}: 
Color magnitude diagram of narrow-band excess color ($z'-NB921$)
vs. narrow-band magnitude ($NB921$) for selecting $z=6.6$ LAEs in the
SXDS field \citep{ouchietal2010}, which will be applied to the LAE
selection of our PFS survey.  Black dots present colors of all the
detected objects, while black filled circles and triangles denote the
$z=6.6$ LAEs.  Red and blue open symbols mark
spectroscopically-identified objects in the redshift range of LAEs
($z=6.45-6.65$) and interlopers, respectively.  The green lines
indicate $2\sigma$ errors of the color of $z'-NB921$.  Dashed and
dotted lines represent the color cut for the narrow-band excess and
the $1\sigma$ limit of $z'$ data, respectively.
{\it Right}: 
Two color diagram of objects down to $i'\le 26.85$ to select $z\sim 4$
LBGs in the Subaru Deep Field \citep{yoshidaetal2006}.  The horizontal
sequence along $B-R = 0.62$ shows objects which are fainter than the
$1\sigma$ magnitude in both $B$ and $R$.  The colored symbols show
objects with spectroscopic redshifts, where cyan, green, violet, and
red represent objects in the range $z<3.0$, $3.0\le z<3.7$, $3.7\le
z<4.0$, and $4.0\le z<4.5$, respectively.  The thick orange line
indicates the boundary which are adopted for the selection of $z\sim
4$ LBGs. Our PFS survey will use a broad-band set of $g'r'i'$
different from this $BRi'$ bands for selecting $z\sim 4$ LBGs.
However, the contamination rates of our $z\sim 4$ LBG sample will be
comparable, as demonstrated by another $z\sim 4$ LBG selection with
$GRI$ bands (Steidel et al. 1999).
\label{fig:LAE_LBG_selections}}
\end{figure*}
%%%%%%%%%%%%%%%%%%%%%%%%%%%%%%%%%%%%%%%%%%%%%%%%%%%%%%

%\begin{table}[t]
%\begin{center}
%{Galaxy Survey Requirements}\\
%\begin{tabular}{ll}\hline\hline
\begin{deluxetable*}{ll}
\tablewidth{0pt}
\tabletypesize{\footnotesize}
\tablecaption{Level 2 Galaxy Survey Requirements
\label{tab:gal_req}
}
\startdata \hline\hline 
Number of science fibers & 2000  \\
Field of view                     & 1.3 deg. diameter \\
Exposure time for flux limits & 3 hrs \\
Fraction of resolution elements above flux limit & $75\%$ \\
\hline
{\bf Blue Arm}                           &   \\
Wavelength coverage (\AA)        & 3800-6700 \\
Spectral Resolution, continuum  & $R \approx 400$  \\
5$\sigma$ Flux limit, continuum     & $24.5$ AB mag \\
Spectral Resolution, emission lines  & $R \approx 1300$  \\
5$\sigma$ Flux limit, line ($\sigma_{\rm gas} = 100$~\kms) & 
$10^{-17}$~erg~s$^{-1}$~cm$^{-2}$ \\
\hline
{\bf Red Arm}                           &   \\
Wavelength coverage (\AA)        & 6500-10000 \\
Spectral Resolution, continuum   & $R \approx 300$  \\
5$\sigma$ Flux limit, continuum     & $24.5$ AB mag \\
Spectral Resolution, emission lines & $R \approx 1300$  \\
5$\sigma$ Flux limit, line ($\sigma_{\rm gas} = 100$~\kms) & 
$7 \times 10^{-18}$~erg~s$^{-1}$~cm$^{-2}$ \\
\hline
{\bf NIR Arm}                           &   \\
Wavelength coverage (\AA)        & 10000-12600 \\
Spectral Resolution, continuum  & $R \approx 300$  \\
5$\sigma$ Flux limit, continuum     & 24 AB mag \\
Spectral Resolution, emission lines  & $R \approx 1300$ for line calculations \\
5$\sigma$ Flux limit, line ($\sigma_{\rm gas} = 100$~\kms) & 
$10^{-17}$~erg~s$^{-1}$~cm$^{-2}$ \\
\hline
Sky continuum                           & 21.56 mag AB arcsec$^{-2}$ at 1$\micron$ at zenith \\
OH line brightness                     & 16.6 mag AB arcsec$^{-2}$ at $J-$band at zenith \\
\hline
%Systematic in sky subtraction $\delta \lambda$        & 0.01 pixel \\
Systematic in sky subtraction $\delta$ sky flux        & $0.5-2\%$  \\
Overall Sky level determination                                             & $0.5\%$ \\
Scattered light                            & $23-19.75$ AB mag 
%\\ \hline\hline
\enddata
%\end{tabular}
%\caption{
\tablecomments{
PFS instrument parameters that we assume in calculating simulated spectra and 
assessing feasibility.  We show only parameters that (a) are important in our 
calculations and (b) differ from the assumptions in Table~\ref{tab:pfs_spec}. 
%of the SRD-Cosmology 
%document.
%\label{tab:gal_req}
}
\end{deluxetable*}

\section{Science Requirements}
\label{chap:sr}

We now summarize the science requirements arising from the
earlier sections. These requirements are necessary for deriving 
lower level instrument requirements and in understanding the 
flow of requirements from the science to detailed engineering.  
Establishing the science requirements flow is likewise essential
in order to track the science impact of potential changes to the instrument
configuration in the future.  The top level science requirements and 
observing requirements are intended to be directly verifiable items
in the project. For convenience, we list the overall drivers, item by item, 
both in tabular form and in more detail below.

The top-level science requirements for the PFS surveys described in the
previous chapters are summarized in Table~\ref{tab:sr_all}\footnote{Our
instrumental requirements for the different scientific goals outlined in
this document are somewhat different.  In particular, the galaxy
science, which requires the deepest exposures, have the strongest
requirements on systematics such as stray light and sky subtraction.
The overall  requirements listed in Table~\ref{tab:sr_all} represent a
compromise of sorts between the needs of the different scientific goals.
As the instrument design becomes more solid, we will see to what extent
the more stringent requirements of the galaxy evolution science are
possible.}. Note that we gave the requirements for the MR mode 
in Table~\ref{tab:GA_requirements}, which is derived from the GA science. 

\begin{table*}[t]
\begin{center}
\caption{\bf Top-Level Science Requirements}
\label{tab:sr_all}
{\small 
\begin{tabular}{l|l|l}\hline\hline
{\bf Requirement} & {\bf Value} & {\bf Main Driver(s)} \\ \hline
Wavelength Coverage & 380 -- 1260nm & All surveys \\ \hline 
Spectral Resolution & Blue: $R\sim 2500$& Matched survey efficiency \\
& Red: $R\sim 3000$ & sky subtraction,  stellar metallicities 
\\
& NIR: $R\sim 4000$ &   radial velocity precision
\\ \hline
Fiber Density & $\ge 2400$ per 1.3 deg. diameter 
 & $\bar{n}_g P_g \simeq $ a few
	 $@k=0.1~h/{\rm Mpc}$ \\ 
& hexagonally-shaped field & requirement in BAO survey \\ \hline
Throughput$^a$ & Average: $\ge$22\% (blue)& Matched survey efficiency
\\
&\hspace{4em}$\ge $22\% (red) &\\
&\hspace{4em}$\ge 24$\% (NIR) &\\
& Worst part of band: $\ge$20\% (blue)& \\
&\hspace{4em}$\ge $20\% (red) &\\
&\hspace{4em}$\ge 18$\% (NIR) &\\ \hline
Fiber Reconfig. Time &$\simlt 3$min in total &  15min exposure in BAO
	 survey  \\ \hline 
Sky Subtraction Accuracy &$\simlt 1$\% of sky background 
& in all surveys \\ 
& per 4-pixel
     resolution element
& \\ \hline 
Stray Light & $\simlt $ a few \% of total sky brightness & in all
	 surveys \\ 
& spread over
     detector 
& \\ \hline 
Read Out & 
$\le 3$ (Red) $e^{-}$ rms per pixel 
& in all surveys \\ 
& $\le 4$ (NIR) $e^{-}$ & \\ \hline 
Wavelength Calibration & $\simlt 0.1$\AA~ & 
for velocity precision 
\\
& \hspace{0.5em}($\sim$1/20th of a resolution
     element) 
& 
\\ \hline
Fiber Diameter & 1.13$^{\prime\prime}$ deg. diameter
 at the field center& Optimal $S/N$ for galaxy survey
 \\
&$1.03^{\prime\prime}$ at the edge &
\\ \hline
Pre-Imaging Data & $gri$ to 25mag over 1400 deg$^2$& Color selection of 
ELGs for BAO survey\\ 
& $griz$+$NB515$ 
survey to $V\simeq 22.5$mag & Selection of stars in Milky Way/M31\\
&$grizy$+NBs+$J$ to $\sim 27$mag over $16$ deg$^2$&  
Color selection for galaxy survey 
\\ \hline
Survey Area & $\sim$1400 deg$^2$ for BAO & 
Statistical requirements, cosmic \\
& $\sim $500 deg$^2$ for GA survey & variance, matching HSC data \\
& $\sim 16$ deg$^2$ for galaxy survey & \\
 \hline\hline
\multicolumn3l{\scriptsize $^{\rm a}$
Excludes atmosphere, central obscuration + WFC vignetting, and fiber
 aperture
effect. 
} \\
\end{tabular}
} 
\end{center}
\end{table*}

We briefly discuss the individual entries in Table~\ref{tab:sr_all}.

\begin{itemize}
\item{}{\it Wavelength Coverage:} The triple-armed spectrograph design offers complete 
and simultaneous spectral coverage from 380 through 1260 nm and clearly this represents
a unique feature of PFS. The driver for this requirement arises from all 3 surveys discussed
in Chapters \ref{chap:cosmology}--\ref{chap:galaxy}
%2-4 
but particularly the cosmology and galaxy surveys. The cosmology survey
will enable the first BAO and RSD tests beyond $z\simeq$2 and the galaxy survey will 
track emission line properties of an unprecedented sample of galaxies with no `gaps'
in redshift coverage.

\item{}{\it Spectral Resolution:} The resolutions of the three arms of each spectrograph are
designed to ensure optimal efficiency in the study of stars and galaxies over the wide
wavelength range. The resolution must be higher in the red and near-infrared arm to
ensure optimal sky subtraction in the dense OH forest. The minimum effective resolution
in the red arm is set by our requirement to measure accurate velocities and stellar
metallicities for the Galactic surveys. 

\item{}{\it Fiber Density:} The fiber density is set by the field of view of the provided Subaru prime
focus corrector and our requirement that we optimally sample the 0.8$<z<$2.4 galaxy
population for the cosmology survey at the wavenumber $k$ where the BAO signal is
to be recovered. Formally, this sets a requirement that the mean number
     density of target galaxies $\bar{n}_g$ and
the galaxy power spectrum $P_g(k)$ should satisfy $\bar{n}_gP_g\simeq $1 (or a few) at $k\simeq$0.1 $h^{-1}$ Mpc.

\item{}{\it Throughput:} The throughput requirement is driven by a combination of the S/N required 
for the various surveys and the available observing time, and is consistent with a preliminary 
assessment of the instrument performance. The values in the table exclude atmosphere, 
central obscuration, vignetting, and the fiber aperture effect. 

\item{}{\it Fiber Reconfiguration Time:} The reconfiguration time is set by the
cosmology survey. 
It has a short exposure time, because the cosmology survey
must
survey a large volume by targeting strong line emitting galaxies. If the exposure time is $\simeq$15 minutes,
a reconfiguration time (including slewing, acquisition, guiding, fiber positioning
and verification) should be completed in under 3 minutes to maximize 
survey efficiency. 
%minimize survey inefficiency.

\item{}{\it Sky Subtraction Accuracy:} This is set by the requirement for high
quality spectra at faint limits in all surveys. It is particularly driven by the need
for accurate absorption line measures in the red and near-infrared spectral
regions where the OH forest is dense.

\item{}{\it Stray Light:} As for sky subtraction, given the intensity of the OH
night sky emission.

\item{}{\it Detector Read Out:} This is set by the signal/noise requirements
of each survey, noting their likely photon and sky-noise components.

\item{}{\it Wavelength Calibration:} This is determined by the radial velocity
accuracy requirement for the Galactic surveys.

\item{}{\it Fiber Diameter:} This is determined by the optimal signal/noise
requirements of the galaxy and cosmology survey, noting the distribution
of line emission in these extended sources.

\item{}{\it Pre-Imaging Data:} This is determined by each individual
survey and the various color-criteria to be adopted in optimally selecting
either galaxies in a particular redshift range with certain star formation
characteristics or stars of a given luminosity class in the Milky Way
or M31. The aim is to ensure that all the relevant data not currently in hand
is taken by HSC.

\item{}{\it Survey Area:} This varies from one survey to another but
is driven by statistical and/or cosmic variance requirements and the availability of HSC
imaging data.
\end{itemize}

%\chapter{OUTSTANDING ISSUES}
\section{Outstanding Issues}
\label{chap:issues}

Our primary goal has been to present the science case for PFS in the
context of an anticipated 300-night Subaru Strategic Program (SSP)
and, thereby, to define the scientific requirements for the instrument as summarized in the previous
section. The successful review of this science case and its associated technical
document has led to an agreement to begin PFS construction with an anticipated
first light in early 2017.

Science planning will naturally continue 
%through to the Preliminary Design
%Review (anticipated in early 2013) and 
and here we briefly describe those activities in the
context of instrumental choices, the design of the survey and the broader impact of 
PFS beyond the initial SSP.

It should be understood that PFS planning involves two rather
different communities: the international partnership focused
scientifically via the PFS Survey Committee, and the Japanese
astronomical community, whose interests are coordinated by the National
Astronomical Observatory of Japan and its appointed national committees.

In the following we briefly introduce some of the items which will form
the basis of the project discussions.

\subsection{Requirements Flowdown}

As discussed in Section~\ref{chap:sr},
%5, 
for a project with the size and complexity of PFS, formal science and
instrument requirements are necessary to guide the design and
construction of the system, and allow the rapid assessment of design or
performance changes during the course of the project. They will also
guide the instrument test plan to be developed at a later stage in the
project; each requirement will include the method for validating that it
has been met. The full documentation of formal and tracked requirements
%will be completed by the PDR, and 
will define the required performance
of the instrument system and the subsystems in order to achieve the
science goals.

The requirements can be broken into several levels, and we will use the
following definitions.  The Level 1 (L1) requirements are the top level
science requirements that comprise the reason for building the
instrument. The Level 2 (L2) requirements are ones that are imposed on
the entire instrument or multiple subsystems.  Requirements on a single
subsystem are Level 3 (L3) requirements.  Level 4 (L4) requirements are
the requirements created by the subsystem designers in order to
implement the Level 3 requirements.

Although the top level science requirements, L2, are summarized in
Section~\ref{chap:sr},
%5,
they will be refined 
%and formalized by PDR 
(the L1 requirements are described in detail in
Sections~\ref{chap:cosmology} -- \ref{chap:galaxy} for each science case).  
%L2 requirements are also given in Chapter~\ref{chap:sr}.
%5. 
These focus on the science driven requirements and may not yet be
complete. Specifically, they do not yet include all of the instrument
level operational and interface requirements. The L3 and L4 requirements
are in the process of being developed. Although there has been extensive
discussion of these requirements, they are not yet ready for this
documentation.

The PFS project office will lead the effort to refine, formalize, and
document the L1-L4 requirements, in coordination with the science and
instrument working groups. This process will be complete for the L1-L3
requirements soon.
%, while the definition of L4 requirements may evolve after the PDR.  
The documentation will consist of a formal list
of individual requirements with a description, rationale, and validation
method listed for each requirement. The rationale will consist of a
short paragraph explaining why the requirement is needed, any
assumptions that were made, and what design effort drove the
requirement.  It puts the requirement in context, and helps reduce
ambiguities and misunderstandings.  All of the instrument requirements
are intended to be verified during the acceptance testing of the
instrument, under a written test plan.

\subsection{Planning the Subaru Strategic Program}

The three Working Groups have, thus far, largely worked independently in defining their
science requirements and there has been little discussion of how to integrate
the 3 surveys into a single coherent observing program. Ultimately
the PFS partnership will have to take the programs described in
Sections~\ref{chap:cosmology} --
\ref{chap:galaxy}
%2-4 
forward and write one or more Subaru Strategic Programs for consideration
by the NAOJ Subaru Time Allocation Committee.

Issues that the PFS team will address in discussions with NAOJ and the HSC team 
will include:

\begin{itemize}

\item{} Selecting fields within the HSC imaging survey

\item{} Science prioritization of the various survey components

\item{} Target sharing and exposure time optimization

\item{} Bright and dark time balance on the telescope

\end{itemize}

The bright and dark time balance will depend critically on the question
of how the Subaru Observatory intends to operate PFS alongside HSC and
the bright time use of PFS will, in turn, depend on whether there is a
feasible intermediate dispersion capability.

\subsection{Legacy Value}

The Subaru archive is formally a responsibility of NAOJ, but the quality
and delivery of the survey products are that of the current PFS science team. 
No negotiations have yet taken placed on this important issue, but
we expect
 a full agreement on responsibilities, proprietary periods,
international access etc.
% by the time of the PDR.

\subsection{Synergy with Other Facilities}

PFS will have a significant impact beyond
the surveys discussed in this document in two ways.

Firstly, the survey fields will deliver spectroscopic data
of value to extant imaging facilities. During the pre-LSST era 
we can expect inquiries from those international teams for access to PFS data.
Such access could be dealt with largely under the auspices
of the Subaru archive discussed above, but there may be
cases where new collaborations will be valuable, either to the
PFS team, the Japanese community or both. Such opportunities
will need to be considered and the roles and rights of the 
PFS team, NAOJ etc fully understood.

A potentially more interesting opportunity is the likelihood
of new collaborations with both current imaging survey telescopes
and future facilities such as LSST, Euclid and perhaps ultimately TMT, 
beyond the scope of the Subaru Strategic Program (SSP) introduced
here. Formally, beyond the SSP, PFS will become the property
of NAOJ but the current partnership may have the opportunity
to participate and contribute in future surveys conducted in
conjunction with other facilities.

\section{Summary}

Our intention via this document is to describe in as much detail
as practical, the exciting scientific programs we have designed for
the Subaru Prime Focus Spectrograph, as well as to present our
first iteration for a Subaru Strategic Program (SSP) of $\simeq$300 nights
that would realize these goals. The flow-down from these science
requirements has been summarized in the context of the top-level
instrument requirements which provide important constraints on
the technical design of each component of PFS.

PFS will likely be the first massively-multiplexed spectrograph on a
large aperture telescope to achieve first light. Its unique capabilities on an 
8 meter platform at an excellent site promise exciting discoveries in 3 broad
areas: the nature of dark energy through an ambitious survey
of emission line galaxies over an unprecedented redshift
range 0.8$<z<$2.4, the assembly history of the Milky Way
and M31 through strategically-designed programs that
exploit astrometric data from the Gaia mission, and galaxy
evolution over a wide range in redshift (1$<z<7$) exploiting
PFS' unique wide wavelength coverage.

Although our initial goal is to construct the instrument and
conduct a refined version of the SSP presented in this
article, we are strongly motivated to consider the longer
term role of this survey instrument in the era of LSST, Euclid and
TMT.

\bigskip
This work is
supported in part 
by the JSPS
Core-to-Core Program ``International Research Network for Dark
Energy'',
by World Premier
International Research Center Initiative (WPI Initiative), MEXT, Japan, 
and by the FIRST program
``Subaru Measurements of Images and Redshifts (SuMIRe)'', CSTP, Japan.

\appendix

\section{PFS collaboration}
\label{app:PFSteam}

\noindent{\bf Principal Investigator (PI)}\\
\noindent\begin{tabular}{ll}
Hitoshi Murayama & Kavli Institute of the Physics and \\
& Mathematics for the
	      Universe (Kavli IPMU, WPI), Japan;  \\ 
& Physics Department, University of California, Berkeley; \\
& Lawrence Berkeley National Laboratory, Berkley, USA
\end{tabular}
\bigskip

\noindent{\bf Co-Chairs of the PFS Survey Committee}\\
\noindent\begin{tabular}{ll}
Richard Ellis  & Caltech, USA \\
Masahiro Takada & Kavli IPMU, Japan
\end{tabular}
\bigskip

\noindent{\bf Co-Chairs of the PFS Science Working Groups}\\
\noindent\begin{tabular}{ll}
\hspace{1em}{\bf PFS Cosmology WG} \\
\hspace{4em} Masahiro Takada & Kavli IPMU, Japan \\ 
\hspace{4em}  Christopher Hirata & Caltech, USA \\
\hspace{4em}  Jean-Paul Kneib & LAM, France \\
\hspace{1em}{\bf PFS Galactic Archaeology WG} \\
\hspace{4em} Masashi Chiba & Tohoku University, Japan \\ 
\hspace{4em}  Judith Cohen & Caltech, USA \\
\hspace{4em}  Rosemary Wyse & JHU, USA \\
\hspace{1em}{\bf PFS Galaxy WG} \\
\hspace{4em} Jenny Greene & Princeton University, USA \\ 
\hspace{4em} Kevin Bundy  & Kavli IPMU, Japan\\
\hspace{4em} John Silverman  & Kavli IPMU, Japan\\
\hspace{4em} Masami Ouchi  & The University of Tokyo, Japan\\
\hspace{1em}{\bf PFS AGN/QSO WG} \\
\hspace{4em} Tohru Nagao & Kyoto University, Japan\\
\hspace{4em} Michael Strauss  & Princeton University, USA\\
\end{tabular}

\bigskip

\noindent{\bf PFS Steering Committee}

\noindent 
H. Aihara (U. Tokyo/Kavli IPMU), N. Arimoto (NAOJ),
R. Ellis (Caltech), T. Heckman (JHU), P. Ho
(ASIAA), O. Le Fevre (LAM), H. Murayama (Kavli IPMU), L. Sodre Jr. (Sau
Paulo), 
M. Seiffert (JPL/Caltech),
D.~N.~Spergel (Princeton), Y. Suto (U. Tokyo), 
H. Takami (NAOJ)

\bigskip

\noindent{\bf Members in the PFS Science Working Groups}

\noindent L. R. Abramo (Sao Paulo),
H. Aihara (U. Tokyo/Kavli IPMU),
W. Aoki (NAOJ),
N. Arimoto (NAOJ),
C. Bennett (JHU),
T.-T Chang (AIAA),
M.-Y. Chou (ASIAA),
J. Coupon (ASIAA),
R. Dekany (Caltech),
C.~M. de Oliveira (Sao Paulo),
O. Dore (JPL/Caltech),
T. Goto (Hawaii),
J. Gunn (Princeton),
K. Hayashi (Tohoku),
S. Hayashi (NAOJ),
T. Heckman (JHU),
C. Hikage (Nagoya U.),
B.-C. Hsieh (ASIAA),
K. Ichikawa (Kyoto U.),
K. Ichiki (Nagoya U.),
M. Imanishi (NAOJ),
M. Ishigaki (NAOJ),
M. Iye (NAOJ),
I. Iwata (NAOJ),
O. LeFevere (LAM),
L. Lin (ASIAA),
Y. T. Lin (ASIAA),
I. Kayo (Toho U.),
H. Karoji (Kavli IPMU),
N. Katayama (Kavli IPMU),
D. Kawata (UCL),
E. Kirby (Caltech),
T. Kodama (NAOJ),
Y. Komuro (Tohoku U.),
K. Maeda (Kavli IPMU),
R. Mandelbaum (Princeton),
T. Marriage (JHU),
T. Matsubara (Nagoya U.),
K. Matsuoka (Ehime U.),
B. Menard (JHU),
S. Mineo (U. Tokyo),
H. Miyatake (U. Tokyo),
A. More (Kavli IPMU),
S. More (Kavli IPMU),
T. Morokuma (U. Tokyo),
K. Motohara (U. Tokyo),
L. Moustakas (JPL),
Y. Matsuda (Caltech),
S. S. Murray (JHU),
F. Nakamura (NAOJ),
T. Nishimichi (Kavli IPMU),
A. Nishizawa (Kavli IPMU),
M. Oguri (Kavli IPMU),
S. Okamoto (Beijing),
Y. Okura (NAOJ),
Y. Ono (U. Tokyo),
P. Price (Princeton),
A. R. Pullen (JPL),
R. Quimby (Kavli IPMU),
A. Raccanelli (JPL),
%A. Reccanelli (JPL),
M. Sato (Nagoya U.),
M. Seiffert (JPL),
A. Shimono (Kavli IPMU),
D. N. Spergel (Princeton),
D. K. Stern (JPL),
H. Sugai (Kavli IPMU),
S. Saito (Berkeley),
T. Saito (Kavli IPMU),
T. Storchi-Bergmann (IF-UFRGS, Brazil),
Y. Suto (U. Tokyo),
N. Suzuki (LBL),
A. Szalay (JHU),
R. Takahashi (Hirosaki U.),
N. Takato (NAOJ),
M. Tanaka (Kavli IPMU),
M. Tanaka (Tohoku)
A. Taruya (U. Tokyo),
N. Tominaga (Konan U.),
Y. Urata (NCU),
K. Yamamoto (Hiroshima U.),
K. Wada (Kagoshima U.),
W.-H. Wang (ASIAA),
R. Wyse (JHU),
T. Yamada (Tohoku U.),
N. Yasuda (Kavli IPMU),
N. Zakamska (JHU)

\bigskip

\noindent{\bf PFS Project Office}\\
\noindent H. Karoji (Kavli IPMU),
H.-H. Ling (ASIAA),
Y. Ohyama (ASIAA),
H. Sugai (Kavli IPMU),
A. Shimono (Kavli IPMU),
N. Takato (NAOJ),
N. Tamura (Kavli IPMU),
A. Ueda (NAOJ)

\end{document}